\documentclass[12pt,english,floatfix,nofootinbib,superscriptaddress,aps,prd,preprint]{revtex4}
\usepackage[utf8]{inputenc}
\usepackage{float}
\usepackage{array}
\usepackage{bbold}
\usepackage{lipsum}
\usepackage{dsfont}
\usepackage{graphicx}
\usepackage{amsmath,amsthm,amsfonts,amssymb}
\usepackage{graphicx}
\usepackage[english]{babel} 
\usepackage{color}
\usepackage{tensor}
\usepackage{esint}
\usepackage[dvips]{epsfig}
\usepackage[dvips]{graphicx}
\usepackage{float}
\usepackage{units}
\usepackage{textcomp}
\usepackage{mathrsfs}
\usepackage{amsmath}
\usepackage[makeroom]{cancel}
\usepackage{amssymb}
\usepackage{amsbsy}
\usepackage{amsfonts}
\usepackage{amssymb,mathrsfs,xcolor}
\usepackage{esint}
\usepackage{braket}
\usepackage{array}
\usepackage{graphicx}

\usepackage{wasysym}
\usepackage{multirow}
\usepackage{wrapfig}
\usepackage{subfig}

\usepackage{stmaryrd}
\usepackage{upgreek}

\makeatletter

\makeatletter\usepackage{babel}

\usepackage{hyperref}
\hypersetup{
    colorlinks,
    citecolor=blue,
    filecolor=green,
    linkcolor=purple,
    urlcolor=red,
}

\usepackage{slashed}

\newcommand{\ie}{\begin{equation}}
\newcommand{\fe}{\end{equation}}
\newcommand{\se}{\begin{eqnarray}}
\newcommand{\ff}{\end{eqnarray}}

\begin{document}

\title{How does non-metricity affect particle creation and evaporation in bumblebee gravity?}

%%%%%%%%%%%%%%%%%%%%%%%%%%%%%%%%%%%%%%%%%%%%%%%%%%%%%%%%%%%%%%%%%%%%%%
\author{A. A. Ara\'{u}jo Filho}
\email{dilto@fisica.ufc.br}

\affiliation{Departamento de Física, Universidade Federal da Paraíba, Caixa Postal 5008, 58051-970, João Pessoa, Paraíba,  Brazil.}

%%%%%%%%%%%%%%%%%%%%%%%%%%%%%%%%%%%%%%%%%%%%%%%%%%%%%%%%%%%%%%%%%%%%%%%%%%%%%%%%%%%%%%%%%%%%%%%%%%%%%%%%%%%%%%%%%%%%%%%%%%%%%%%%%%%%%%%%%%%%%%%%%%%%%%%%%%%%%%%%%%%%%%%%%%%%%%%%%%%%%%%%%%%%%%%%%%%%%%%%%%%%%%%%%%%%%%%%%%%%%%%%%%%%%%%%%%%%%%%%%%%%%%%%%%%%%%%%%%%%%%%%%%%%%%%%%%%%%%%%%%%%%%%%%%%%%%%%%%%%%%%%%%%%

\date{\today}

\begin{abstract}

In this work, we analyze the impact of non--metricity on particle creation and the evaporation process of black holes within the framework of bumblebee gravity. In general lines, we compare black holes in the \textit{metric} formalism \cite{14} and the \textit{metric--affine} approach \cite{filho2023vacuum}. Initially, we focus on bosonic particle modes to investigate Hawking radiation. Using the Klein--Gordon equation, we compute the Bogoliubov coefficients and derive the Hawking temperature. Subsequently, we examine Hawking radiation as a tunneling process, resolving divergent integrals through the residue method. The analysis is then extended to fermionic particle modes, also within the tunneling framework. Particle creation densities are calculated for both bosonic and fermionic cases. Additionally, greybody bounds are estimated for bosonic and fermionic particles. Furthermore, we explore the evaporation process, considering the final state of the black holes and we also investigate the correlation between the greybody factors and the quasinormal modes. Finally, constraints on the Lorentz--violating parameters $\ell$ (for the \textit{metric} case) and $X$ (for the \textit{metric--affine} case) are established using recent astrophysical data on black hole lifetimes.
In a general panorama, non--metricity (except for the tensor perturbations) in bumblebee gravity raises particle density for bosons while reducing it for fermions, increases greybody factors (for both bosons and fermions), amplifies the emission rate, and accelerates the evaporation process.

\end{abstract}

%\keywords{*****}
\maketitle

\tableofcontents     
     
%%%%%%%%%%%%%%%%%%%%%%%%%%%%%%%%%%%%%%%%%%%%%%%%%%%%%%%%%%%%%%%%%%%%%%%%%%%%%%%%%%%%%%%%%%%%%%%%%%%%%%%%%%%%%%%%%%%%%%%%%%%%%%%%%%%%%%%%%%%%%%%%%%%%%%%%%%%%%%%%%%%%%%%%%%%%%%%%%%%%%%%%%%%%%%%%%%%%%%%%%%%%%%%%%%%%%%%%%%%%%%%%%%%%%%%%%%%%%%%%%%%%%%%%%%%%%%%%%%%%%%%%%%%%%%%%%%%%%%%%%%%%%%%%%%%%%%%%%%%%%%%%%%%%%%%%%%%%%%%%%%%%%%%%%%%%%%%%%%%%%%%%%%%%%%%%%%%%%%%%%%%%%%%%%%%%%%%%%%%%%%%%%%%%%%%%%%%%%%%%%%%%%%%%%

\section{Introduction}

Lorentz symmetry, a foundational concept in modern physics, asserts that physical laws remain consistent across all inertial reference frames. Although this principle has been rigorously confirmed through experiments and observations, some high--energy theoretical frameworks suggest it might not hold universally. Models such as Hořava--Lifshitz,gravity \cite{3}, massive gravity \cite{6}, string theory \cite{1}, Einstein-aether theory \cite{5}, loop quantum gravity \cite{2}, \(f(T)\) gravity \cite{7}, and very special relativity \cite{8} explore scenarios where Lorentz invariance could be violated.

Lorentz symmetry breaking (LSB) can manifest in two distinct forms: explicit and spontaneous \cite{bluhm2006overview}. Explicit breaking arises when the Lagrangian density lacks Lorentz invariance, leading to variations in physical laws between certain reference frames. In contrast, spontaneous breaking occurs when the Lagrangian density preserves Lorentz symmetry, but the system's vacuum state does not maintain it \cite{bluhm2008spontaneous}.

Studies on the phenomenon of spontaneous Lorentz symmetry breaking \cite{KhodadiPoDU2023,13,12,10,9} are often framed within the context of the Standard Model Extension. Among the simplest theoretical constructs in this domain are bumblebee models \cite{KhodadiEPJC20232,12,KhodadiEPJC2023,10,1,KhodadiPRD2022,13,11,CapozzielloJCAP2023}, where a vector field, referred to as the bumblebee field, develops a non--zero vacuum expectation value (VEV). It introduces a preferred spatial direction, resulting in the violation of local Lorentz invariance at the particle level. Such symmetry breaking significantly influences various aspects of physics, including thermodynamic behavior \cite{araujo2021thermodynamic,anacleto2018lorentz,araujo2021higher,aa2022particles,araujo2022thermal,aaa2021thermodynamics,aa2021lorentz,araujo2022does,petrov2021bouncing2,reis2021thermal,paperrainbow}.

Ref. \cite{14} presents an exact solution describing a static, spherically symmetric spacetime within the framework of bumblebee gravity. Similarly, a Schwarzschild--like solution has been extensively studied from multiple perspectives, taking into account gravitational lensing phenomena \cite{15}, the dynamics of accretion \cite{18,17}, quasinormal mode behavior \cite{19}, and the characteristics of Hawking radiation \cite{16}.

Expanding on previous studies, Maluf et al. proposed an (A)dS--Schwarzschild--like solution that relaxes the requirement for strict vacuum conditions \cite{20}. Similarly, Xu et al. developed new classes of static, spherically symmetric bumblebee black hole solutions by including a background bumblebee field with a non--zero time component. This modification enabled an analysis of their thermodynamic behavior and potential observational characteristics, as detailed in Refs. \cite{23,21,24,22}.

Hawking's seminal work established a crucial connection between gravity and quantum mechanics, providing the basis for advancing quantum gravity theories \cite{o1,o11,o111}. He revealed that black holes emit radiation with a thermal spectrum, a process now termed Hawking radiation, which leads to their gradual loss of mass and eventual evaporation \cite{eeeKuang:2017sqa,gibbons1977cosmological,eeeOvgun:2019ygw,eeeOvgun:2019jdo,eeeOvgun:2015box,eeeKuang:2018goo}. This phenomenon, derived through quantum field theory calculations near event horizons, has had a transformative impact on the study of black hole thermodynamics and quantum effects in intense gravitational fields \cite{o7,aa2024implications,o9,o8,sedaghatnia2023thermodynamical,o4,o6,o3,araujo2023analysis,araujo2024dark}. Subsequent contributions by Kraus and Wilczek \cite{o10}, and later by Parikh and Wilczek \cite{o13,o12,011}, redefined Hawking radiation as a semi--classical tunneling mechanism. This tunneling model has since been applied broadly to different black hole scenarios, offering new perspectives on their quantum and thermodynamic characteristics \cite{asasas1,vanzo2011tunnelling,asasas2,calmet2023quantum,del2024tunneling,zhang2005new,asasas3,mirekhtiary2024tunneling,senjaya2024bocharova,johnson2020hawking,mitra2007hawking,medved2002radiation,silva2013quantum,anacleto2015quantum}.

The \textit{metric--affine} formulation, also referred to as the Palatini approach, departs from the standard \textit{metric} approach by treating the \textit{metric} and \textit{affine} connection as distinct geometric entities, each with its own dynamical role. This separation has led to a wide array of theoretical implications, as shown in several comprehensive studies \cite{Ghil1,Ghil2,gonzalo1,gonzalo2,gonzalo3,gonzalo4}. Nevertheless, investigations involving Lorentz symmetry breaking (LSB) remain limited within this framework. Recent contributions have started to bridge this gap, particularly by embedding spontaneous LSB mechanisms into the \textit{metric--affine} setting through bumblebee gravity models \cite{Paulo2,Lambiase:2023zeo, Jha:2023vhn,Filho:2023etf,Paulo3,AraujoFilho:2024ykw,Paulo4,Filho:2022yrk}.

This study investigates how non--metricity influences particle creation and black hole evaporation in the context of bumblebee gravity. In order to address the question posed in the title of this manuscript, we specifically compare black holes described within the \textit{metric} formalism \cite{14} to those in the \textit{metric--affine} framework \cite{filho2023vacuum}. The analysis begins with bosonic particle modes, focusing on Hawking radiation. Scalar field is used to compute Bogoliubov coefficients and determine the Hawking temperature. Hawking radiation is also examined as a tunneling phenomenon, where divergent integrals are resolved using the residue method. The study then extends to fermionic particle modes, applying the tunneling approach as well to calculate particle creation densities. For bosons and fermions, greybody bounds are also evaluated. Furthermore, the study also explores the evaporation process, emphasizing the end stage of black hole evolution. In addition, it examines how greybody factors are related to quasinormal modes. Constraints on the Lorentz--violating parameters—$\ell$ in the \textit{metric} framework and $X$ in the \textit{metric--affine} formulation—are derived by confronting the model with current astrophysical observations of black hole lifetime. In general lines, non--metricity (except for the tensor perturbations) in bumblebee gravity increases particle density for bosons while decreasing it for fermions, raises greybody factors for both, intensifies the emission rate, and reduces the evaporation timescale.

%%%%%%%%%%%%%%%%%%%%%%%%%%%%%%%%%%%%%%%%%%%%%%%%%%%%%%%%%%%%%%%%%%%%%%%%%%%%%%%%%%%%%%%%%%%%%%%%%%%%%%%%%%%%%%%%%%%%%%%%%%%%%%%%%%%%%%%%%%%%%%%%%%%%%%%%%%%%%%%%%%%%%%%%%%%%%%%%%%%%%%%%%%%%%%%%%%%%%%%%%%%%%%%%%%%%%%%%%%%%%%%%%%%%%%%%%%%%%%%%%%%%%%%%%%%%%%%%%%%%%%%%

\section{The general panorama}

Recently, the literature has introduced two distinct black hole solutions. For clarity, the first of these will be referred to as the bumblebee solution in the \textit{metric} formalism \cite{14}
\ie
\begin{split}
\label{model2}
\mathrm{d}s^{2} = & - \left( 1 - \frac{2M}{r}   \right) \mathrm{d}t^{2} + \frac{(1 + \ell)}{1 - \frac{2M}{r} } \, \mathrm{d}r^{2} + r^{2}\mathrm{d}\theta^{2} \\
& + r^{2} \sin^{2}\mathrm{d}\varphi^{2}.
\end{split}
\fe
and in the \textit{metric--affine} approach \cite{filho2023vacuum}
\ie
\begin{split}
\label{model1}
  \mathrm{d}s^2= &- \frac{\left(1-\frac{2M}{r}\right)\mathrm{d}t^2}{\sqrt{\left(1+\frac{3X}{4}\right)\left(1-\frac{X}{4}\right)}}+\frac{\mathrm{d}r^2}{\left(1-\frac{2M}{r}\right)}\sqrt{\frac{\left(1+\frac{3X}{4}\right)}{\left(1-\frac{X}{4}\right)^3}}\\
  & +r^{2}\left(\mathrm{d}\theta^2 +\sin^{2}{\theta}\mathrm{d}\phi^2\right).
\end{split}
\fe
In particular, the bumblebee black hole within the \textit{metric} framework has been extensively studied across a wide range of applications, including gravitational lensing via the Gauss--Bonnet theorem \cite{Ovgun:2018ran}, matter accretion processes \cite{Yang:2018zef}, black hole thermodynamics \cite{Gomes:2018oyd,Ovgun:2019ygw}, and its generalization with GUP corrections \cite{Kanzi:2019gtu}. Further extensions encompass Ricci dark energy models \cite{Jesus:2019nwi}, Kerr--like solutions \cite{Ding:2019mal,Liu:2019mls}, circular orbits and additional gravitational lensing studies \cite{Li:2020wvn}, black hole configurations with a cosmological constant \cite{20}, particle motion in non--commutative backgrounds \cite{KumarJha:2020ivj}, solutions incorporating topological defects \cite{Gullu:2020qzu}, quasinormal mode analysis \cite{Oliveira:2021abg}, gravitational wave polarization effects \cite{Liang:2022hxd}, connections to Kasner cosmology \cite{Neves:2022qyb}, compact star models \cite{Neves:2024ggn}, and the degradation of quantum entanglement \cite{Liu:2024wpa}.

Furthermore, the bumblebee solution in the \textit{metric--affine} formalism has been the subject of several investigations as well. These include studies on gravitational effects such as time delays, quasinormal modes, and the bending angle \cite{gravitationaltraces}, gravitational lensing in the strong deflection limit \cite{araujo2024gravitational}, and its generalization to a Kerr--like solution \cite{AraujoFilho:2024ykw}. Additional analyses cover quasiperiodic oscillations in galactic microquasars, black hole shadows \cite{Gao:2024ejs}, the properties of strange quark stars and condensate dark stars \cite{Panotopoulos:2024jtn}, scattering phenomena \cite{heidari2024scattering}, and the dynamics of accretion disks \cite{Lambiase:2023zeo}. Other topics include the deflection angle, greybody bounds (for bosonic particles), and neutrino propagation \cite{Lambiase:2023zeo}.

To ensure clarity, the subsequent sections will separately analyze black holes within the \textit{metric} and \textit{metric--affine} formalisms. This approach facilitates a detailed examination of particle creation processes for both bosons and fermions. Additionally, greybody bounds and the evaporation process will be thoroughly investigated for both of them.

%%%%%%%%%%%%%%%%%%%%%%%%%%%%%%%%%%%%%%%%%%%%%%%%%%%%%%%%%%%%%%%%%%%%%%%%%%%%%%%%%%%%%%%%%%%%%%%%%%%%%%%%%%%%%%%%%%%%%%%%%%%%%%%%%%%%%%%%%%%%%%%%%%%%%%%%%%%%%%%%%%%%%%%%%%%%%%%%%%%%%%%%%%%%%%%%%%%%%%%%%%%%%%%%%%%%%%%%%%%%%%%%%%%%%%%%%%%%%%%%%%%%%%%%%%%%%%%%%%%%%%%%

\section{The \textit{metric} case}

This section explores the particle creation properties of the bumblebee black hole within the \textit{metric} formalism as introduced in Ref. \cite{14}. The analysis begins with bosonic particle modes, utilizing the tunneling framework. To facilitate the calculations, the metric coordinates are transformed into the Painlevé--Gullstrand form, effectively removing the divergence at the event horizon. The resulting divergent integrals, specifically the imaginary part of the action, \(\mathcal{S}\), denoted as \(\text{Im}\,\mathcal{S}_{\text{metric}}\), are resolved using the residue method, enabling the estimation of the particle density for bosons, \(n_{\text{metric}}\).

The study then shifts to fermionic particle modes, which are also analyzed using the tunneling method. In this case, the near--horizon approximation \cite{araujo2023analysis} is employed to simplify the calculations and determine the particle density for fermions, \(n_{\psi_{\text{metric}}}\).

Following this, the greybody factors for bosonic and fermionic particles are computed. Finally, the evaporation timescale of the black hole is derived analytically, allowing for a comparison with the Schwarzschild black hole and a recent configuration in Kalb--Ramond gravity discussed in the literature \cite{Liu:2024oas}.

%%%%%%%%%%%%%%%%%%%%%%%%%%%%%%%%%%%%%%%%%%%%%%%%%%%%%%%%%%%%%%%%%%%%%%%%%%%%%%%%%%%%%%%%%%%%%%%%%%%%%%%%%%%%%%%%%%%%%%%%%%%%%%%%%%%%%%%%%%%%%%%%%%%%%%%%%%%%%%%%%%%%%%%%%%%%%%%%%%%%%%%%%%%%%%%%%%%%%%%%%%%%%%%%%%%%%%%%%%%%%%%%%%%%%%%%%%%%%%%%%%%%%%%%%%%%%%%%%%%%%%%%%%%%%%%%%%%%%%%%%%%%%%%%%%%%%%%%%%%%%%%%%%%%%%%%%%%%%%%%%%%%%%%%%%%%%%%%%%%%%%%%%%%%%%%%%%%%%%%%%%%%%%%%%%%%%%%%%%%%%%%%%%%%%%%%%%%%%%%%%%%%%%%%%%%%%%%%%%%%%%%%%%%%%%%%%%%%%%%%%%%%%%%%%%%%%%%%%%%%%%%%%%%%%%%%%%%%%%%%%%%%%%%%%%%%%%%%%%%%%%%%%%%%%%%%%%%%%%%%%%%%%%%%%%%%%%%%%%%%%%%%%%%%%%%%%%%%%%%%%%%%%%%%%%%%%%%%%%%%%%%%%%%%%%%%%%%%%%%%%%%%%
\subsection{Bosonic modes}

%%%%%%%%%%%%%%%%%%%%%%%%%%%%%%%%%%%%%%%%%%%%%%%%%%%%%%%%%%%%%%%%%%%%%%%%%%%%%%%%%%%%%%%%%%%%%%%%%%%%%%%%%%%%%%%%%%%%%%%%%%%%%%%%%%%%%%%%%%%%%%%%%%%%%%%%%%%%%%%%%%%%%%%%%%%%%%%%%%%%%%%%%%%%%%%%%%%%%%%%%%%%%%%%%%%%%%%%%%%%%%%%%%%%%%%%%%%%%%%%%%%%%%%%%%%%%%%%%%%%%%%%%%%%%%%%%%%%%%%%%%%%%%%%%%%%%%%%%%%%%%%%%%%%%%%%%%%%%%%%%%%%%%%%%%%%%%%%%%%%%%%%%%%%%%%%%%%%%%%%%%%%%%%%%%%%%%%%%%%%%%%%%%%%%%%%%%%%%%%%%%%%%%%%%%%%%%%%%%%%%%%%%%%%%%%%%%%%%%%%%%%%%%%%%%%%%%%%%%%%%%%%%%%%%%%%%%%%%%%%%%%%%%%%%%%%%%%%%%%%%%%%%%%%%%%%%%%%%%%%%%%%%%%%%%%%%%%%%%%%%%%%%%%%%%%%%%%%%%%%%%%%%%%%%%%%%%%%%%%%%%%%%%%%%%%%%%%%%%%%%%%%%

\subsubsection{The Hawking radiation}

The analysis begins with a general spherically symmetric spacetime
\ie
\label{mainmetric}
\mathrm{d} s^2  = - f_{\text{m}}(r)\mathrm{d}t^2 + \frac{1}{g_{\text{m}}(r)}\mathrm{d}r^2 + r^2\mathrm{d}\Omega^2,
\fe    
in which 
\ie
f_{\text{m}}(r) =  1 - \frac{2M}{r},
\fe
and
\ie
g_{\text{m}}(r) = \frac{ 1 - \frac{2M}{r}}{1 + \ell}.
\fe 
Here, the indices ${\text{m}}$ denote the metric components corresponding to the \textit{metric} case. Using this framework, we investigate the effect of Lorentz violation, characterized by $\ell$, on the emission of Hawking particles. In his seminal work \cite{hawking1975particle}, Hawking studied the wave function of a scalar field, $\Phi$, expressed below
\ie
\label{kleinggordoneq}
\frac{1}{\sqrt{-\bar{g}}}\partial_{\mu}(\bar{g}^{\mu\nu}\sqrt{-\bar{g}} \, \partial_{\nu}\Phi) = 0.
\fe
The metric tensor, $\bar{g}$, corresponds to the bumblebee black hole within the \textit{metric} formalism framework. To further investigate, the curved spacetime setting is considered under the Schwarzschild solution. In this scenario, the field operator takes the following form:
\ie
\Phi = \sum_{i} \left (f_i  a_i + \bar{f}_{i} a^{\dagger}_{i} \right) = \sum_{i} \left( p_{i} b_{i} + \bar{p}_{i} b^{\dagger}_{i} + q_{i}  c_{i} + \bar{q}_{i}  c^{\dagger}_{i} \right ) .
\fe

In this framework, the solutions $f_{i}$ and $ \bar{f}_{i}$ (with  $\bar{f}_{i}$ being the complex conjugate) represent components of the wave equation that are purely ingoing. Similarly, $p_{i}$ and  $\bar{p}_{i}$ describe solutions that are entirely outgoing, whereas $q_{i}$ and $\bar{q}_{i}$ correspond to components devoid of any outgoing contributions. The operators $a_{i}$, $b_{i}$, and $c_{i}$ function as annihilation operators, while their counterparts $a_{i}^{\dagger}$, $b_{i}^{\dagger}$, and $c_{i}^{\dagger}$ act as creation operators. The objective here is to demonstrate that the solutions $ f_{i}$, $\bar{f}_{i}$, $p_{i}$, $\bar{p}_{i}$, $ q_{i} $, and $\bar{q}_{i}$ are influenced by the presence of Lorentz violation. In particular, the focus is on examining how the Lorentz--violating parameter modifies the original solutions proposed by Hawking.

Spherical symmetry characterizes both the classical Schwarzschild spacetime and the framework of bumblebee gravity, enabling the representation of incoming and outgoing wave solutions through spherical harmonics. In the region outside the black hole, these solutions can be expressed as follows \cite{asasas1,calmet2023quantum,heidari2024quantum}:
\ie
\begin{split}
f_{\omega^\prime l m} & =  \frac{1}{\sqrt{2 \pi \omega^\prime} r }  \mathcal{F}_{\omega^\prime}(r) e^{i \omega^\prime v} Y_{lm}(\theta,\phi)\ , \\ 
p_{\omega l m} & = \frac{1}{\sqrt{2 \pi \omega} r }  \mathcal{P}_\omega(r) e^{i \omega u} Y_{lm}(\theta,\phi). 
\end{split}
\fe
In this framework, the advanced coordinate $v$ and the retarded coordinate $u$ are utilized. For the scenario under consideration, this can be expressed as
\ie
v = t + r^{*} = t  + r \sqrt{1 + \ell} + 2 \sqrt{1 + \ell} \, M \ln |r - 2M|,
\fe
and
\ie
u = t - r^{*} = t - r \sqrt{1 + \ell} - 2 \sqrt{1 + \ell} \, M \ln |r - 2M| .
\fe

To determine the Lorentz--violating modifications arising from these coordinate functions, one effective strategy involves examining the trajectory of a particle moving along a geodesic in the given spacetime background. The motion is parametrized by an affine parameter $\lambda$, enabling the particle's momentum to be characterized through expression below
\ie
p_{\mu}=\bar{g}_{\mu\nu}\frac{\mathrm{d}x}{\mathrm{d}\lambda}^\nu.
\fe
The momentum is preserved throughout the particle's motion along the geodesic. Additionally, we have the formulation
\ie
\mathcal{L} = \bar{g}_{\mu\nu} \frac{\mathrm{d}x^\mu}{\mathrm{d}\lambda} \frac{\mathrm{d}x^\nu}{\mathrm{d}\lambda}.
\fe
This quantity remains invariant along geodesic trajectories. For particles with nonzero mass, the conditions $\mathcal{L} = -1$ and $\lambda = \tau$ are imposed, where $\tau$ is the proper time of the particle. Conversely, for massless particles, which are the primary subject of this investigation, we use $\lambda$ as an arbitrary affine parameter. By employing a stationary, spherically symmetric metric as outlined in the referenced framework and focusing on radial geodesics ($p_\varphi = L = 0$) within the equatorial plane ($\theta = \pi/2$), the associated expressions can be derived
\ie
E =  f_{\text{m}}(r) \dot{t},
\fe
with $E = -p_t$, and the dot represents the derivative with respect to the affine parameter $\lambda$, i.e., $\mathrm{d}/\mathrm{d}\lambda$. Proceeding with this formulation, we derive the following:
\ie
\label{fgfgfg}
    \left( \frac{\mathrm{d}r}{\mathrm{d}\lambda} \right)^2 = \frac{E^2}{f_{\text{m}}(r)g_{\text{m}}(r)^{-1}},
\fe
so that after some algebraic procedures, it reads
\ie
\frac{\mathrm{d}}{\mathrm{d}\lambda}\left(t\mp r^{*}\right) = 0,
\fe
in which $r^{*}$ is the so--called tortoise coordinate, which is written below
\ie
\mathrm{d}r^{*} = \frac{\mathrm{d}r}{\sqrt{f_{\text{m}}(r)g_{\text{m}}(r)}}.
\fe

Reformulating the expression for the retarded coordinate leads us to:
\ie
\label{nmnma}
\frac{\mathrm{d}u}{\mathrm{d}\lambda}=\frac{2E}{f_{\text{m}}(r)}.
\fe
For an ingoing geodesic parametrized by $\lambda$, the advanced coordinate $u$ is described as a function $u(\lambda)$. Determining this relation requires two key steps: first, expressing $r$ in terms of $\lambda$, and then performing the integral outlined in Eq. \eqref{nmnma}. The specific form of $u(\lambda)$ directly influences the derived Bogoliubov coefficients, which are fundamental in characterizing the black hole’s quantum emission. To proceed, the functions $f_{\text{m}}(r)$ and $g_{\text{m}}(r)$ are employed, integrating the square root in Eq. \eqref{fgfgfg} across the interval $\tilde{r} \in [r_h, r]$, corresponding to $\tilde{\lambda} \in [0, \lambda]$, where $r_h$ denotes the event horizon. Following this approach, we obtain:
\ie
\label{rvv}
r = 2M - \frac{E\lambda}{1 + \ell}.
\fe
To obtain this result, the negative sign in the square root of Eq. \eqref{fgfgfg} was chosen, reflecting the ingoing geodesic trajectory.

The integration is then performed using $r(\lambda)$, leading to:
\ie
u(\lambda) = -4 \sqrt{1 + \ell}\, M \ln \left( \frac{\lambda}{C} \right).
\fe
Notice that there exists a constant of integration, denoted by $C$, appearing in the solution. Additionally, the relationship between ingoing and outgoing null coordinates is established through geometric optics. This connection is expressed as $\lambda = (v_0 - v)/D$, where $v_0$ identifies the advanced coordinate corresponding to the reflection point at the horizon ($\lambda = 0$), while $D$ represents a proportionality constant \cite{calmet2023quantum}.

Building upon these initial steps, the outgoing solutions to the modified Klein--Gordon equation, which include the Lorentz--violating parameter $\ell$, are now determined. The derived expressions take the following form:
\ie
p_{\omega} =\int_0^\infty \left ( \alpha_{\omega\omega^\prime} f_{\omega^\prime} + \beta_{\omega\omega^\prime} \bar{ f}_{\omega^\prime}  \right)\mathrm{d} \omega^\prime,
\fe
where $\alpha_{\omega\omega^\prime}$ and $\beta_{\omega\omega^\prime}$ represent the so--called Bogoliubov coefficients \cite{parker2009quantum, hollands2015quantum, wald1994quantum, fulling1989aspects}
\begin{equation}
\begin{split}
    \alpha_{\omega\omega^\prime}=& -i K e^{i\omega^\prime v_0}e^{\pi \left[2 M \sqrt{1 + \ell} \right]\omega} \int_{-\infty}^{0} \,\mathrm{d}x\,\Big(\frac{\omega^\prime}{\omega}\Big)^{1/2}e^{\omega^\prime x} \\ 
    & \times e^{i\omega\left[4M\sqrt{1 + \ell})\right]\ln\left(\frac{|x|}{CD}\right)},
    \end{split}
\end{equation}
and
\begin{equation}
\begin{split}
    \beta_{\omega\omega'} &= i Ke^{-i\omega^\prime v_0}e^{-\pi \left[2 M \sqrt{1 + \ell} \right]\omega}
    \int_{-\infty}^{0} \,\mathrm{d}x\,\left(\frac{\omega^\prime}{\omega}\right)^{1/2}e^{\omega^\prime x} \\
    & \times e^{i\omega\left[4M\sqrt{1 + \ell})\right]\ln\left(\frac{|x|}{CD}\right)}.
    \end{split}
\end{equation}
This shows that Lorentz--violating corrections, encapsulated by \(\ell\) in the metric, influence the quantum amplitude for particle production. Through this mechanism, information ``scape'' from the black hole becomes possible.

Interestingly, despite the influence of the quantum gravitational correction on the quantum amplitude, the power spectrum retains its blackbody nature at this stage. Confirming this requires calculating:
\ie
|\alpha_{\omega\omega'}|^2 = e^{\big(8\pi M \sqrt{1 + \ell} \big)\omega}|\beta_{\omega\omega'}|^2\,.
\fe
Furthermore, by analyzing the flux of outgoing particles within the frequency interval $[\omega, \omega + \mathrm{d}\omega]$ \cite{o10}, we obtain:
\ie
\mathcal{P}(\omega, \ell)=\frac{\mathrm{d}\omega}{2\pi}\frac{1}{\left \lvert\frac{\alpha_{\omega\omega^\prime}}{\beta_{\omega\omega^\prime}}\right \rvert^2-1}\, ,
\fe
or, in other words, 
\ie
\mathcal{P}(\omega, \ell)=\frac{\mathrm{d}\omega}{2\pi}\frac{1}{e^{\left(8\pi M \sqrt{1 + \ell}\right)\omega}-1}\,.
\fe
An important point to highlight is that comparing the expression above with the Planck distribution reveals that
\ie
\mathcal{P}(\omega, \ell)=\frac{\mathrm{d}\omega}{2\pi}\frac{1}{e^{\frac{\omega}{T}}-1}.
\fe
In this manner, we can properly obtain
\ie
\label{hawtempmetricase}
    T_{\text{metric}} = \frac{1}{8 \pi  \sqrt{1 + \ell} M}.
\fe

\begin{figure}
    \centering
      \includegraphics[scale=0.55]{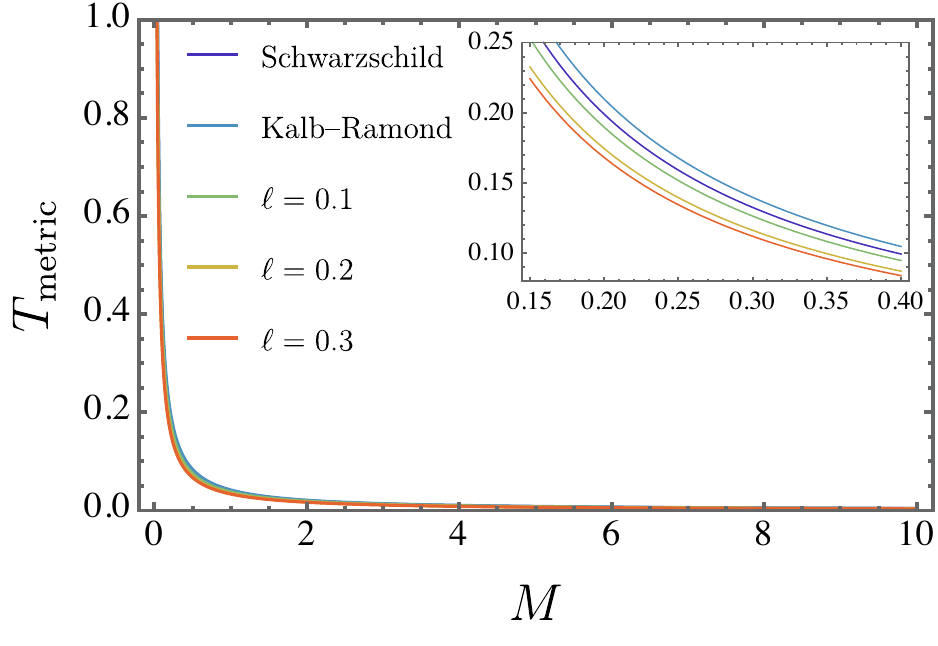}
    \caption{The Hawking temperature $T_{\text{metric}}$ as a function of mass \( M \) for various values of \( \ell \), being compared with the Schwarzschild and Kalb--Ramond cases.}
    \label{tmetric}
\end{figure}

As we shall see in the evaporation subsection, the Hawking temperature obtained from Eq. (\ref{hawtempmetricase}) matches perfectly the temperature calculated using surface gravity in Eq. (\ref{sufacceeee}), as expected. To illustrate this thermal characteristic, its behavior is depicted in Fig. \ref{tmetric}, which also contrasts it with the standard Schwarzschild scenario and the Kalb--Ramond solution. In general, larger values of $\ell$ lead to an increase in the Hawking temperature.

In other words, this suggests that a black hole governed by a Lorentz--violating metric emits radiation analogous to that of a greybody, with an effective temperature $T$ specified by Eq. \eqref{hawtempmetricase}.

Energy conservation for the entire system has not been fully addressed up to this point. With every radiation emission, the black hole’s mass diminishes, leading to a gradual reduction in its size. To account for this phenomenon, the next section will utilize the tunneling framework proposed by Parikh and Wilczek \cite{011}.

%%%%%%%%%%%%%%%%%%%%%%%%%%%%%%%%%%%%%%%%%%%%%%%%%%%%%%%%%%%%%%%%%%%%%%%%%%%%%%%%%%%%%%%%%%%%%%%%%%%%%%%%%%%%%%%%%%%%%%%%%%%%%%%%%%%%%%%%%%%%%%%%%%%%%%%%%%%%%%%%%%%%%%%%%%%%%%%%%%%%%%%%%%%%%%%%%%%%%%%%%%%%%%%%%%%%%%%%%%%%%%%%%%%%%%%%%%%%%%%%%%%%%%%%%%%%%%%%%%%%%%%%%%%%%%%%%%%%%%%%%%%%%%%%%%%%%%%%%%%%%%%%%%%%%%%%%%%%%%%%%%%%%%%%%%%%%%%%%%%%%%%%%%%%%%%%%%%%%%%%%%%%%%%%%%%%%%%%%%%%%%%%%%%%%%%%%%%%%%%%%%%%%%%%%%%%%%%%%%%%%%%%%%%%%%%%%%%%%%%%%%%%%%%%%%%%%%%%%%%%%%%%%%%%%%%%%%%%%%%%%%%%%%%%%%%%%%%%%%%%%%%%%%%%%%%%%%%%%%%%%%%%%%%%%%%%%%%%%%%%%%%%%%%%%%%%%%%%%%%%%%%%%%%%%%%%%%%%%%%%%%%%%%%%%%%%%%%%%%%%%%%%%%%%%%%%%%%%%%%%%%%%%%%%%%%%%%%%%%%%%%%%%%%%%%%%%%%%%%%%

\subsubsection{The tunneling process}

In order to take into account energy conservation, the calculation of the black hole’s radiation spectrum, we adopt the methodology detailed in Ref. \cite{011, vanzo2011tunnelling, parikh2004energy, calmet2023quantum}. By transitioning to Painlevé--Gullstrand coordinates, the metric is expressed as $\mathrm{d}s^2 = -f_{\text{m}}(r)\mathrm{d}t^2 + 2h_{\text{m}}(r)\mathrm{d}t\mathrm{d}r + \mathrm{d}r^2 + r^2\mathrm{d}\Omega^2$, where $h_{\text{m}}(r) = \sqrt{f_{\text{m}}(r)\big(g_{\text{m}}(r)^{-1} - 1\big)}$. The rate of tunneling is connected to the imaginary part of the particle’s action \cite{parikh2004energy, vanzo2011tunnelling, calmet2023quantum}.

The action $\mathcal{S}_{\text{metric}}$ describing a particle's motion in a curved spacetime is given by $\mathcal{S}_{\text{metric}} = \int p_\mu \, \mathrm{d}x^\mu$. In calculating $\text{Im} \, \mathcal{S}_{\text{metric}}$, only the term $p_r \mathrm{d}r$ contributes, as the component $p_t \mathrm{d}t = -\omega \mathrm{d}t$ remains entirely real and does not affect the imaginary part. As a result
\ie
\text{Im}\,\mathcal{S}_{\text{metric}} = \text{Im}\,\int_{r_i}^{r_f} \,p_r\,\mathrm{d}r=\text{Im}\,\int_{r_i}^{r_f}\int_{0}^{p_r} \,\mathrm{d}p_r'\,\mathrm{d}r.
\fe
Applying Hamilton's equations to a system governed by the Hamiltonian $H = M - \omega'$, we determine that $\mathrm{d}H = -\mathrm{d}\omega'$, where $0 \leq \omega' \leq \omega$, with $\omega$ representing the energy of the particle emitted. This leads to the following result
\ie
\begin{split}
\text{Im}\, \mathcal{S}_{\text{metric}} & = \text{Im}\,\int_{r_i}^{r_f}\int_{M}^{M-\omega} \,\frac{\mathrm{d}H}{\mathrm{d}r/\mathrm{d}t}\,\mathrm{d} r \\
& =\text{Im}\,\int_{r_i}^{r_f}\,\mathrm{d}r\int_{0}^{\omega} \,-\frac{\mathrm{d}\omega'}{\mathrm{d}r/\mathrm{d}t}\,.
\end{split}
\fe
Reordering the integration sequence and applying the substitution
\ie
\frac{\mathrm{d}r}{\mathrm{d}t} = -h_{\text{m}}(r)+\sqrt{f_{\text{m}}(r) + h_{\text{m}}(r)^2} = 1 - \sqrt{\frac{\Delta(r)}{r}}, 
\fe
with $\Delta(r)= \frac{\ell r + 2(M-\omega^{\prime})}{\ell+1}$. Therefore, we can write
\ie
\label{ims}
\text{Im}\, \mathcal{S}_{\text{metric}} =\text{Im}\,\int_{0}^{\omega} -\mathrm{d}\omega'\int_{r_i}^{r_f}\,\frac{\mathrm{d}r}{\sqrt{1 + \ell}\left(1-\sqrt{\frac{\Delta(r,\,\omega^\prime)}{r}}\right)}.
\fe

\begin{figure}
    \centering
      \includegraphics[scale=0.55]{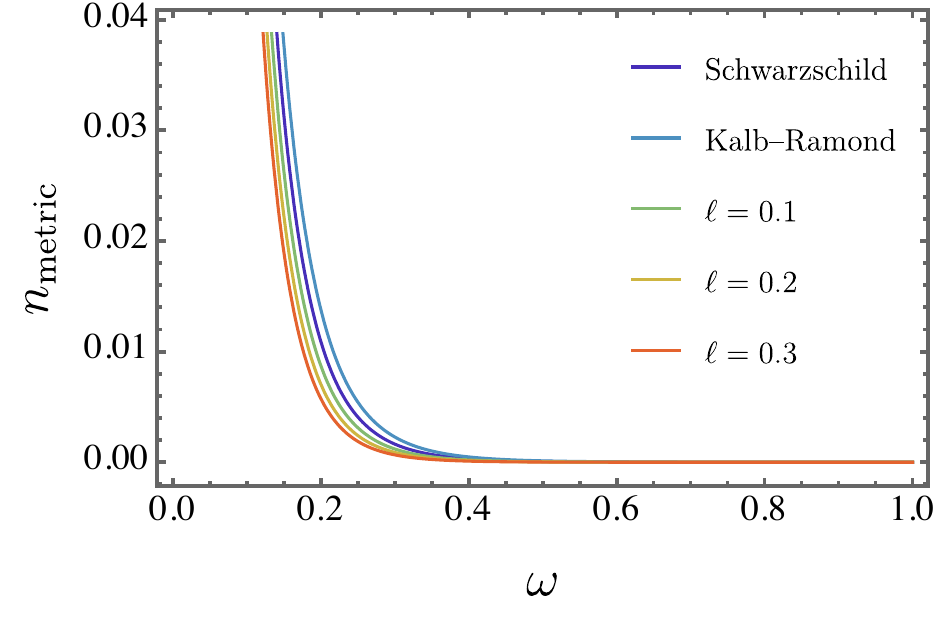}
    \caption{The particle density \( n_{\text{metric}} \) is shown for different values of \( \ell \) for the metric case. The Schwarzschild and the Kalb--Ramond cases are also compared.}
    \label{ndensitymetric}
\end{figure}

Replacing $M$ with $(M - \omega')$ in the metric modifies the function $\Delta(r)$, making it explicitly dependent on $\omega'$. This change introduces a singularity at the new horizon location. Calculating the contour integral around this singularity in a counterclockwise orientation gives
\begin{eqnarray}
    \text{Im}\, \mathcal{S}_{\text{metric}}  = 4\pi \sqrt{1 + \ell} \, \omega \left( M - \frac{\omega}{2} \right)  .
\end{eqnarray}
As described in \cite{vanzo2011tunnelling}, the Lorentz--violating corrections modify the emission rate of a Hawking particle, which can be represented as
\ie
\Gamma_{\text{metric}} \sim e^{-2 \, \text{Im}\, S_{\text{metric}}}=e^{-8 \sqrt{1 + \ell} \, \omega \left( M - \frac{\omega}{2} \right)} .
\fe
In the limit $\omega \to 0$, the emission spectrum simplifies to the familiar Planckian distribution initially derived by Hawking. In this way, the spectrum can be expressed as
\begin{equation}
    \mathcal{P}_{\text{metric}}(\omega)=\frac{\mathrm{d}\omega}{2\pi}\frac{1}{e^{8 \pi \sqrt{1 + \ell} \,\omega \left( M - \frac{\omega}{2} \right)
    }-1}.
\end{equation}
The spectrum of emitted radiation, shaped by its dependence on $\omega$, diverges from the standard blackbody distribution, a difference that becomes clear upon analysis. For small $\omega$, the spectrum resembles a Planck--like distribution, though with an altered Hawking temperature. Moreover, the particle number density is directly linked to the tunneling rate and can be written as:
\ie
n_{\text{metric}} = \frac{\Gamma_{\text{metric}}}{1 - \Gamma_{\text{metric}}} = \frac{1}{e^{8 \pi  \sqrt{1 + \ell} \, \omega  \left(M-\frac{\omega }{2}\right)} - 1}.
\fe
Fig. \ref{ndensitymetric} illustrates the influence of the Lorentz--violating parameter $\ell$ on $n_{\text{metric}}$. The plot indicates that as $\ell$ increases, the particle number density decreases. For comparison, $n_{\text{metric}}$ is evaluated alongside the Schwarzschild and Kalb--Ramond cases. The results show that the Kalb--Ramond solution exhibits the highest particle creation density, the Schwarzschild case occupies an intermediate position, and the bumblebee black hole yields the lowest values.

The results imply that the radiation emitted by a black hole reveals details about its internal properties. The Lorentz--violating parameter $\ell$ influences the Hawking amplitudes, and the power spectrum, incorporating these corrections, deviates from the standard blackbody form when the effects of energy conservation are included.

%%%%%%%%%%%%%%%%%%%%%%%%%%%%%%%%%%%%%%%%%%%%%%%%%%%%%%%%%%%%%%%%%%%%%%%%%%%%%%%%%%%%%%%%%%%%%%%%%%%%%%%%%%%%%%%%%%%%%%%%%%%%%%%%%%%%%%%%%%%%%%%%%%%%%%%%%%%%%%%%%%%%%%%%%%%%%%%%%%%%%%%%%%%%%%%%%%%%%%%%%%%%%%%%%%%%%%%%%%%%%%%%%%%%%%%%%%%%%%%%%%%%%%%%%%%%%%%%%%%%%%%%

\subsection{Fermionic modes}

Black holes, possessing an intrinsic temperature, are known to emit radiation resembling blackbody radiation, although this emission typically does not account for greybody factors. The resulting spectrum encompasses particles of various spins, including fermions. Investigations by Kerner and Mann \cite{o69}, with additional studies \cite{o75,o72,o71,o74,o73,o70}, have shown that both massless fermions and bosons are emitted at the same temperature. Moreover, research on spin--$1$ bosons has revealed that the Hawking temperature remains unaffected, even when higher--order quantum corrections are considered \cite{o77,o76}.

The action for fermions is typically associated with the phase of the spinor wave function, which satisfies the Hamilton--Jacobi equation. Other formulations of the action have been explored in studies such as \cite{o83, o84, vanzo2011tunnelling}. Corrections arising from the interaction between the particle’s spin and the spacetime’s spin connection do not create singularities at the horizon. These effects are minimal, mainly influencing the spin’s precession, and are not considered significant in this context.  Moreover, the contribution of emitted particle spins to the black hole’s angular momentum is extremely small, especially for black holes that lack rotation and have masses much larger than the Planck scale \cite{vanzo2011tunnelling}. The symmetric emission of particles with opposite spins ensures that, on average, the black hole’s angular momentum will not be modified.

\begin{figure}
    \centering
      \includegraphics[scale=0.55]{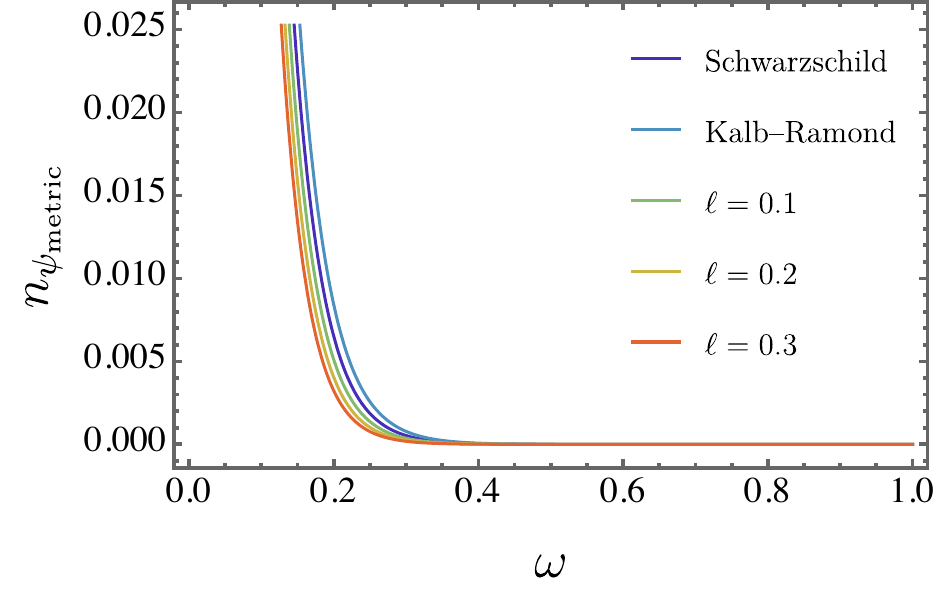}
    \caption{The particle density $n_{\psi_{\text{metric}}}$ is shown for various values of \( \ell \). The Schwarzschild and the Kalb--Ramond cases are compared.}
    \label{nfermionsmetric}
\end{figure}

Building on earlier work, we analyze the process by which fermionic particles tunnel through the event horizon in this black hole configuration. The emission rate is derived using Schwarzschild--like coordinates, which are known to exhibit singular behavior at the horizon. Alternative methods, including those based on generalized Painlevé--Gullstrand or Kruskal--Szekeres coordinates, have been explored in prior studies \cite{o69}. To frame this investigation, we start by introducing a general metric, expressed as:
\ie
\mathrm{d}s^{2} = \mathcal{A}(r) \mathrm{d}t^{2} + [1/\mathcal{B}(r)]\mathrm{d}r^{2} + \mathcal{C}(r)[\mathrm{d}\theta^{2} + r^{2}\sin^{2}\theta ]\mathrm{d}\varphi^{2}.
\fe
The Dirac equation, when extended to curved spacetime, is written as
\ie
\left(\gamma^\mu \nabla_\mu + \frac{m}{\hbar}\right) \Psi(t,r,\theta,\varphi) = 0
\fe
in which
\ie
\nabla_\mu = \partial_\mu + \frac{i}{2} {\Gamma^\alpha_{\;\mu}}^{\;\beta} \,\Sigma_{\alpha\beta}
\fe and 
\ie
\Sigma_{\alpha\beta} = \frac{i}{4} [\gamma_\alpha,  \gamma_\beta].
\fe
The $\gamma^\mu$ matrices satisfy the conditions of the Clifford algebra, defined by
\ie
\{\gamma_\alpha,\gamma_\beta\} = 2 g_{\alpha\beta} \mathbb{1},
\fe
where $\mathbb{1}$ is the $4 \times 4$ identity matrix. Within this context, the chosen representation for the $\gamma$ matrices is
\begin{eqnarray*}
 \gamma ^{t} &=&\frac{i}{\sqrt{\mathcal{A}(r)}}\left( \begin{array}{cc}
\vec{1}& \vec{ 0} \\ 
\vec{ 0} & -\vec{ 1}%
\end{array}%
\right), \;\;
\gamma ^{r} =\sqrt{\mathcal{B}(r)}\left( 
\begin{array}{cc}
\vec{0} &  \vec{\sigma}_{3} \\ 
 \vec{\sigma}_{3} & \vec{0}%
\end{array}%
\right), \\
\gamma ^{\theta } &=&\frac{1}{r}\left( 
\begin{array}{cc}
\vec{0} &  \vec{\sigma}_{1} \\ 
 \vec{\sigma}_{1} & \vec{0}%
\end{array}%
\right), \;\;
\gamma ^{\varphi } =\frac{1}{r\sin \theta }\left( 
\begin{array}{cc}
\vec{0} &  \vec{\sigma}_{2} \\ 
 \vec{\sigma}_{2} & \vec{0}%
\end{array}%
\right),
\end{eqnarray*}%
where $\vec{\sigma}$ represents the Pauli matrices, which satisfy the standard commutation relations:
\ie
 \sigma_i  \sigma_j = \vec{1} \delta_{ij} + i \varepsilon_{ijk} \sigma_k, \,\, \text{in which}\,\, i,j,k =1,2,3. 
\fe 
The $\gamma^5$ matrix, on the other hand, is written
\begin{equation*}
\gamma ^{5}=i\gamma ^{t}\gamma ^{r}\gamma ^{\theta }\gamma ^{\varphi }=i\sqrt{\frac{\mathcal{B}(r)}{\mathcal{A}(r)}}\frac{1}{r^{2}\sin \theta }\left( 
\begin{array}{cc}
\vec{ 0} & - \vec{ 1} \\ 
\vec{ 1} & \vec{ 0}%
\end{array}%
\right)\:.
\end{equation*}
To model a Dirac field with its spin oriented upward along the positive $r$--axis, the ansatz adopted is\cite{vagnozzi2022horizon}:
\begin{equation}
\Psi^{+}(t,r,\theta ,\varphi ) = \left( \begin{array}{c}
\mathcal{H}(t,r,\theta ,\varphi ) \\ 
0 \\ 
\mathcal{Y}(t,r,\theta ,\varphi ) \\ 
0%
\end{array}%
\right) \exp \left[ \frac{i}{\hbar }\psi^{+}(t,r,\theta ,\varphi )\right]\;.
\label{spinupbh} 
\end{equation} 
Our analysis centers on the spin--up ($+$) configuration, while acknowledging that the spin--down ($-$) case, corresponding to orientation along the negative $r$--axis, can be treated through an analogous procedure. By inserting the ansatz (\ref{spinupbh}) into the Dirac equation, we obtain:
\ie
\begin{split}
-\left( \frac{i \,\mathcal{H}}{\sqrt{\mathcal{A}(r)}}\,\partial _{t} \psi_{+} + \mathcal{Y} \sqrt{\mathcal{B}(r)} \,\partial_{r} \psi_{+}\right) + \mathcal{H} m &=0, \\
-\frac{\mathcal{Y}}{r}\left( \partial _{\theta }\psi_{+} +\frac{i}{\sin \theta } \, \partial _{\varphi }\psi_{+}\right) &= 0, \\
\left( \frac{i \,\mathcal{Y}}{\sqrt{\mathcal{A}(r)}}\,\partial _{t}\psi_{+} - \mathcal{H} \sqrt{\mathcal{B}(r)}\,\partial_{r}\psi_{+}\right) + \mathcal{Y} m & = 0, \\
-\frac{\mathcal{H}}{r}\left(\partial _{\theta }\psi_{+} + \frac{i}{\sin \theta }\,\partial _{\varphi }\psi_{+}\right) &= 0,
\end{split}
\fe%
At the leading order in \(\hbar\), the action is represented as
$
\psi_{+}=- \omega\, t + \chi(r) + L(\theta ,\varphi )  $
in a such way that
\cite{vanzo2011tunnelling}
\begin{eqnarray}
\left( \frac{i\, \omega\, \mathcal{H}}{\sqrt{\mathcal{A}(r)}} - \mathcal{Y} \sqrt{\mathcal{B}(r)}\,  \chi^{\prime }(r)\right) +m\, \mathcal{H} &=&0,
\label{bhspin5} \\
-\frac{\mathcal{H}}{r}\left( L_{\theta }+\frac{i}{\sin \theta }L_{\varphi }\right) &=&0,
\label{bhspin6} \\
-\left( \frac{i\,\omega\, \mathcal{Y}}{\sqrt{\mathcal{A}(r)}} + \mathcal{H}\sqrt{\mathcal{B}(r)}\, \mathcal \chi^{\prime }(r)\right) +\mathcal{Y}\,m &=&0,
\label{bhspin7} \\
-\frac{\mathcal{H}}{r}\left( L_{\theta } + \frac{i}{\sin \theta }L_{\varphi }\right) &=& 0.
\label{bhspin8}
\end{eqnarray}
The forms of $\mathcal{H}$ and $\mathcal{Y}$ are irrelevant to the conclusion that Eqs. (\ref{bhspin6}) and (\ref{bhspin8}) impose the condition $L_{\theta} + i(\sin \theta)^{-1} L_{\varphi} = 0$, which ensures that $L(\theta, \varphi)$ is inherently complex. This condition applies equally to both the outgoing and incoming cases. As a result, when evaluating the ratio of outgoing to incoming probabilities, the terms involving $L$ cancel each other out, allowing us to disregard $L$ in further calculations. In the case of a massless particle, Eqs. (\ref{bhspin5}) and (\ref{bhspin7}) provide two possible solutions:
\ie
\mathcal{H} = -i \mathcal{Y}, \qquad \chi^{\prime }(r) = \chi_{\text{out}}' = \frac{\omega}{\sqrt{\mathcal{A}(r)\mathcal{B}(r)}},
\fe
\ie
\mathcal{H} = i \mathcal{Y}, \qquad \chi^{\prime }(r) = \chi_{\text{in}}' = - \frac{\omega}{\sqrt{\mathcal{A}(r)\mathcal{B}(r)}}.
\fe
In this context, \(\chi_{\text{out}}\) and \(\chi_{\text{in}}\) denote the outgoing and incoming solutions, respectively \cite{vanzo2011tunnelling}. The tunneling probability is then determined as \(\Gamma_{\psi_{\text{metric}}} \sim e^{-2 \, \text{Im} \, (\chi_{\text{out}} - \chi_{\text{in}})}\). Accordingly,
\ie
\mathcal \chi_{ \text{out}}(r)= - \mathcal \chi_{ \text{in}} (r) = \int \mathrm{d} r \,\frac{\omega}{\sqrt{\mathcal{A}(r)\mathcal{B}(r)}}\,.
\fe
It is important to note that, based on the dominant energy condition and the Einstein field equations, the functions \(\mathcal{A}(r)\) and \(\mathcal{B}(r)\) have identical zeros. Therefore, near \(r = r_h\), we can approximate these functions to first order as:
\ie
\mathcal{A}(r)\mathcal{B}(r) = \mathcal{A}'(r_{h})\mathcal{B}'(r_{h})(r - r_{h})^2 + \dots \, .
\fe
This reveals the existence of a simple pole with a well--defined coefficient. Applying Feynman’s method, we arrive at:
\ie
2\mbox{ Im}\;\left( \mathcal \chi_{ \text{out}} -  \chi_{ \text{in}} \right) = \mbox{Im}\int \mathrm{d} r \,\frac{4\omega}{\sqrt{\mathcal{A}(r)\mathcal{B}(r)}}=\frac{2\pi\omega}{\kappa},
\fe
where the surface gravity is defined as 
\ie
\kappa = \frac{1}{2} \sqrt{\mathcal{A}'(r_{h}) \mathcal{B}'(r_{h})} .
\fe 
In this framework, the particle density \(n_{\psi_{\text{metric}}}\) for the black hole solution is given by \(\Gamma_{\psi_{\text{metric}}} \sim e^{-\frac{2 \pi \omega}{\kappa}}\)
\ie
n_{\psi_{\text{metric}}} = \frac{\Gamma_{\psi{\text{metric}}}}{1+\Gamma_{\psi{\text{metric}}}}  = \frac{1}{e^{8 \pi  \sqrt{1+ \ell} \, M \omega }+1}.
\fe
Fig. \ref{nfermionsmetric} displays the behavior of $n_{\psi_{\text{metric}}}$ for varying values of \(\ell\), while also comparing it to the standard Schwarzschild and Kalb--Ramond scenarios. Overall, increasing \(\ell\) leads to a decrease in particle density. Among the curves, the Schwarzschild case occupies an intermediate position, the Kalb--Ramond solution reaches the highest values, and the bumblebee model within the metric approach exhibits the lowest particle density.

%%%%%%%%%%%%%%%%%%%%%%%%%%%%%%%%%%%%%%%%%%%%%%%%%%%%%%%%%%%%%%%%%%%%%%%%%%%%%%%%%%%%%%%%%%%%%%%%%%%%%%%%%%%%%%%%%%%%%%%%%%%%%%%%%%%%%%%%%%%%%%%%%%%%%%%%%%%%%%%%%%%%%%%%%%%%%%%%%%%%%%%%%%%%%%%%%%%%%%%%%%%%%%%%%%%%%%%%%%%%%%%%%%%%%%%%%%%%%%%%%%%%%%%%%%%%%%%%%%%%%%%%%%%%%%%%%%%%%%%%%%%%%%%%%%%%%%%%%%%%%%%%%%%%%%%%%%%%%%%%%%%%%%%%%%%%%%%%%%%%%%%%%%%%%%%%%%%%%%%%%%%%%%%%%%%%%%%%%%%%%%%%%%%%%%%%%%%%%%%%%%%%%%%%%%%%%%%%%%%%%%%%%%%%%%%%%%%%%%%%%%%%%%%%%%%%%%%%%%%%%%%%%%%%%%%%%%%%%%%%%%%%%%%%%%%%%%%%%%%%%%%%%%%%%%%%%%%%%%%%%%%%%%%%%%%%%%%%%%%%%%%%%%%%%%%%%%%%%%%%%%%%%%%%%%%%%%%%%%%%%%%%%%%%%%%%%%%%%%%%%%%%%

\subsection{Greybody factors for bosons}

%%%%%%%%%%%%%%%%%%%%%%%%%%%%%%%%%%%%%%%%%%%%%%%%%%%%%%%%%%%%%%%%%%%%%%%%%%%%%%%%%%%%%%%%%%%%%%%%%%%%%%%%%%%%%%%%%%%%%%%%%%%%%%%%%%%%%%%%%%%%%%%%%%%%%%%%%%%%%%%%%%%%%%%%%%%%%%%%%%%%%%%%%%%%%%%%%%%%%%%%%%%%%%%%%%%%%%%%%%%%%%%%%%%%%%%%%%%%%%%%%%%%%%%%%%%%%%%%%%%%%%%%%%%%%%%%%%%%%%%%%%%%%%%%%%%%%%%%%%%%

\subsubsection{Scalar perturbations}

The partial wave equation is derived by revisiting the Klein--Gordon equation in the context of a spherically symmetric curved spacetime, as presented earlier in Eq. (\ref{kleinggordoneq}). Employing the separation of variables method, the equation is reformulated as
\ie
\label{sai}
{\Psi _{\omega lm}}(r,t) = \frac{{{\psi_{\omega l}}(r)}}{r}{Y_{lm}}(\theta ,\varphi ){e^{ - i \omega t}}.
\fe
In the case of a spherically symmetric spacetime, the tortoise coordinate ($r^{*}$) is introduced using the metric components associated with time and radial coordinates, defined as
\ie
{\rm{d}}r^{*} = \frac{{\rm{d}}r}{\sqrt {\mathcal{A}(r)\mathcal{B}(r)} },
\fe
which this transformation reformulates the Klein--Gordon equation into a wave equation resembling the Schr\"{o}dinger equation
\ie
\left[\frac{{{\rm{d}^2}}}{{\mathrm{d}{{r^*}^2}}} + ({\omega ^2} - {\mathcal{V}^{\,\text{s}}_{{\text{metric}}}})\right]\psi_{\omega l} (r) = 0.
\fe
The term $\mathcal{V}^{\,\text{s}}_{{\text{metric}}}$ represents the effective potential governing scalar perturbations, expressed as
\ie
\begin{split}
\label{scalarpotentialmetric}
\mathcal{V}^{\,\text{s}}_{{\text{metric}}} & = \mathcal{A}(r)\left[\frac{{l(l + 1)}}{{{r^2}}} + \frac{1}{{r\sqrt {{\mathcal{A}(r)}{\mathcal{B}(r)^{ - 1}}} }}\frac{\mathrm{d}}{{\mathrm{d}r}}\sqrt {{\mathcal{A}(r)}\mathcal{B}(r)}\right]\\
& = \frac{\left(1-\frac{2 M}{r}\right) \left(l (l+1) r+\frac{2 M}{\ell+1}\right)}{r^3}.
\end{split}
\fe

\begin{figure}
    \centering
      \includegraphics[scale=0.55]{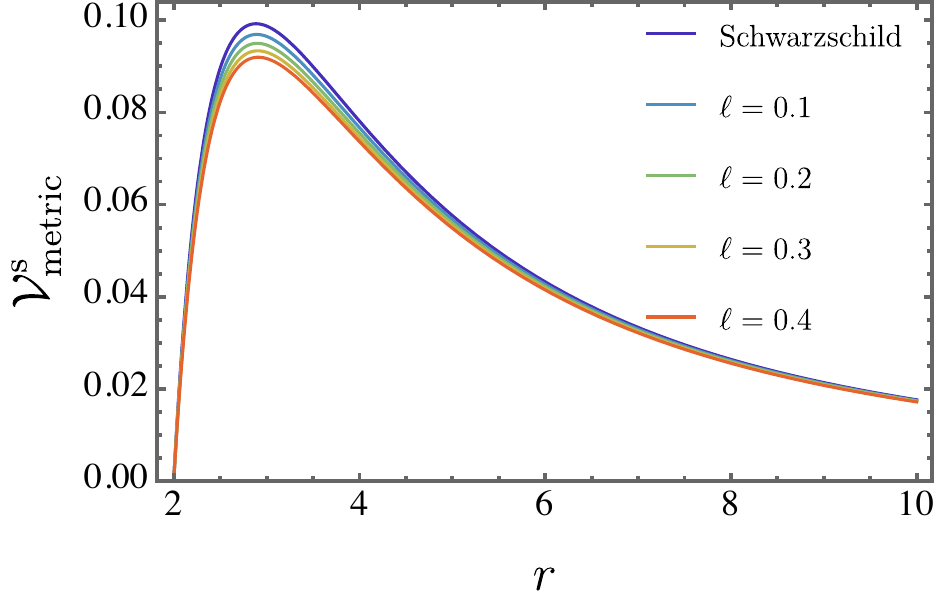}
    \caption{The effective potential $\mathcal{V}^{\,\text{s}}_{{\text{metric}}}$ is shown for different values of $\ell$ for $l=1$. Also, the Schwarzschild case is compared in this analysis.}
    \label{scalarpotential}
\end{figure}

Fig. \ref{scalarpotential} illustrates the behavior of the effective potential $\mathcal{V}^{\,\text{s}}_{{\text{metric}}}$ associated with scalar perturbations. As the parameter $\ell$ grows, the potential's magnitude decreases correspondingly. For comparative purposes, the Schwarzschild scenario is also plotted, serving as a reference for this analysis.

Hawking radiation from black holes is modified by spacetime curvature as it travels from the event horizon to infinity. The deviation from a pure black body spectrum is characterized by the greybody factor. This section examines it for a massless spin--0 field using semi--analytic methods \cite{sakalli2022topical,boonserm2019greybody,ovgun2024shadow,al2024fermionic,boonserm2008transmission}. The bound for the greybody factor, represented as $T_b$, is given by
\ie
\label{tgmetric}
T^{\,\text{s}}_{b_{\text{metric}}} \ge {\mathop{\rm sech}\nolimits} ^2 \left(\int_\infty^ {+\infty} {\mathfrak{G} \,\rm{d}}r^{*}\right),
\fe
with 
\ie
\mathfrak{G} = \frac{{\sqrt {{{(\xi')}^2} + {{({\omega ^2} - \mathcal{V}^{\,\text{s}}_{{\text{metric}}} - {\xi^2})}^2}} }}{{2\xi}}.
\fe
The function $\xi$ is positive and fulfills the conditions $\xi(+\infty) = \xi(-\infty) = \omega$. Assigning $\xi$ the value $\omega$ simplifies Eq. \eqref{tgmetric} to
\ie
\begin{split}
& T^{\,\text{s}}_{b_{\text{metric}}}  \ge {\mathop{\rm sech}\nolimits} ^2 \left[\int_{-\infty}^ {+\infty} \frac{\mathcal{V}^{\,\text{s}}_{{\text{metric}}}} {2\omega}\mathrm{d}r^{*}\right] \\
& ={\mathop{\rm sech}\nolimits} ^2 \left[\int_{r_{ h}}^ {+\infty} \frac{\mathcal{V}^{\,\text{s}}_{{\text{metric}}}} {2\omega\sqrt{\mathcal{A}(r)\mathcal{B}(r)}}\mathrm{d} r\right] \\
& ={\mathop{\rm sech}\nolimits} ^2 \left[ \frac{1}{2\omega} \left(\frac{2 l (l+1) (\ell+1)+1}{4 \sqrt{\ell+1} M} \right)  \right].
\end{split}
\fe

\begin{figure}
    \centering
      \includegraphics[scale=0.51]{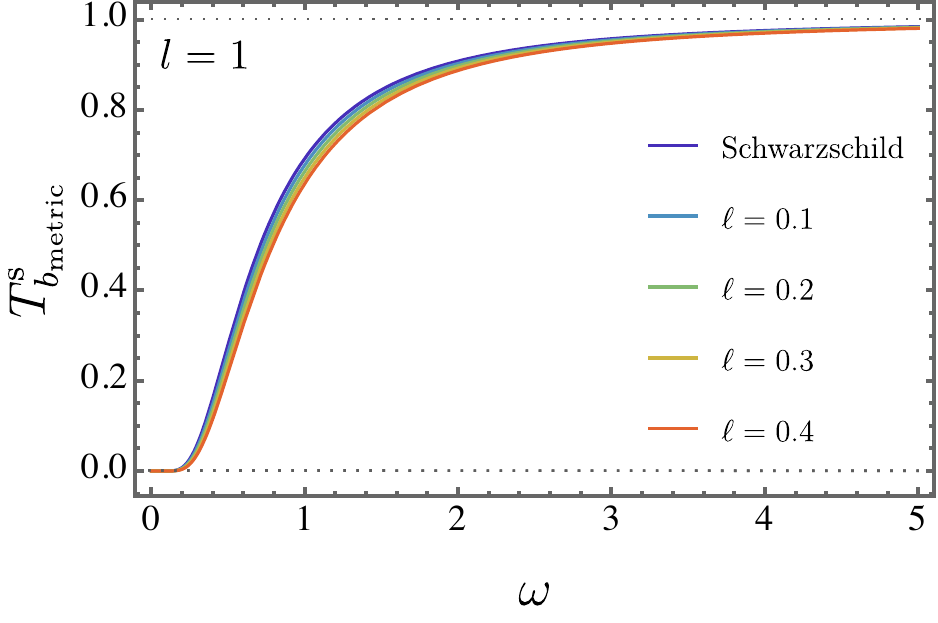}
      \includegraphics[scale=0.51]{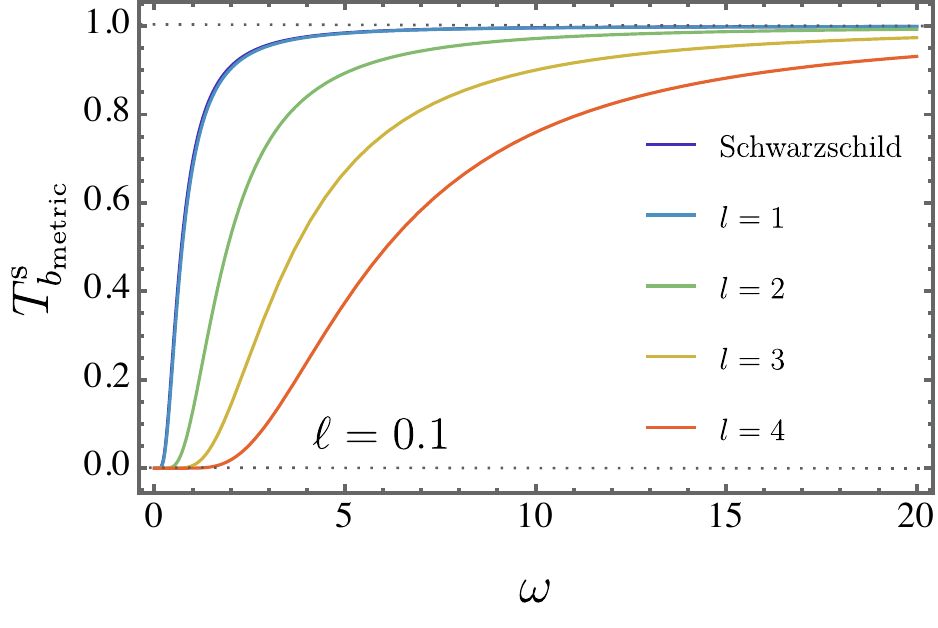}
    \caption{The greybody factors $T^{\,\text{s}}_{b_{\text{metric}}}$ is displayed for different values of $\ell$ when keeping $l=1$ (the the top panel) and for different values of $l$ for a fixed value of $\ell = 0.1$. For both cases, the Schwarzschild case is compared.}
    \label{greybodymetricbosons}
\end{figure}

Fig. \ref{greybodymetricbosons} presents the greybody factors for bosons, $T^{\,\text{s}}_{b_{\text{metric}}}$, under two scenarios: varying $\ell$ while fixing $l=1$ (top panel) and varying $l$ with $\ell$ held constant at 0.1. In both cases, the results are compared with the Schwarzschild scenario.

%%%%%%%%%%%%%%%%%%%%%%%%%%%%%%%%%%%%%%%%%%%%%%%%%%%%%%%%%%%%%%%%%%%%%%%%%%%%%%%%%%%%%%%%%%%%%%%%%%%%%%%%%%%%%%%%%%%%%%%%%%%%%%%%%%%%%%%%%%%%%%%%%%%%%%%%%%%%%%%%%%%%%%%%%%%%%%%%%%%%%%%%%%%%%%%%%%%%%%%%%%%%%%%%%%%%%%%%%%%%%%%%%%%%%%%%%%%%%%%%%%%%%%%%%%%%%%%%%%%%%%%%%%%%%%%%%%%%%%%%%%%%%%%%%%%%%%%%%%%%

\subsubsection{Vector perturbations}

To examine electromagnetic perturbations, we employ the tetrad formalism, as outlined in \cite{chandrasekhar1998mathematical, Bouhmadi-Lopez:2020oia, Gogoi:2023kjt}. This method defines a tetrad basis \(\mathrm{e}_\mu^{a}\) corresponding to the black hole metric \(\bar{g}_{\mu\nu}\), ensuring the following conditions are met:
\ie
\begin{split}
& \mathrm{e}^{a}_\mu \mathrm{e}^\mu_{b} = \delta^{a}_{b}, \, \, \, \,
\mathrm{e}^{a}_\mu \mathrm{e}^\nu_{a} = \delta^{\nu}_{\mu}, \, \, \, \,\\
& \mathrm{e}^{a}_\mu = \bar{g}_{\mu\nu} \eta^{a b} \mathrm{e}^\nu_{b}, \, \, \, \,
\bar{g}_{\mu\nu} = \eta_{a b}\mathrm{e}^{a}_\mu \mathrm{e}^{b}_\nu = \mathrm{e}_{a\mu} \mathrm{e}^{a}_\nu.
\end{split}
\fe

For electromagnetic perturbations analyzed through the tetrad formalism, the Bianchi identity for the field strength, $\mathcal{F}_{[ab|c]} = 0$, results in
\begin{align}
\left( r \sqrt{\mathcal{A}(r)}\, \mathcal{F}_{t \phi}\right)_{,r} + r \sqrt{\mathcal{B}(r)}\,
\mathcal{F}_{\phi r, t} &=0,  \label{edem1} \\
\left( r \sqrt{\mathcal{A}(r)}\, \mathcal{F}_{ t \phi}\sin\theta\right)_{,\theta} + r^2
\sin\theta\, \mathcal{F}_{\phi r, t} &=0.  \label{edem2}
\end{align}

As a result, the conservation equation takes the following form:
\ie
\eta^{b c}\! \left( \mathcal{F}_{a b} \right)_{|c} =0.
\fe

In spherical polar coordinates, this equation can be re--expressed in the form:
\ie  \label{edem3}
\left( r \sqrt{\mathcal{A}(r)}\, \mathcal{F}_{\phi r}\right)_{,r} + \sqrt{\mathcal{A}(r) \mathcal{B}(r)}%
\, \mathcal{F}_{\phi \theta,\theta} + r \sqrt{\mathcal{B}(r)}\, \mathcal{F}_{t \phi, t} = 0.
\fe

Here, the vertical bar and comma represent intrinsic and directional derivatives associated with the tetrad indices. Using Eqs. \eqref{edem1} and \eqref{edem2}, along with the time derivative of Eq. \eqref{edem3}, the following result is obtained
\ie  \label{edem4}
\begin{split}
& \left[ \sqrt{\mathcal{A}(r) \mathcal{B}(r)^{-1}} \left( r \sqrt{\mathcal{A}(r)}\, \mathcal{F}
\right)_{,r} \right]_{,r} \\
& + \dfrac{\mathcal{A}(r) \sqrt{\mathcal{B}(r)}}{r} \left( \dfrac{%
\mathcal{F}_{,\theta}}{\sin\theta} \right)_{,\theta}\!\! \sin\theta - r 
\sqrt{\mathcal{B}(r)}\, \mathcal{F}_{,tt} = 0.
\end{split}
\fe
Let \( F = \mathcal{F}_{t \phi } \sin\theta \). Utilizing Fourier decomposition \((\partial_t \rightarrow -i \omega)\) and redefining the field as \(F(r,\theta) = F(r) Y_{,\theta} / \sin\theta\), where \(Y(\theta)\) denotes the Gegenbauer function \cite{g1,g2,g3,g5,g6}, Eq. \eqref{edem4} can be recast as: 
\ie  
\begin{split}
\label{edem5}
& \left[ \sqrt{\mathcal{A}(r) \mathcal{B}(r)^{-1}} \left( r \sqrt{\mathcal{A}(r)}\, F
\right)_{,r} \right]_{,r} \\
& + \omega^2 r \sqrt{\mathcal{B}(r)}\, F -
\mathcal{A}(r) \sqrt{\mathcal{B}(r)} r^{-1} l (l + 1)\, F = 0.
\end{split}
\fe

Defining $\psi^{\, \text{v}}_{\text{metric}} \equiv r \sqrt{\mathcal{A}(r)} \, F$, Eq. \eqref{edem5} is transformed into a Schrödinger--like equation, which takes the form
\ie
\partial^2_{r_*} \psi^{\, \text{v}}_{\text{metric}} + \omega^2 \psi^{\, \text{v}}_{\text{metric}} = \mathcal{V}^{\, \text{v}}_{\text{metric}}(r) \psi^{\, \text{v}}_{\text{metric}},
\fe
in a such way that the effective potential associated with the vectorial perturbation is then expressed below
\ie  
\label{vectorpotentialmetric}
\mathcal{V}^{\, \text{v}}_{\text{metric}}(r) = \mathcal{A}(r) \, \dfrac{l ( l + 1 )}{r^2}.
\fe

It is worth noting that $\mathcal{V}^{\, \text{v}}_{\text{metric}}(r)$ is not plotted here, as the Lorentz--violating contributions introduce no changes, leaving it identical to the Schwarzschild case for vector perturbations. Furthermore, the greybody factors are expressed as:
\ie
T^{\,\text{v}}_{b_{\text{metric}}} = {\mathop{\rm sech}\nolimits} ^2 \left[ \frac{1}{2\omega} \left(\frac{l (l+1) \sqrt{\ell+1}}{2 M} \right)  \right].
\fe

\begin{figure}
    \centering
      \includegraphics[scale=0.51]{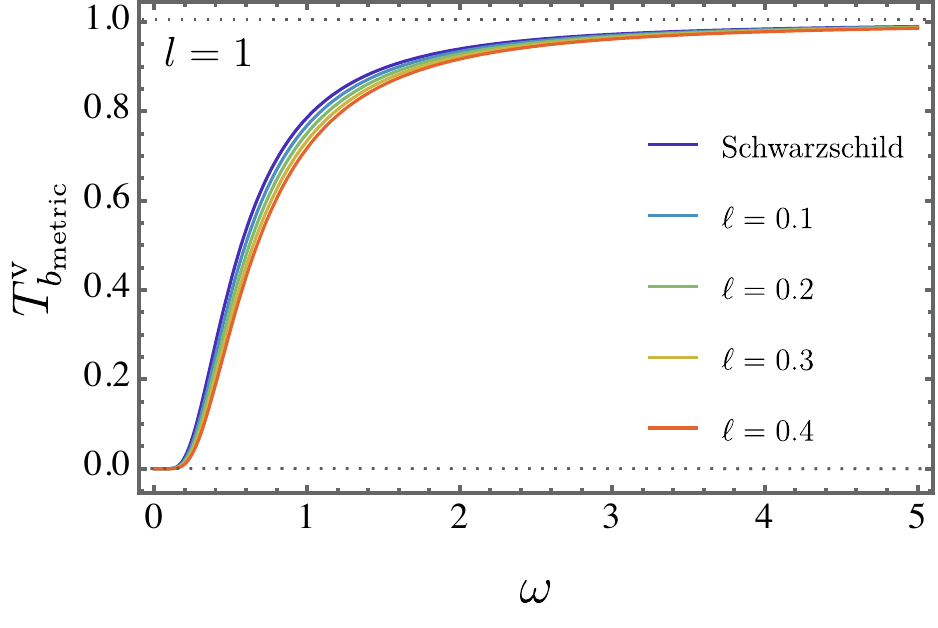}
      \includegraphics[scale=0.51]{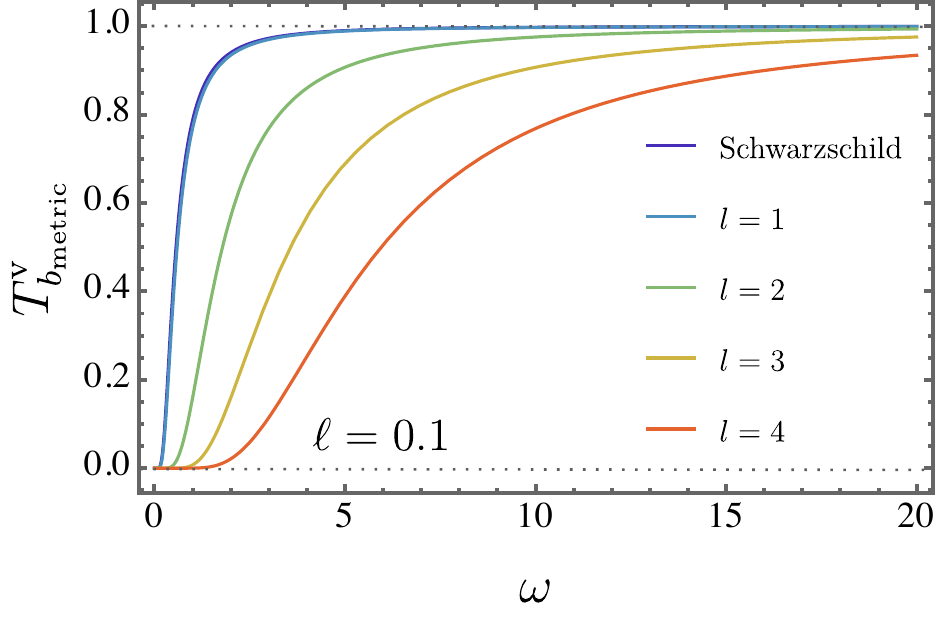}
    \caption{The greybody factors $T^{\,\text{v}}_{b_{\text{metric}}}$ is displayed for different values of $\ell$ when keeping $l=1$ (the the top panel) and for different values of $l$ for a fixed value of $\ell = 0.1$. For both cases, the Schwarzschild case is compared.}
    \label{greybodymetricbosonsvector}
\end{figure}

Fig. \ref{greybodymetricbosonsvector} presents the greybody factors for vectorial perturbations, $T^{\,\text{v}}_{b_{\text{metric}}}$, under two scenarios: varying $\ell$ while keeping $l=1$ (top panel) and varying $l$ with $\ell$ fixed at $0.1$. In both cases, comparisons are made with the Schwarzschild scenario.

%%%%%%%%%%%%%%%%%%%%%%%%%%%%%%%%%%%%%%%%%%%%%%%%%%%%%%%%%%%%%%%%%%%%%%%%%%%%%%%%%%%%%%%%%%%%%%%%%%%%%%%%%%%%%%%%%%%%%%%%%%%%%%%%%%%%%%%%%%%%%%%%%%%%%%%%%%%%%%%%%%%%%%%%%%%%%%%%%%%%%%%%%%%%%%%%%%%%%%%%%%%%%%%%%%%%%%%%%%%%%%%%%%%%%%%%%%%%%%%%%%%%%%%%%%%%%%%%%%%%%%%%%%%%%%%%%%%%%%%%%%%%%%%%%%%%%%%%%%%%

\subsubsection{Tensor perturbations}

The formulation of the master equations was carried out without anchoring the analysis to any specific foundational theory, relying solely on the presumed applicability of the Klein--Gordon and Maxwell dynamics. Yet, in gravitational settings where the matter sector is not minimally linked to the spacetime geometry described by $g_{\mu\nu}$, the usual conservation properties associated with these fields can break down.

Axial (odd--parity) gravitational perturbations demand variations not only in the gravitational field equations but also in the stress-energy content. When no fundamental theory is explicitly defined, a different strategy becomes necessary. Here, the analysis proceeds by interpreting the background through the lens of the Einstein equation, with modifications introduced via an effective stress--energy contribution. Comparable procedures have been employed across various scenarios in previous works \cite{AraujoFilho:2025rwr,ashtekar2018quantum,ashtekar2018quantum2,asasas2,baruah2025quasinormal}. From a phenomenological standpoint, the stress--energy distribution corresponding to the black hole can be effectively described using an anisotropic fluid representation
\ie
T_{\mu\nu}=\left(\rho + p_2\right)u_\mu u_\nu+\left(p_1-p_2\right)x_\mu x_\nu+p_2 g_{\mu\nu}.\label{anisopf}
\fe

In this setup, $\rho$ denotes the energy density measured in the rest frame of the fluid. The vector $u^\mu$ refers to the fluid’s timelike four--velocity, while $x^\mu$ is a unit spacelike vector orthogonal to both $u^\mu$ and the angular basis directions. Eq. \eqref{anisopf} features $p_1$ and $p_2$, which characterize the pressures along the radial and tangential directions, respectively. Moreover, the vectors $u^\mu$ and $x^\mu$ obey the following normalization and orthogonality conditions:
\ie
u_\mu u^\mu=-1\,,\qquad x_\mu x^\mu=1\,.\label{fourvelocity}
\fe

The metric $g_{\mu\nu}$ serves as the operator for index manipulation throughout this analysis. In the rest frame of the fluid, the temporal and radial components of the basis vectors simplify to $u^\mu = (u^t, 0, 0, 0)$ for the four--velocity and $x^\mu = (0, x^r, 0, 0)$ for the associated spacelike direction. Taking Eq. (\ref{fourvelocity}) into account, this configuration leads to the relation:
\ie
u_t^2 = g_{tt}(r )u_tu^t = -g_{tt}(r)\,,\qquad x_r^2 = g_{rr}(r)x_rx^r = g_{rr}(r)\,.
\fe

At the background level, the components of the stress--energy tensor take the following form:
\begin{align}
T_{tt}&= - g_{tt}(r)\rho\,,\qquad T_t^t=-\rho\,,\\
T_{rr}& = g_{rr}(r)p_1\,,\qquad T_r^r=p_1\,,\\
&T_{\theta}^{\theta}=T_\varphi^\varphi=p_2\,.
\end{align}
Notice that the radial profiles of $\rho$, $p_1$, and $p_2$ are determined by the structure of the spacetime and emerge from a direct computation of the Einstein tensor components.

To study the greybody facotrs in a static, spherically symmetric black hole background, a perturbation is introduced that alters the geometry into a time-dependent, axisymmetric form. This deformation leads to the following expression for the perturbed metric \cite{chen2019gravitational}:
\begin{align}
\mathrm{d}s^2=&-e^{2\nu}\left(\mathrm{d}x^0\right)^2+e^{2\psi}\left(\mathrm{d}x^1-\sigma \mathrm{d}x^0-q_2\mathrm{d}x^2-q_3\mathrm{d}x^3\right)^2\nonumber\\&+e^{2\mu_2}\left(\mathrm{d}x^2\right)^2+e^{2\mu_3}\left(\mathrm{d}x^3\right)^2\,.\label{metricg}
\end{align}

Here, it is important to highlight that the functions $\nu$, $\psi$, $\mu_2$, $\mu_3$, $\sigma$, $q_2$, and $q_3$ are taken to depend explicitly on the coordinates $t = x^0$, $r = x^2$, and $\theta = x^3$, while remaining independent of the azimuthal direction $\varphi = x^1$, reflecting the axisymmetric character of the perturbed geometry. This coordinate and notation scheme follows the framework adopted in Ref.~\cite{chen2019gravitational}. In the limit where the background spacetime remains static and spherically symmetric, the functions $\sigma$, $q_2$, and $q_3$ identically vanish. Therefore, in a linear perturbative analysis, these terms are introduced only at first order.

The next step involves reformulating the problem using the tetrad approach, where an orthonormal frame adapted to the spacetime described by metric \eqref{metricg} is introduced. This formalism streamlines the perturbative analysis by expressing tensorial quantities in terms of locally flat basis vectors. For further details on the implementation of this method, see Ref.~\cite{chen2019gravitational}
\begin{align}
e^{\mu}_{0}&=\left(e^{-\nu},\sigma e^{-\nu},0,0\right)\,,\nonumber\\
e^{\mu}_{1}&=\left(0, e^{-\psi}, 0,0\right)\,,\nonumber\\
e^{\mu}_{2}&=\left(0, q_2e^{-\mu_2}, e^{-\mu_2},0\right)\,,\nonumber\\
e^{\mu}_{3}&=\left(0, q_3e^{-\mu_3}, 0, e^{-\mu_3}\right)\,.\label{tetradbasis111}
\end{align}

Within this approach, indices corresponding to the tetrad basis are written in parentheses to clearly separate them from those tied to the coordinate system. The method involves translating all physical and geometrical quantities—originally formulated in terms of the metric $g_{\mu\nu}$—into a local inertial frame defined by the metric $\eta_{ab}$, through the application of a tetrad basis. Conventionally, $\eta_{ab}$ is identified with the Minkowski metric to simplify calculations. This redefinition allows vectors and tensors to be described using their components relative to the orthonormal frame rather than the coordinate basis
\begin{align}
A_{\mu}&=e_{\mu}^{a}A_{a}\,,\quad A_{a}=e_{a}^{\mu}A_{\mu}\,,\nonumber\\
B_{\mu\nu}&=e_{\mu}^{a}e_{\nu}^{b}B_{ab}\,,\quad B_{ab}=e_{a}^{\mu}e_{b}^{\nu}B_{\mu\nu}\,.
\end{align}

In the context of the tetrad formalism, the perturbed stress--energy tensor describing an anisotropic fluid is written in terms of components projected onto the orthonormal frame, yielding the following structure:
\begin{align}
\delta T_{ab}=&\,(\rho+p_2)\delta(u_{a}u_{b})+(\delta\rho+\delta p_2)u_{a}u_{b}\nonumber\\
&+(p_1-p_2)\delta(x_{a}x_{b})+(\delta p_1-\delta p_2)x_{a}x_{b}\nonumber\\&+\delta p_2\eta_{ab}.
\end{align}
By imposing the normalization and orthogonality conditions on the vectors $u^\mu$ and $x^\mu$—specifically, Eq.~\eqref{fourvelocity} together with $u^\mu x_\mu = 0$—one obtains that all axial components of the perturbed stress--energy tensor vanish when expressed in the tetrad frame
\begin{align}
\delta T_{10}&=\delta T_{12}=\delta T_{13}=0\,.
\end{align}

In the tetrad formalism, the Einstein field equations are recast in terms of tetrad components, leading to the following representation
\ie
R_{ab}-\frac{1}{2}\eta_{ab}R=8\pi T_{ab}\,.\label{eineq}
\fe

Since the axial components of the perturbed stress--energy tensor vanish, the equation governing axial perturbations arises directly from the requirement $R_{ab}|_{\text{axial}} = 0$. Through a sequence of algebraic manipulations—outlined in the Appendix of Ref.~\cite{chen2019gravitational}—one arrives at the corresponding master equation. The resulting effective potential for the axial gravitational perturbations is given by \cite{baruah2025quasinormal}
\ie
\mathcal{V}^{\, \text{t}}_{\text{metric}}(r) = g_{tt}(r) \left[ \dfrac{2}{r^2} \left( \dfrac{1}{g_{rr}(r)} - 1 \right) + \dfrac{l(l+1)}{r^2} - \dfrac{1}{r \sqrt{g_{tt}(r) g_{rr}(r)}} \left( \dfrac{\mathrm{d}}{\mathrm{d}r} \sqrt{g_{tt}(r) g_{rr}^{-1}(r)} \right) \right],
\fe
or, more explicitly
\ie
\label{tensorpotentialmetric}
\mathcal{V}^{\, \text{t}}_{\text{metric}}(r) = \left(1-\frac{2 M}{r}\right) \left(\frac{l (l+1)}{r^2}-\frac{2 (\ell r+3 M)}{(\ell+1) r^3}\right).
\fe

\begin{figure}
    \centering
      \includegraphics[scale=0.51]{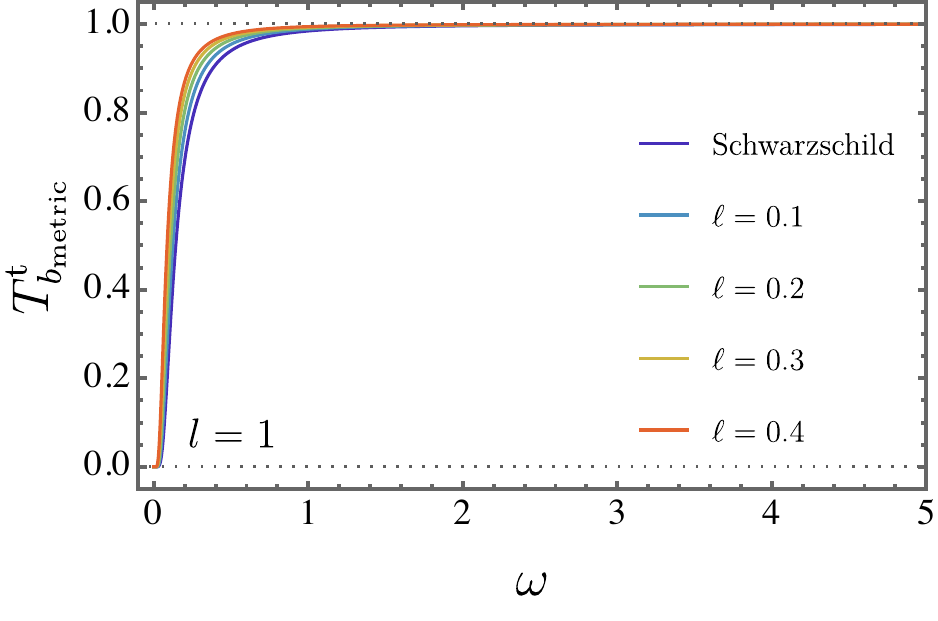}
      \includegraphics[scale=0.515]{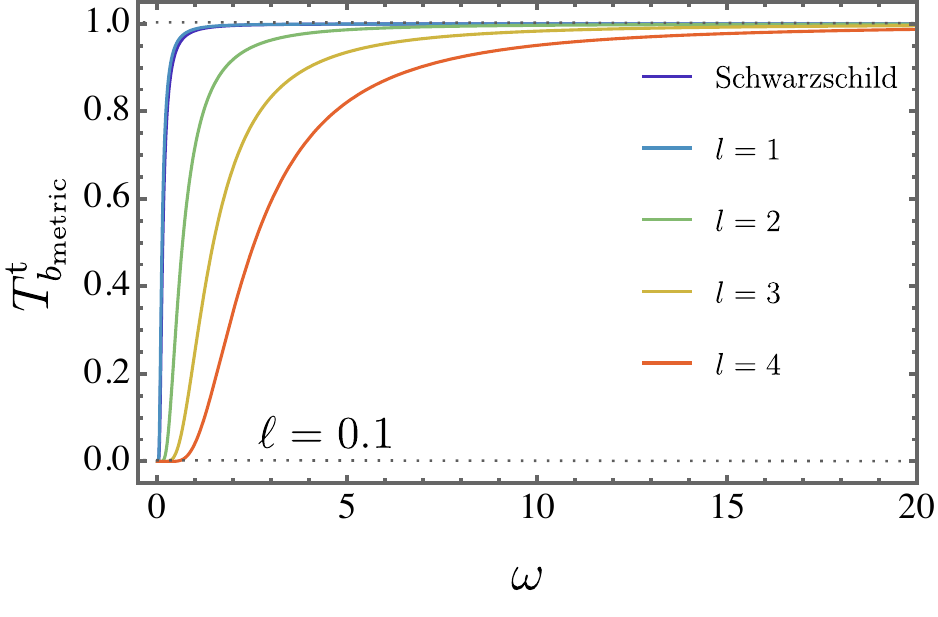}
    \caption{\label{greybodyfactorsfortensormetric}The greybody factors $T^{\,\text{t}}_{b_{\text{metric}}}$ is displayed for different values of $\ell$ when keeping $l=1$ (the the top panel) and for different values of $l$ for a fixed value of $\ell = 0.1$. For both cases, the Schwarzschild case is compared.}
    \label{greybodymetricbosonstensor}
\end{figure}

It is worth noting that in the limit $\ell \to 0$, the effective potential for odd--parity tensor perturbations recovers the familiar form associated with the Schwarzschild solution. With this groundwork in place, the study of greybody factors for tensorial perturbations can now be carried out as follows
\ie
T^{\,\text{t}}_{b_{\text{metric}}} = \text{sech}^2\left[\frac{l (l+1) (\ell+1)-2 \ell-1}{(2 \omega ) \left(4 (\ell+1)^{3/2} M\right)}\right].
\fe
Fig. \ref{greybodymetricbosonstensor} displays the greybody factors associated with tensor perturbations, $T^{\,\text{t}}_{b_{\text{metric}}}$, under two distinct conditions: the top panel illustrates the behavior as $\ell$ varies with $l$ fixed at 1, while the bottom panel shows results for different values of $l$ with $\ell$ held constant at 0.1. In each case, the outcomes are compared against the corresponding Schwarzschild baseline.

\begin{figure}
    \centering
      \includegraphics[scale=0.55]{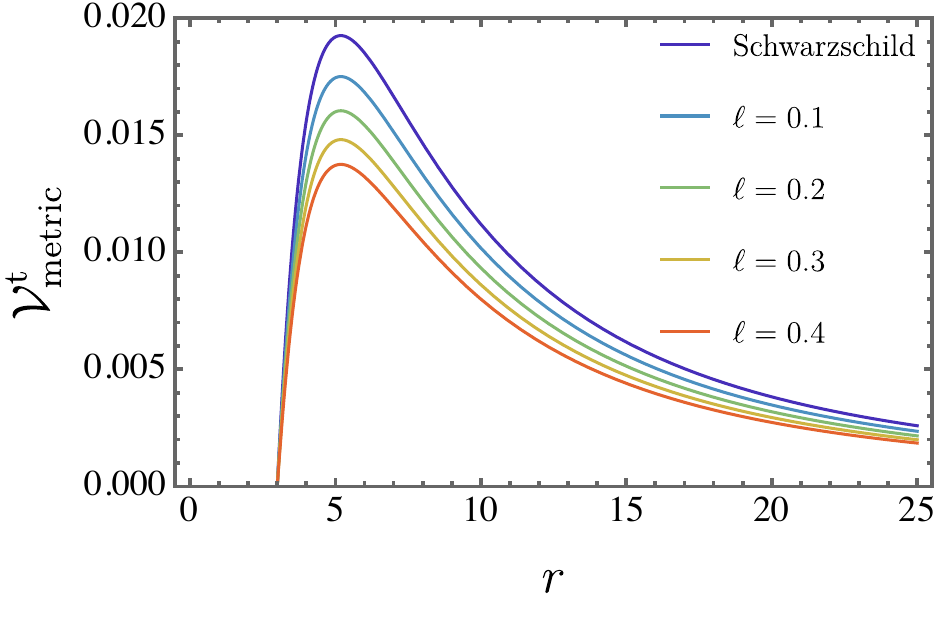}
    \caption{The effective potential $\mathcal{V}^{\,\text{t}}_{{\text{metric}}}$ is shown for different values of $\ell$ for $l=1$. Also, the Schwarzschild case is compared in this analysis.}
    \label{tensorpotential}
\end{figure}

%%%%%%%%%%%%%%%%%%%%%%%%%%%%%%%%%%%%%%%%%%%%%%%%%%%%%%%%%%%%%%%%%%%%%%%%%%%%%%%%%%%%%%%%%%%%%%%%%%%%%%%%%%%%%%%%%%%%%%%%%%%%%%%%%%%%%%%%%%%%%%%%%%%%%%%%%%%%%%%%%%%%%%%%%%%%%%%%%%%%%%%%%%%%%%%%%%%%%%%%%%%%%%%%%%%%%%%%%%%%%%%%%%%%%%%%%%%%%%%%%%%%%%%%%%%%%%%%%%%%%%%%%%%%%%%%%%%%%%%%%%%%%%%%%%%%%%%%%%%%%%%%%%%%%%%%%%%%%%%%%%%%%%%%%%%%%%%%%%%%%%%%%%%%%%%%%%%%%%%%%%%%%%%%%%%%%%%%%%%%%%%%%%%%%%%%%%%%%%%%%%%%%%%%%%%%%%%%%%%%%%%%%%%%%%%%%%%%%%%%%%%%%%%%%%%%%%%%%%%%%%%%%%%%%%%%%%%%%%%%%%%%%%%%%%%%%%%%%%%%%%%%%%%%%%%%%%%%%%%%%%%%%%%%%%%%%%%%%%%%%%%%%%%%%%%%%%%%%%%%%%%%%%%%%%%%%%%%%%%%%%%%%%%%%%%%%%%%%%%%%%%%%

\subsection{Greybody factors for fermions}

In this analysis, we explore the behavior of massless Dirac perturbations within a static and spherically symmetric black hole spacetime. To study the dynamics of the massless spin--1/2 field, we adopt the Newman--Penrose formalism as our primary approach. The corresponding Dirac equations, which govern this scenario, are formulated as \cite{newman1962approach, chandrasekhar1984mathematical}:
\begin{align}
(D + \epsilon - \rho) \psi_1 +( \bar{\delta} + \pi - \alpha) \psi_2 &= 0, \\
(\Delta + \mu - \gamma) \psi_2 + (\delta + \beta - \tau) \psi_1 &= 0.
\end{align}
In this framework, the Dirac spinors $\psi_1$ and $\psi_2$ are introduced, with the directional derivatives $D = l^\mu \partial_\mu$, $\Delta = n^\mu \partial_\mu$, $\delta = m^\mu \partial_\mu$, and $\bar{\delta} = \bar{m}^\mu \partial_\mu$ associated with the chosen null tetrad.

To proceed, the null tetrad basis vectors are constructed based on the metric components and are expressed as follows:
\ie
\begin{split}
l^\mu &= \left(\frac{1}{\mathcal{A}(r)}, \sqrt{\frac{\mathcal{B}(r)}{\mathcal{A}(r)}}, 0, 0\right), \\
n^\mu & = \frac{1}{2} \left(1, -\sqrt{\mathcal{A}(r) \mathcal{B}(r)}, 0, 0\right), \\
m^\mu &= \frac{1}{\sqrt{2} r} \left(0, 0, 1, \frac{i}{\sin \theta}\right), \\
\bar{m}^\mu &= \frac{1}{\sqrt{2} r} \left(0, 0, 1, \frac{-i}{\sin \theta}\right).
\end{split}
\fe

Based on these definitions, the non--zero components of the spin coefficients are determined as:
\ie
\begin{split}
 & \rho = -\frac{1}{r} \frac{\mathcal{B}(r)}{\mathcal{A}(r)},  \quad
\mu = -\frac{\sqrt{\mathcal{A}(r) \mathcal{B}(r)}}{2r}, \\ & \gamma = \frac{\mathcal{A}(r)'}{4}\sqrt{\frac{\mathcal{B}(r)}{\mathcal{A}(r)}},  \quad
\beta = -\alpha = \frac{\cot{\theta}}{2\sqrt{2}r}. 
\end{split}
\fe

By decoupling the equations governing the dynamics of a massless Dirac field, a single equation of motion for $\psi_1$ is obtained, encapsulating its behavior
\begin{align}
\left[(D - 2\rho)(\Delta + \mu - \gamma) - (\delta + \beta) (\bar{\delta}+\beta)\right] \psi_1 = 0.
\end{align}

By substituting the explicit forms of the directional derivatives and spin coefficients, the equation can be rewritten in the following form
\ie
\begin{split}
&\left[ \frac{1}{2\mathcal{A}(r)} \partial_t^2 - \left( \frac{\sqrt{\mathcal{A}(r)\mathcal{B}(r)}}{2r} +\frac{\mathcal{A}(r)'}{4}\sqrt{\frac{\mathcal{B}(r)}{\mathcal{A}(r)}}\right)\frac{1}{\mathcal{A}(r)}\partial_t \right. \\
& \left. - \frac{\sqrt{\mathcal{A}(r)\mathcal{B}(r)}}{2} \sqrt{\frac{\mathcal{B}(r)}{\mathcal{A}(r)}}\partial_r^2 \right. \\
& \left. -\sqrt{\frac{\mathcal{B}(r)}{\mathcal{A}(r)}} \partial_r \left( \frac{\sqrt{\mathcal{A}(r)\mathcal{B}(r)}}{2} + \frac{\mathcal{A}(r)'}{4}{\sqrt{\frac{\mathcal{B}(r)}{\mathcal{A}(r)}}} \right) \right] \psi_1 \\ + 
&\left[ \frac{1}{\sin^2\theta} \partial_\phi^2 + i \frac{\cot \theta}{\sin \theta}\partial_\phi \right. \\
& \left.+ \frac{1}{\sin \theta}\partial_\theta \left( \sin \theta \partial_\theta \right) - \frac{1}{4} \cot^2 \theta + \frac{1}{2} \right] \psi_1 = 0.
\end{split}
\fe

To achieve separation of the equations into radial and angular parts, the wave function reads
\begin{align}
\psi_1 = \Psi(r) Y_{lm}(\theta, \phi) e^{-i \omega t},
\end{align}
so that
\begin{align}
&\left[  \frac{-\omega^2}{2\mathcal{A}(r)} - \left(\frac{\sqrt{\mathcal{A}(r)\mathcal{B}(r)}}{2r}+\frac{\mathcal{A}(r)'}{4} + \sqrt{\frac{\mathcal{B}(r)}{\mathcal{A}(r)}}\right)\frac{- i\omega}{\mathcal{A}(r)} \right. \\
& \left. - \frac{\sqrt{\mathcal{A}(r)\mathcal{B}(r)}}{2} \sqrt{\frac{\mathcal{B}(r)}{\mathcal{A}(r)}}\partial_r^2 -\lambda_{lm} \right. \\
& \left. - \sqrt{\frac{\mathcal{B}(r)}{\mathcal{A}(r)}}\partial_r \left(\frac{\sqrt{\mathcal{A}(r)\mathcal{B}(r)}}{2r} + \frac{\mathcal{A}(r)'}{4}\sqrt{\frac{\mathcal{B}(r)}{\mathcal{A}(r)}}\right) \right] \Psi(r) = 0.
\end{align}

\begin{figure}
    \centering
      \includegraphics[scale=0.55]{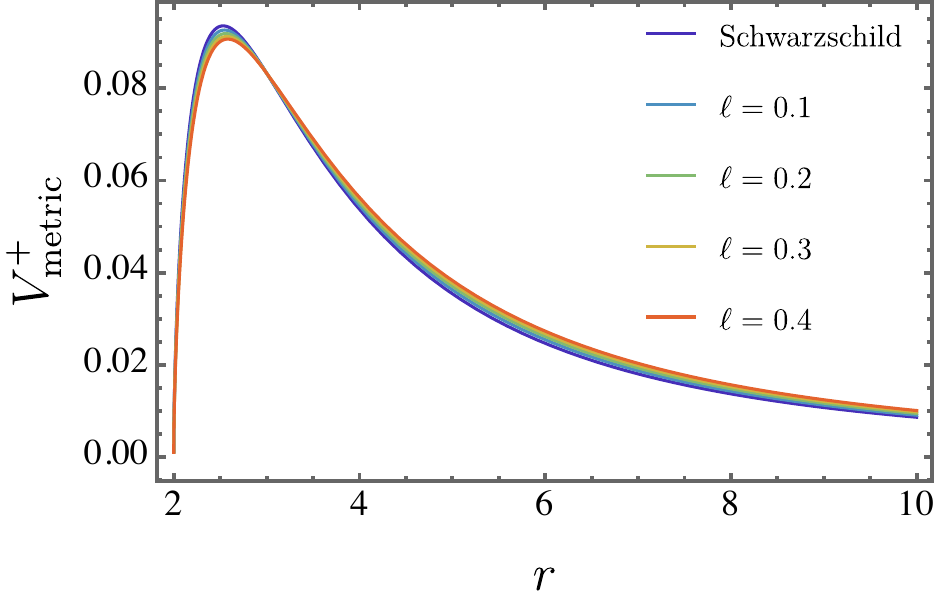}
    \caption{The effective potential $V_{\text{metric}}^{+}$ is shown for different values of $\ell$. Also, the Schwarzschild case is compared in this analysis.}
    \label{vmetric}
\end{figure}

In this context, $\lambda_{lm}$ functions as the separation constant. Utilizing the generalized tortoise coordinate $r^*$, the radial wave equation is converted into a Schrödinger--like form, given by:
\begin{align}
\left[\frac{\mathrm{d}^2 }{\mathrm{d}r_*^2} +( \omega^2 - V_{\text{metric}}^{\pm}) \right]\Psi_{\pm}(r) = 0.
\end{align}
Furthermore, the potentials $V_{\text{metric}}^{\pm}$ associated with the massless spin--1/2 field are defined as \cite{albuquerque2023massless, al2024massless,arbey2021hawking}
\ie
\begin{split}\label{Vpm}
& V_{\text{metric}}^{\pm} = \frac{(l + \frac{1}{2})^2}{r^2} \mathcal{A}(r) \\
& \pm \left(l + \frac{1}{2}\right) \sqrt{\mathcal{A}(r) \mathcal{B}(r)} \partial_r \left(\frac{\sqrt{\mathcal{A}(r)}}{r}\right).
\end{split}
\fe
For this analysis, we choose the potential $V_{\text{metric}}^+$ without any loss of generality. A parallel approach can be applied to $V_{\text{metric}}^-$; however, as the qualitative behavior of $V_{\text{metric}}^-$ is analogous to that of $V_{\text{metric}}^+$ \cite{albuquerque2023massless,devi2020quasinormal}, the focus will remain on $V_{\text{metric}}^+$. To illustrate the characteristics of $V_{\text{metric}}^+$, we provide Fig. \ref{vmetric}. As expected, $V_{\text{metric}}^+$ approaches zero in the limit $r \to \infty$.

In other words, using the Dirac effective potential from Eq. \eqref{Vpm}, we derive the greybody factor bounds for bumblebee in the metric formalism. They can be expressed in a simplified form as
\ie
\begin{split}
	T_{b_{\text{metric}}} & \ge \mathrm{sech}^2 {\left(\frac{1}{2\omega}\int_{2M}^ {+\infty} \frac{V_{\text{metric}}^{+}} {\sqrt{\mathcal{A}(r)\mathcal{B}(r)}} \rm{d}r\right) }\\
 & =  \mathrm{sech}^2 \left[\frac{1}{2\omega} \left(  \frac{(2 l+1)^2 \sqrt{\ell+1}}{8 M} \right)\right].
\end{split}
\fe

\begin{figure}
    \centering
      \includegraphics[scale=0.51]{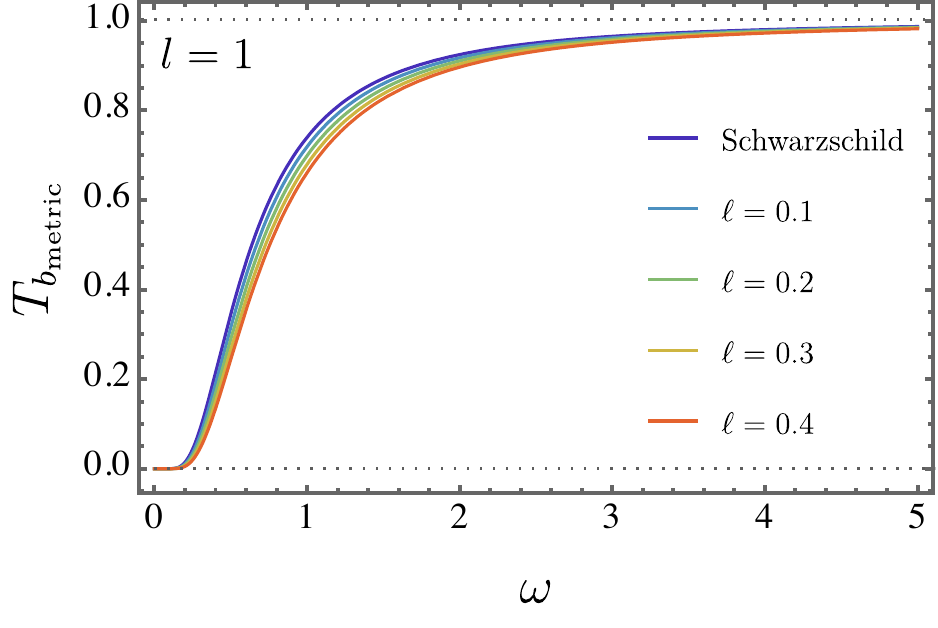}
      \includegraphics[scale=0.51]{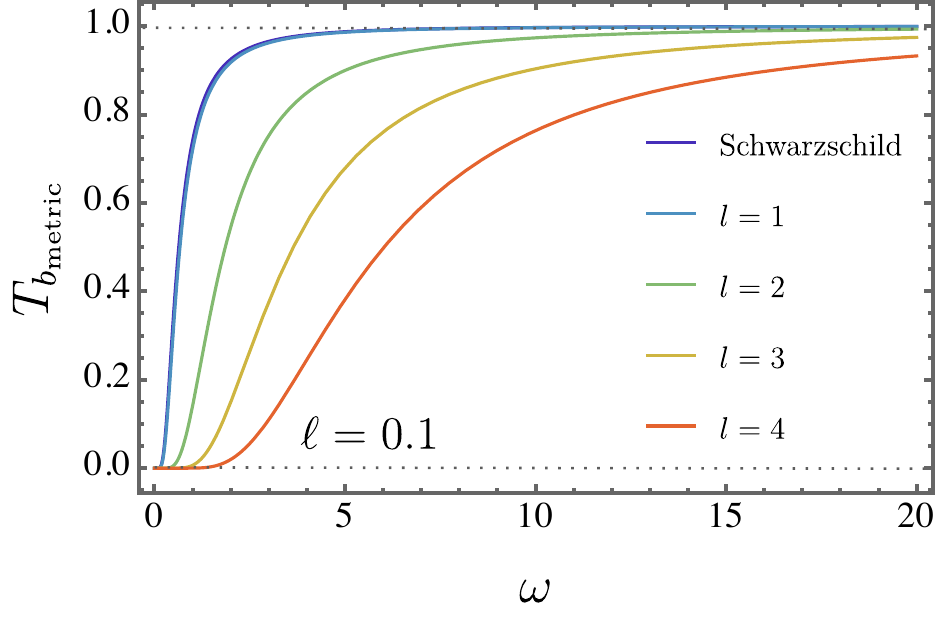}
    \caption{The greybody factors $T_{b_{\text{metric}}}$ is displayed for different values of $\ell$ when keeping $l=1$ (the the top panel) and for different values of $l$ for a fixed value of $\ell = 0.1$. For both cases, the Schwarzschild case is compared.}
    \label{greybodymetricfermions}
\end{figure}

Fig. \ref{greybodymetricfermions} illustrates how the greybody factor varies with frequency $\omega$. The top panel displays the behavior of them for different values of $\ell$, while keeping $M = 1$ and $l = 1$ fixed. The plots show that as the Lorentz symmetry--breaking parameter $\ell$ increases, the value of $T_{b_{\text{metric}}}$ decreases. Additionally, the results for the bumblebee black hole (within the \textit{metric} formalism) are consistently lower compared to the Schwarzschild case. In the bottom panel, the variation of $T_{b_{\text{metric}}}$ is shown as $\omega$ changes for different values of $l$ (for a fixed value of $\ell = 0.1$). As $l$ increases, $T_{b_{\text{metric}}}$ decreases. In other words, the Schwarzschild case emerges as the one with the highest intensity.

%%%%%%%%%%%%%%%%%%%%%%%%%%%%%%%%%%%%%%%%%%%%%%%%%%%%%%%%%%%%%%%%%%%%%%%%%%%%%%%%%%%%%%%%%%%%%%%%%%%%%%%%%%%%%%%%%%%%%%%%%%%%%%%%%%%%%%%%%%%%%%%%%%%%%%%%%%%%%%%%%%%%%%%%%%%%%%%%%%%%%%%%%%%%%%%%%%%%%%%%%%%%%%%%%%%%%%%%%%%%%%%%%%%%%%%%%%%%%%%%%%%%%%%%%%%%%%%%%%%%%%%%%%%%%%%%%%%%%%%%%%%%%%%%%%%%%%%%%%%%%%%%%%%%%%%%%%%%%%%%%%%%%%%%%%%%%%%%%%%%%%%%%%%%%%%%%%%%%%%%%%%%%%%%%%%%%%%%%%%%%%%%%%%%%%%%%%%%%%%%%%%%%%%%%%%%%%%%%%%%%%%%%%%%%%%%%%%%%%%%%%%%%%%%%%%%%%%%%%%%%%%%%%%%%%%%%%%%%%%%%%%%%%%%%%%%%%%%%%%%%%%%%%%%%%%%%%%%%%%%%%%%%%%%%%%%%%%%%%%%%%%%%%%%%%%%%%%%%%%%%%%%%%%%%%%%%%%%%%%%%%%%%%%%%%%%%%%%%%%%%%%%%%%%%%%%%%%%%%%%%%%%%%%%%%%%%%%%%%%%%%%%%%%%%%%%%%%%%%%%%%%%%%%%%%%%%%%%%%%%%%%%%%%%%%%%%%%%%%%%%%%%%%%%%%%%%%%

\subsection{The emission rate}

Within black holes, quantum fluctuations near the event horizon lead to the constant creation and annihilation of particles. Through a process referred to as tunneling, particles with positive energy can escape the gravitational pull of the black hole. Over time, this mechanism results in the gradual loss of the black hole's mass, a phenomenon recognized as Hawking radiation, as discussed in the previous subsections. From the viewpoint of a distant observer, the black hole's shadow corresponds to a high--energy absorption cross-section, which stabilizes at an approximate constant value, $\sigma_{lim}$. According to Ref. \cite{decanini2011universality,papnoi2022rotating}, the energy emission rates are expressed as
\ie
\label{emission}
	\frac{{{\mathrm{d}^2}E}}{{\mathrm{d}\omega \mathrm{d}t}} = \frac{{2{\pi ^2}\sigma_{lim}}}{{{e^{\frac{\omega }{T_{\text{metric}}}}} - 1}} {\omega ^3},
\fe
with $\omega$ represents the photon frequency. The constant limiting value $\sigma_{lim}$ is related to the shadow radius by $\sigma_{lim} \approx \pi R_{\text{sh}}^2$. By substituting the expressions for the shadow radius and the Hawking temperature, the energy emission rate becomes 
\ie
\frac{\mathrm{d}^{2}E}{\mathrm{d}\omega \mathrm{d} t} = \frac{54 \pi ^3 M^2 \omega ^3}{e^{8 \pi  \sqrt{\ell + 1} M \omega }-1}.
\fe

\begin{figure}
    \centering
      \includegraphics[scale=0.55]{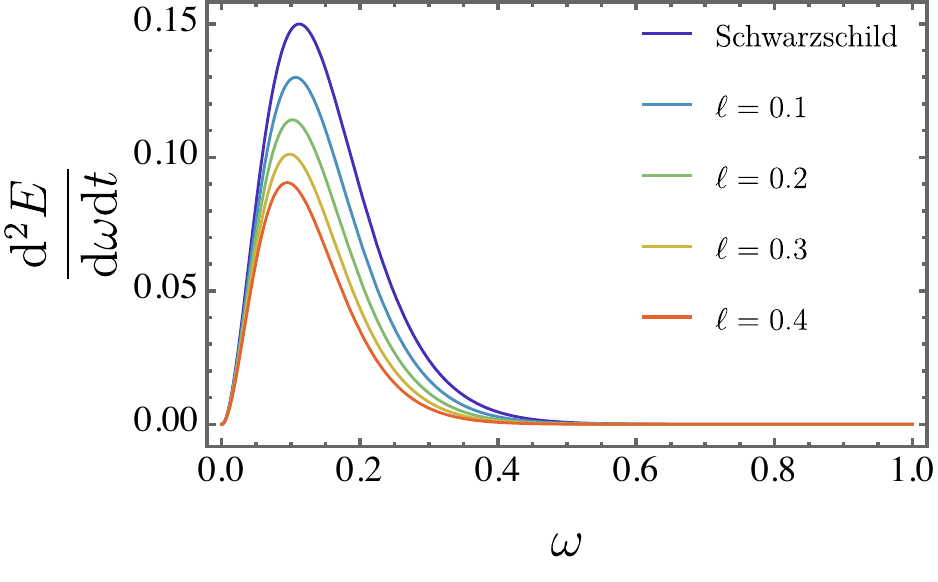}
    \caption{The emission rate for different values of $\ell$ against the frequency $\omega$.}
    \label{emissionratemetric}
\end{figure}

Fig. \ref{emissionratemetric} illustrates the emission rate as a function of $\omega$ for various values of $\ell$. In general lines, as $\ell$ increases, the emission rate decreases in magnitude. For comparison, the Schwarzschild black hole is also included in the analysis. These findings are consistent with the particle density results ($n_{\text{metric}}$) discussed in the previous subsections.

%%%%%%%%%%%%%%%%%%%%%%%%%%%%%%%%%%%%%%%%%%%%%%%%%%%%%%%%%%%%%%%%%%%%%%%%%%%%%%%%%%%%%%%%%%%%%%%%%%%%%%%%%%%%%%%%%%%%%%%%%%%%%%%%%%%%%%%%%%%%%%%%%%%%%%%%%%%%%%%%%%%%%%%%%%%%%%%%%%%%%%%%%%%%%%%%%%%%%%%%%%%%%%%%%%%%%%%%%%%%%%%%%%%%%%%%%%%%%%%%%%%%%%%%%%%%%%%%%%%%%%%%%%%%%%%%%%%%%%%%%%%%%%%%%%%%%%%%%%%%%%%%%%%%%%%%%%%%%%%%%%%%%%%%%%%%%%%%%%%%%%%%%%%%%%%%%%%%%%%%%%%%%%%%%%%%%%%%%%%%%%%%%%%%%%%%%%%%%%%%%%%%%%%%%%%%%%%%%%%%%%%%%%%%%%%%%%%%%%%%%%%%%%%%%%%%%%%%%%%%%%%%%%%%%%%%%%%%%%%%%%%%%%%%%%%%%%%%%%%%%%%%%%%%%%%%%%%%%%%%%%%%%%%%%%%%%%%%%%%%%%%%%%%%%%%%%%%%%%%%%%%%%%%%%%%%%%%%%%%%%%%%%%%%%%%%%%%%%%%%%%%%%%%%%%%%%%%%%%%%%%%%%%%%%%%%%%%%%%%%%%%%%%%%%%%%%%%%%%%%%%%%%%%%%%%%%%%%%%%%%%%%%%%%%%%%%%%%%%%%%%%%%%%%%%%%%%%

\subsection{The evaporation process}

The parameter $\ell$, associated with Lorentz violation, does not impact the radii of the event horizon, photon sphere, or shadow. However, it significantly affects the Hawking temperature, resulting in a distinct evaporation process compared to the Schwarzschild black hole. By applying the surface gravity method \cite{heidari2023gravitational}, the expression is written as
\ie
\label{sufacceeee}
T_{\text{metric}} = \frac{1}{8 \pi  \sqrt{1 + \ell} \, M},
\fe
and using the \textit{Stefan--Boltzmann} law, we have
\ie
\label{aseerrtlarfw}
\frac{\mathrm{d}M}{\mathrm{d}\tau} = - \alpha a \sigma T_{\text{metric}}^{4}.
\fe

In this case, $\alpha$ corresponds to the greybody factor, $a$ is the radiation constant, and $\sigma$ represents the cross--sectional area. Under the geometric optics approximation, $\sigma$ is interpreted as the photon capture cross--section, given by $\sigma = \pi (3\sqrt{3}M)^2$, leading to
\ie
\begin{split}
\int_{0}^{t_{\text{metric}}} \xi \mathrm{d}\tau & = - \int_{M_{i}}^{M_{f}} 
\left[ \frac{27 \xi}{4096 \pi ^3 (1+\ell)^2 M^2}  \right]^{-1} \mathrm{d}M.
\end{split}
\fe
After performing the integration, it reads
\ie
t_{\text{metric}} = - \frac{4096 \pi ^3 (1 + \ell)^2 \left(M_{f}^3 - M_{i}^3\right)}{81 \xi}.
\fe

For this particular black hole configuration, no remnant mass is anticipated. Consequently, it is assumed that the black hole will undergo complete evaporation, with \(M_f \to 0\)
\ie
t_{\text{metric}} = \frac{4096 \pi ^3 (1 + \ell)^2 M_{i}^3}{81 \xi}.
\fe

\begin{figure}
    \centering
      \includegraphics[scale=0.55]{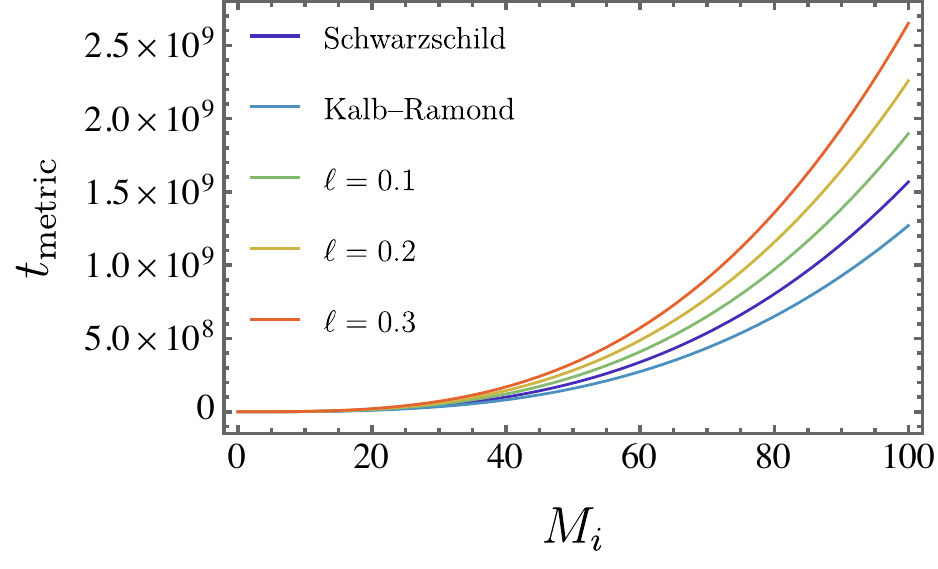}
    \caption{The evaporation time \( t_{\text{metric}} \) is shown for different values of \( \ell \). A comparison with the Schwarzschild and Kalb--Ramond cases is shown.}
    \label{evaporationmetric}
\end{figure}

To provide a clearer understanding of our findings, Fig. \ref{evaporationmetric} presents the evaporation time, $t_{\text{metric}}$, for different values of $\ell$. As $\ell$ increases, the evaporation time also increases. The results are compared to the Schwarzschild solution, revealing that the Lorentz--violating configuration evaporates more slowly than the Schwarzschild black hole. Additionally, when compared to another Lorentz--violating black hole model recently proposed in the context of Kalb--Ramond gravity \cite{Liu:2024oas}, the evaporation times follow this order: $t_{\text{KR}}$ (the fastest), $t_{\text{schw}}$ (intermediate), and $t_{\text{metric}}$ (the slowest).

%%%%%%%%%%%%%%%%%%%%%%%%%%%%%%%%%%%%%%%%%%%%%%%%%%%%%%%%%%%%%%%%%%%%%%%%%%%%%%%%%%%%%%%%%%%%%%%%%%%%%%%%%%%%%%%%%%%%%%%%%%%%%%%%%%%%%%%%%%%%%%%%%%%%%%%%%%%%%%%%%%%%%%%%%%%%%%%%%%%%%%%%%%%%%%%%%%%%%%%%%%%%%%%%%%%%%%%%%%%%%%%%%%%%%%%%%%%%%%%%%%%%%%%%%%%%%%%%%%%%%%%%%%%%%%%%%%%%%%%%%%%%%%%%%%%%%%%%%%%%%%%%%%%%%%%%%%%%%%%%%%%%%%%%%%%%%%%%%%%%%%%%%%%%%%%%%%%%%%%%%%%%%%%%%%%%%%%%%%%%%%%%%%%%%%%%%%%%%%%%%%%%%%%%%%%%%%%%%%%%%%%%%%%%%%%%%%%%%%%%%%%%%%%%%%%%%%%%%%%%%%%%%%%%%%%%%%%%%%%%%%%%%%%%%%%%%%%%%%%%%%%%%%%%%%%%%%%%%%%%%%%%%%%%%%%%%%%%%%%%%%%%%%%%%%%%%%%%%%%%%%%%%%%%%%%%%%%%%%%%%%%%%%%%%%%%%%%%%%%%%%%%%%%%%%%%%%%%%%%%%%%%%%%%%%%%%%%%%%%%%%%%%%%%%%%%%%%%%%%%

\section{The \textit{Metric--affine} case}

This section focuses on examining the same aspects discussed in the previous section, but within the framework of the \textit{metric--affine} approach. Following this analysis, the results will be compared to those obtained for the \textit{metric} case.

%%%%%%%%%%%%%%%%%%%%%%%%%%%%%%%%%%%%%%%%%%%%%%%%%%%%%%%%%%%%%%%%%%%%%%%%%%%%%%%%%%%%%%%%%%%%%%%%%%%%%%%%%%%%%%%%%%%%%%%%%%%%%%%%%%%%%%%%%%%%%%%%%%%%%%%%%%%%%%%%%%%%%%%%%%%%%%%%%%%%%%%%%%%%%%%%%%%%%%%%%%%%%%%%%%%%%%%%%%%%%%%%%%%%%%%%%%%%%%%%%%%%%%%%%%%%%%%%%%%%%%%%

\subsection{Bosonic modes}

%%%%%%%%%%%%%%%%%%%%%%%%%%%%%%%%%%%%%%%%%%%%%%%%%%%%%%%%%%%%%%%%%%%%%%%%%%%%%%%%%%%%%%%%%%%%%%%%%%%%%%%%%%%%%%%%%%%%%%%%%%%%%%%%%%%%%%%%%%%%%%%%%%%%%%%%%%%%%%%%%%%%%%%%%%%%%%%%%%%%%%%%%%%%%%%%%%%%%%%%%%%%%%%%%%%%%%%%%%%%%%%%%%%%%%%%%%%%%%%%%%%%%%%%%%%%

\subsubsection{The Hawking radiation}

Following a similar approach to the one employed in the previous section for the bumblebee model in the \textit{metric} formalism, we now focus on the \textit{metric--affine} framework. Specifically, analogous to Eq. (\ref{rvv}), we write
\ie
\label{sddsaas}
r_{\text{met--aff}} = 2M - \frac{1}{4} E \lambda  \sqrt[4]{4-X} \sqrt[4]{-(X-4)^3} ,
\fe
where the negative solution of the square root in Eq. (\ref{fgfgfg}) is also considered to account for ingoing geodesics. Through this approach, we obtain
\ie
u_{\text{met--aff}}(\lambda) = -\frac{4 M \left(-(X-4)^3\right)^{3/4} \sqrt{3 X+4} }{(4-X)^{11/4}}\ln \left(\frac{\lambda}{C^{\prime}} \right),
\fe
and
\ie
p_{\omega_{\text{met--aff}}} =\int_0^\infty \left ( \alpha_{\omega\omega^\prime_{\text{met--aff}}} f_{\omega^\prime_{\text{met--aff}}} + \beta_{\omega\omega^\prime_{\text{met--aff}}} \bar{f}_{\omega^\prime_{\text{met--aff}}}  \right)\mathrm{d} \omega^\prime,
\fe
such that the Bogoliubov coefficients are expressed below
\begin{equation}
\begin{split}
& \alpha_{\omega\omega^\prime_{\text{met--aff}}} = -i K e^{i\omega^\prime v_0}e^{\pi \left[\frac{2 M \left(-(X-4)^3\right)^{3/4} \sqrt{3 X+4} }{(4-X)^{11/4}} \right]\omega}  \\ 
& \times \int_{-\infty}^{0} \,\mathrm{d}x\,\Big(\frac{\omega^\prime}{\omega}\Big)^{1/2}e^{\omega^\prime x}  e^{i\omega\left[\frac{4 M \left(-(X-4)^3\right)^{3/4} \sqrt{3 X+4} }{(4-X)^{11/4}}\right]\ln\left(\frac{|x|}{C^{\prime}D^{\prime}}\right)},
\end{split}
\end{equation}
and
\begin{equation}
\begin{split}
& \beta_{\omega\omega'_{\text{met--aff}}} = i K e^{-i\omega^\prime v_0}e^{-\pi \left[ \frac{2 M \left(-(X-4)^3\right)^{3/4} \sqrt{3 X+4} }{(4-X)^{11/4}} \right]\omega}\\
& \times  \int_{-\infty}^{0} \,\mathrm{d}x\,\left(\frac{\omega^\prime}{\omega}\right)^{1/2}e^{\omega^\prime x}  e^{i\omega\left[\frac{4 M \left(-(X-4)^3\right)^{3/4} \sqrt{3 X+4} }{(4-X)^{11/4}}\right]\ln\left(\frac{|x|}{C^{\prime}D^{\prime}}\right)}.
\end{split}
\end{equation}
Notice that, after performing algebraic manipulations, the correlation between $\alpha_{\omega\omega^\prime_{\text{met--aff}}}$ and $\beta_{\omega\omega'_{\text{met--aff}}}$ turns out to be
\begin{equation}
    |\alpha_{\omega\omega'_{\text{met--aff}}}|^2 = e^{\big( \frac{8 \pi M \left(-(X-4)^3\right)^{3/4} \sqrt{3 X+4} }{(4-X)^{11/4}} \big)\omega}|\beta_{\omega\omega'_{\text{met--aff}}}|^2\,.
\end{equation}
In a manner similar to the approach taken in the previous sections, we now examine the interval $\omega$ to $\omega + \mathrm{d}\omega$, resulting in
\ie
\mathcal{P}_{\text{met--aff}}(\omega, \ell)=\frac{\mathrm{d}\omega}{2\pi}\frac{1}{\left \lvert\frac{\alpha_{\omega\omega^\prime_{\text{met--aff}}}}{\beta_{\omega\omega^\prime_{\text{met--aff}}}}\right \rvert^2-1}\, ,
\fe
so that
\ie
\mathcal{P}_{\text{met--aff}}(\omega, \ell)=\frac{\mathrm{d}\omega}{2\pi}\frac{1}{e^{\left(\frac{8 \pi M \left(-(X-4)^3\right)^{3/4} \sqrt{3 X+4} }{(4-X)^{11/4}}\right)\omega}-1}\,.
\fe
It is worth mentioning that, when compared to Planck distribution, we have
\begin{equation} \label{fffgg}
    \mathcal{P}_{\text{met--aff}}(\omega, \ell)=\frac{\mathrm{d}\omega}{2\pi}\frac{1}{e^{\frac{\omega}{T}}-1},
\end{equation}
yielding
\ie
\begin{split}
\label{hawtemp}
    T_{\text{metric--affine}} =&  \frac{(4-X)^{11/4}}{8 \pi M \left(-(X-4)^3\right)^{3/4} \sqrt{3 X+4} } \\
    \approx &  \, \frac{1}{8 \pi M} - \frac{X}{16 \pi M}.
    \end{split}
\fe

\begin{figure}
    \centering
      \includegraphics[scale=0.55]{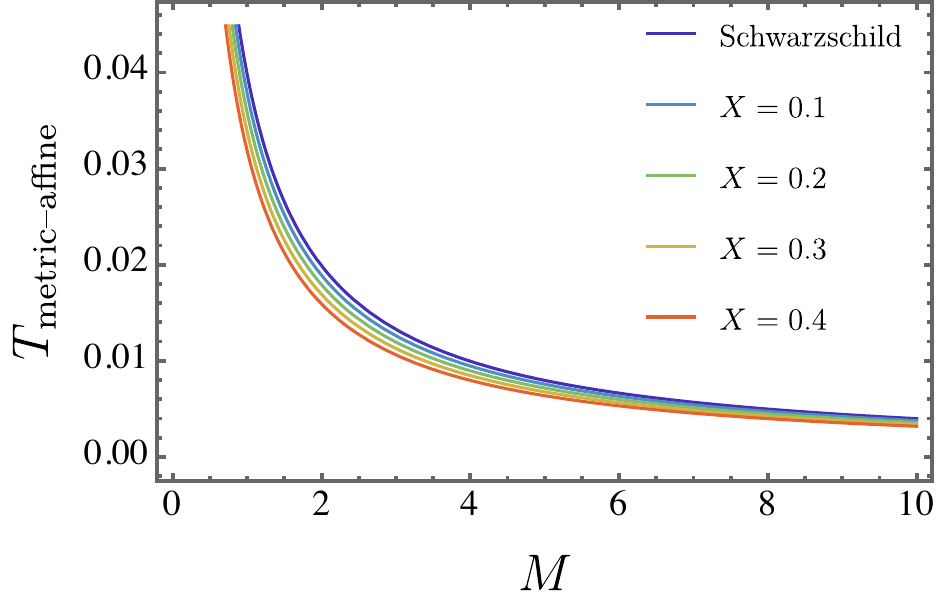}
    \caption{The Hawking temperature $T_{\text{metric--affine}}$ is exhibited for different values of $X$. The Schwarzschild case is compared.}
    \label{hawkingtemperatureformetricaffine}
\end{figure}

Here, we have assumed $X$ to be small. Notably, in the limit $X \to 0$, the Schwarzschild temperature is recovered. It is important to highlight that these findings are consistent with those reported in Ref. \cite{araujo2024gravitational}. In general lines, as the Lorentz--violating parameter $X$ increases, the magnitude of $T_{\text{metric--affine}}$ against $M$ decreases (as it can be seen in Fig. \ref{hawkingtemperatureformetricaffine}). Compared to the Schwarzschild case, the values of $T_{\text{metric--affine}}$ are consistently smaller as $X$ varies. Furthermore, as observed in the previous section, this scenario does not allow for the existence of a remnant mass.

Similar to the case of the bumblebee black hole to the \textit{metric} case, the continuous radiation emitted by the black hole causes a steady reduction in its mass, leading to its gradual contraction. To analyze this effect, we will adopt the tunneling framework proposed by Parikh and Wilczek \cite{011} in the next section. This approach aligns with the methodology applied in the earlier discussions.

%%%%%%%%%%%%%%%%%%%%%%%%%%%%%%%%%%%%%%%%%%%%%%%%%%%%%%%%%%%%%%%%%%%%%%%%%%%%%%%%%%%%%%%%%%%%%%%%%%%%%%%%%%%%%%%%%%%%%%%%%%%%%%%%%%%%%%%%%%%%%%%%%%%%%%%%%%%%%%%%%%%%%%%%%%%%%%%%%%%%%%%%%%%%%%%%%%%%%%%%%%%%%%%%%%%%%%%%%%%%%%%%%%%%%%%%%%%%%%%%%%%%%%%%%%%%%%%%%%%%%%%%%%%%%%%%%%%%%%%%%%%%%%%%%%%%%%%%%%%%%%%%%%%%%%%%%%%%%%%%%%%%%%%%%%%%%%%%%%%%%%%%%%%%%%%%%%%%%%%%%%%%%%%%%%%%%%%%%%%%%%%%%%%%%%%%%%%%%%%%%%%%%%%%%%%%%%%%%%%%%%%%%%%%%%%%%%%%%%%%%%%%%%%%%%%%%%%%%%%%%%%%%%%%%%%%%%%%%%%%%%%%%%%%%%%%%%%%%%%%%%%%%%%%%%%%%%%%%%%%%%%%%%%%%%%%%%%%%%%%%%%%%%%%%%%%%%%%%%%%%%%%%%%%%%%%%%%%%%%%%%%%%%%%%%%%%%%%%%%%%%%%%%%%%%%%%%%%%%%%%%%%%%%%%%%%%%%%%%%%%%%%%%%%%%%%%%%%%%%%

\subsubsection{The tunneling process}

Following the methodology employed in the previous section for the bumblebee black hole, we now focus on:
\ie
\begin{split}
& \Delta(r)_{\text{met--aff}} =\\
& \frac{1}{4} r \left(\frac{2 \sqrt{-(X-4)^3} (M-\omega^{\prime})}{r \sqrt{3 X+4}}-\frac{\sqrt{-(X-4)^3}}{\sqrt{3 X+4}}+4\right)
\end{split}
\fe
in a such way that the integral present in Eq. (\ref{ims}) is cast below
\ie
\begin{split}
\label{ims}
&\text{Im}\, \mathcal{S}_{\text{met--aff}} \\
& =\text{Im}\,\int_{0}^{\omega} -\mathrm{d}\omega'\int_{r_i}^{r_f}\,\frac{\mathrm{d}r}{ 4 \sqrt{\frac{\sqrt{-(X-4)^3}}{(4-X)^{7/2}}}\left( 1-\sqrt{\frac{\Delta(r)_{\text{met--aff}}}{r}}\right)}.
\end{split}
\fe
By substituting $M$ with $(M - \omega')$ in the metric, the function $\Delta(r)_{\text{met--aff}}$ acquires a dependence on $\omega'$. This change introduces a singularity at the location of the modified horizon. A counterclockwise contour integration performed around this singularity results in
\ie
\begin{split}
&\text{Im}\, \mathcal{S}_{\text{met--aff}}  \\
&= \frac{4 \pi  \sqrt{3 X+4} \omega  (M - \frac{\omega}{2} )}{(4-X)^{7/2} \left(\frac{\sqrt{-(X-4)^3}}{(4-X)^{7/2}}\right)^{3/2}}.
\end{split}
\fe
As a result, the emission rate for a Hawking particle, incorporating the Lorentz--violating correction, reads
\ie
\begin{split}
& \Gamma_{\text{met--aff}} \sim e^{-2 \, \text{Im}\, S_{\text{met--aff}}} \\
& = e^{-\frac{8 \pi  \sqrt{3 X+4} \omega  (M - \frac{\omega}{2} )}{(4-X)^{7/2} \left(\frac{\sqrt{-(X-4)^3}}{(4-X)^{7/2}}\right)^{3/2}}} .
\end{split}
\fe
In the limit \(X \to 0\), the standard Schwarzschild case is retrieved, with \(\Gamma = e^{-8 \pi \, \omega \left( M - \frac{\omega}{2} \right)}\). Similarly, when \(\omega \to 0\), the emission spectrum reverts to the original Planckian distribution derived by Hawking. Therefore, ir reads
\ie
    \mathcal{P}_{\text{met--aff}}(\omega)=\frac{\mathrm{d}\omega}{2\pi}\frac{1}{e^{\frac{8 \pi  \sqrt{3 X+4} \omega  (M - \frac{\omega}{2} )}{(4-X)^{7/2} \left(\frac{\sqrt{-(X-4)^3}}{(4-X)^{7/2}}\right)^{3/2}}
    }-1}.
\fe

It, shaped by its additional dependence on \(\omega\), diverges from the conventional blackbody form, a difference that becomes apparent upon closer analysis. At low \(\omega\), the expression simplifies to a Planck--like distribution, albeit with a modified Hawking temperature. Moreover, the particle number density can be derived through the tunneling rate, expressed as:
\ie
n_{\text{met--aff}} = \frac{\Gamma_{\text{met--aff}}}{1 - \Gamma_{\text{met--aff}}} = \frac{1}{e^{\frac{8 \pi  \sqrt{3 X+4} \omega  (M - \frac{\omega}{2} )}{(4-X)^{7/2} \left(\frac{\sqrt{-(X-4)^3}}{(4-X)^{7/2}}\right)^{3/2}}} - 1}.
\fe

\begin{figure}
    \centering
      \includegraphics[scale=0.55]{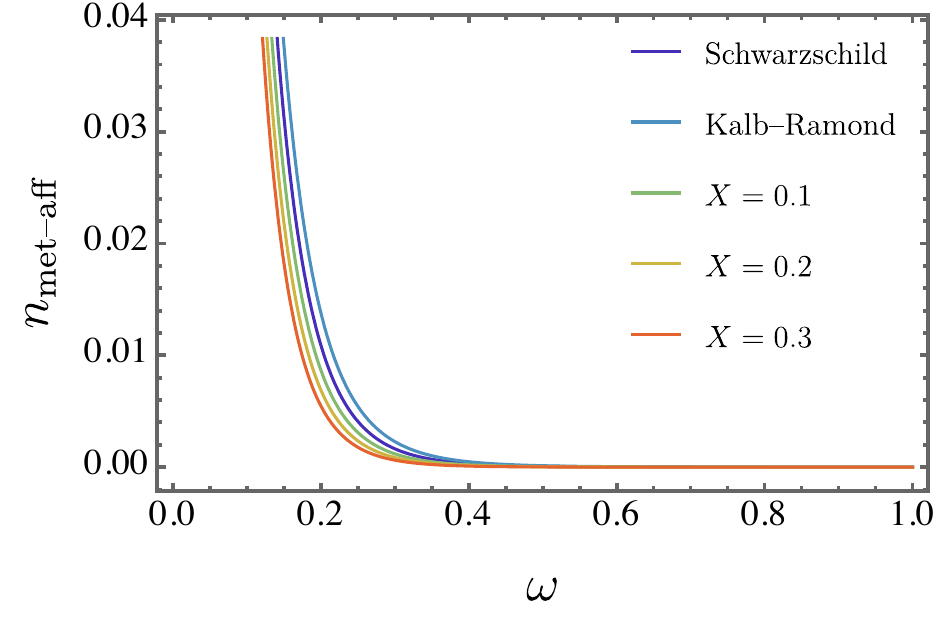}
    \caption{The particle density \( n_{\text{met--aff}} \) is shown for different values of $X$. The Schwarzschild and Kalb--Ramond cases are also compared.}
    \label{pardenmetaffboson}
\end{figure}

To better illustrate the behavior of $n_{\text{met--aff}}$, Fig. \ref{pardenmetaffboson} shows its variation with the Lorentz--violating parameter $X$. As $X$ increases, the particle number density also decreases. Additionally, $n_{\text{met--aff}}$ is compared to the Schwarzschild and Kalb--Ramond cases. The particle density intensities follow the order: $n_{\text{KR}} > n_{\text{Schw}} > n_{\text{met--aff}}$.

\begin{figure}
    \centering
      \includegraphics[scale=0.55]{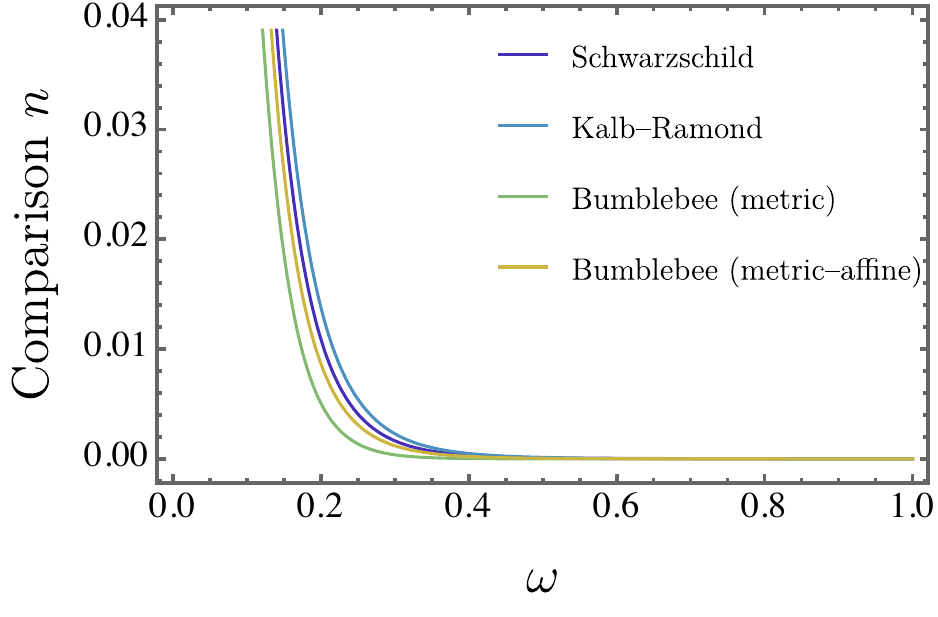}
    \caption{Comparison of \( n \) for Schwarzschild case, Kalb--Ramond, bumblebee in the \textit{metric} formalism and bumblebee with \textit{metric--affine} approach. Here, it is considered $X=0.1 = \ell =0.1$ and $M=1$.}
    \label{comparisonn}
\end{figure}

Further comparisons between the particle densities of the two models developed here for the bosonic case, represented by $n_{\text{metric}}$ and $n_{\text{met--aff}}$, are presented in Fig. \ref{comparisonn}. For reference, the Schwarzschild and Kalb--Ramond cases are included. Overall, the hierarchy of particle density intensities, for $X = \ell = 0.1$, is as follows: $n_{\text{KR}} > n_{\text{Schw}} > n_{\text{met--aff}} > n_{\text{metric}}$.

%%%%%%%%%%%%%%%%%%%%%%%%%%%%%%%%%%%%%%%%%%%%%%%%%%%%%%%%%%%%%%%%%%%%%%%%%%%%%%%%%%%%%%%%%%%%%%%%%%%%%%%%%%%%%%%%%%%%%%%%%%%%%%%%%%%%%%%%%%%%%%%%%%%%%%%%%%%%%%%%%%%%%%%%%%%%%%%%%%%%%%%%%%%%%%%%%%%%%%%%%%%%%%%%%%%%%%%%%%%%%%%%%%%%%%%%%%%%%%%%%%%%%%%%%%%%%%%%%%%%%%%%%%%%%%%%%%%%%%%%%%%%%%%%%%%%%%%%%%%%%%%%%%%%%%%%%%%%%%%%%%%%%%%%%%%%%%%%%%%%%%%%%%%%%%%%%%%%%%%%%%%%%%%%%%%%%%%%%%%%%%%%%%%%%%%%%%%%%%%%%%%%%%%%%%%%%%%%%%%%%%%%%%%%%%%%%%%%%%%%%%%%%%%%%%%%%%%%%%%%%%%%%%%%%%%%%%%%%%%%%%%%%%%%%%%%%%%%%%%%%%%%%%%%%%%%%%%%%%%%%%%%%%%%%%%%%%%%%%%%%%%%%%%%%%

\subsection{Fermionic modes}

\begin{figure}
    \centering
      \includegraphics[scale=0.55]{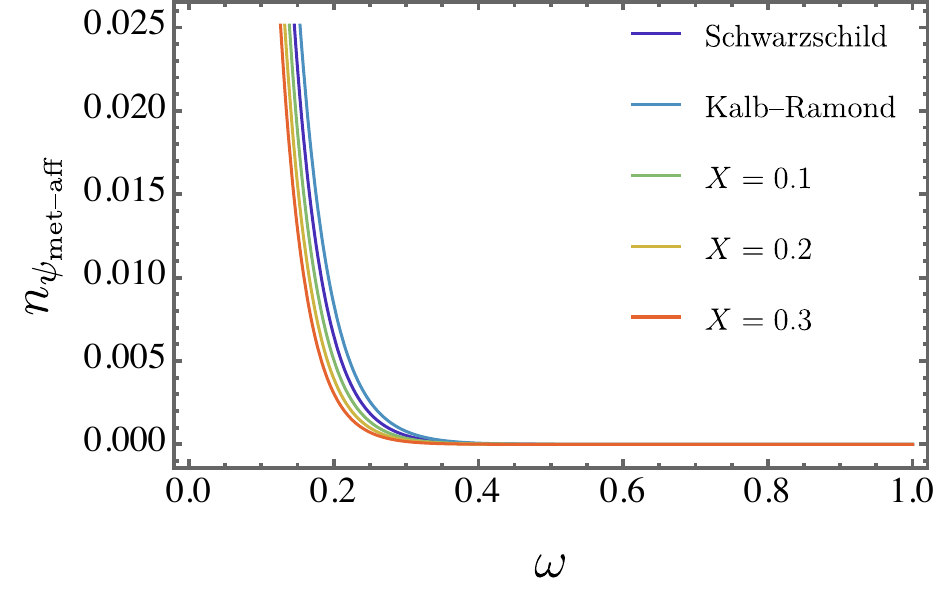}
    \caption{The particle density \( n_{\psi_{\text{met--aff}}} \) is shown for different values of $X$. The Schwarzschild and Kalb--Ramond cases are also compared.}
    \label{nfermmetaff}
\end{figure}

\begin{figure}
    \centering
      \includegraphics[scale=0.5]{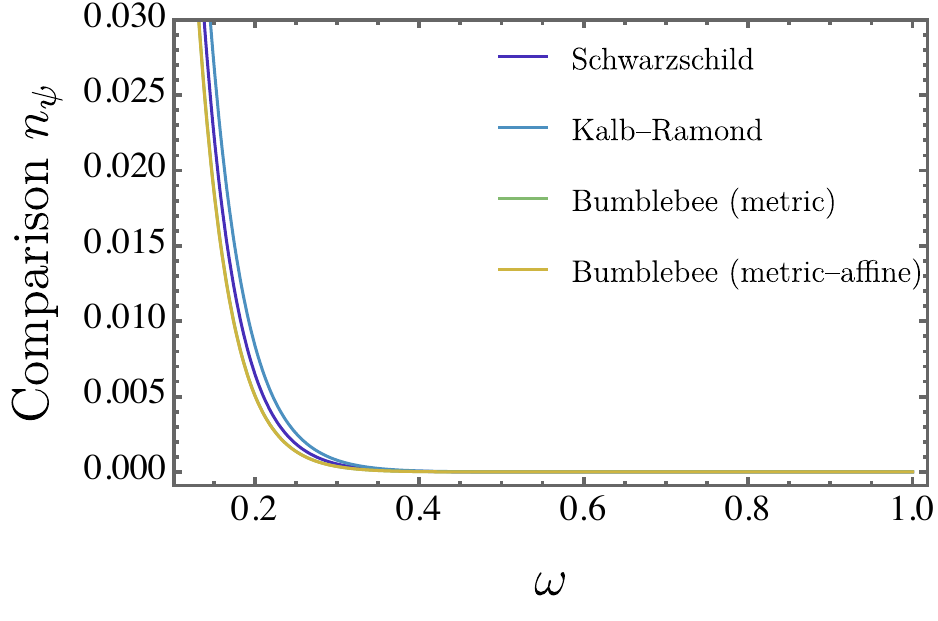}
      \includegraphics[scale=0.52]{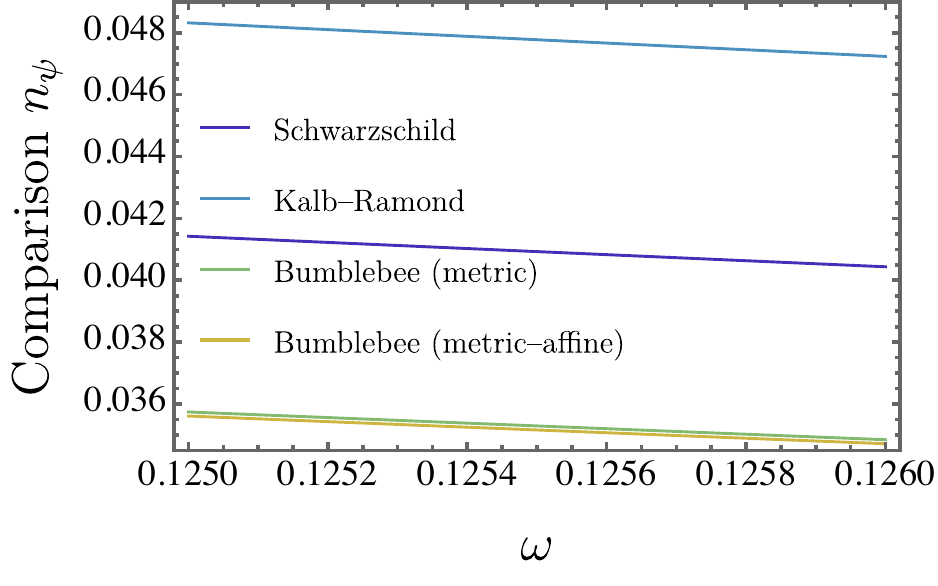}
    \caption{Comparison of \( n_{\psi} \) for the bumblebee black hole in the \textit{metric} and the \textit{metric--affine} formulations (for $X = \ell = 0.1$. In this comparison, the Schwarzschild and the Kalb--Ramond cases are also compared.}
    \label{nfermmetaff1ddd}
\end{figure}

\begin{figure}
    \centering
    \includegraphics[scale=0.51]{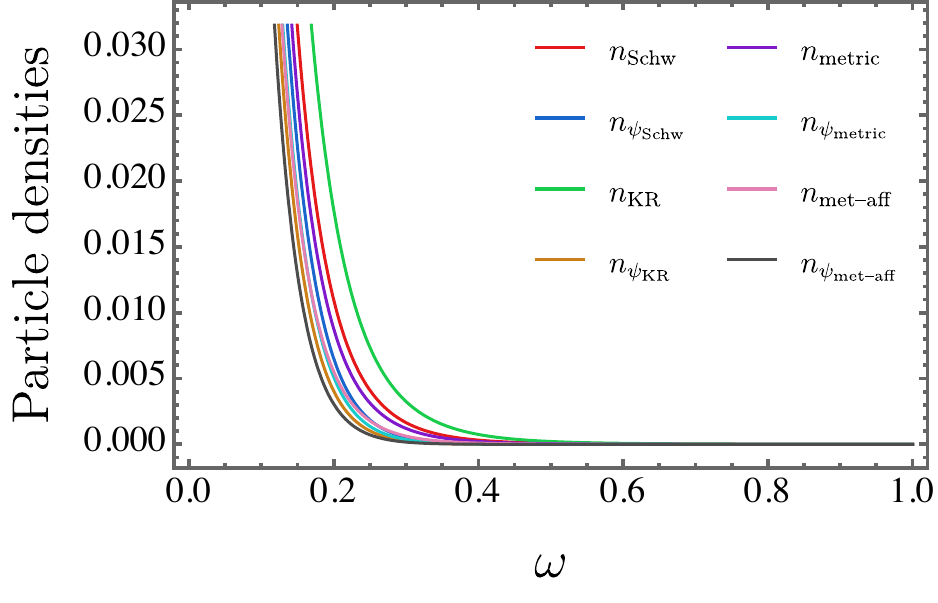}
      \includegraphics[scale=0.51]{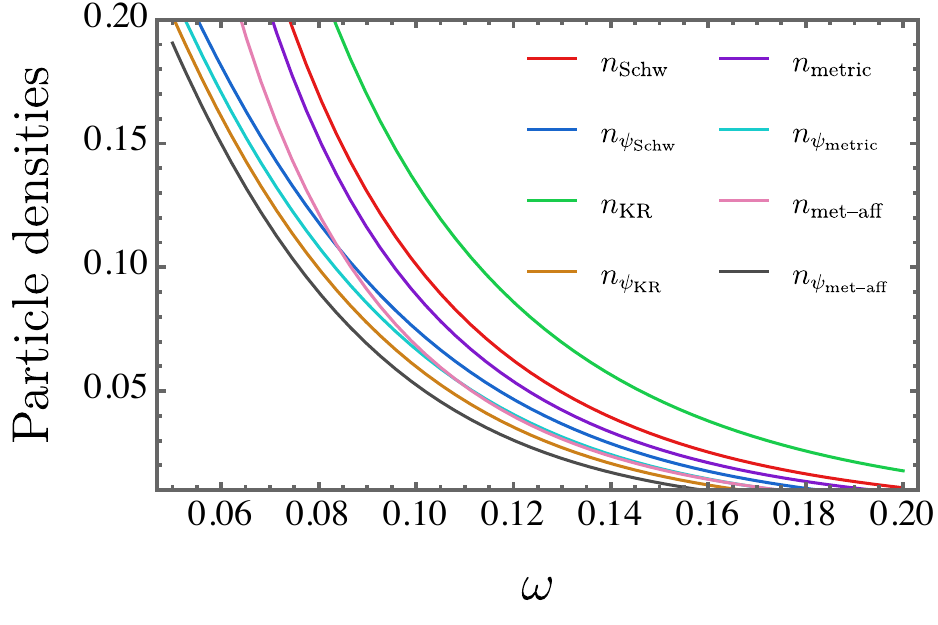}
    \caption{Comparison of \( n \), and \( n_{\psi} \). The configuration employed is the following: $n_{\text{Schw}}$ (Schwarzschild for bosons): $M=1$; $n_{\psi_{\text{Schw}}}$ (Schwarzschild for fermions): $M=1$; $n_{\text{KR}}$ (Kalb--Ramond for bosons): $M=1$, and $\ell = 0.2$; $n_{\psi_{\text{KR}}}$ (Kalb--Ramond for fermions): $M=1$ and $\ell = 0.2$; $n_{\text{metric}}$ (Bumblebee in \textit{metric} formalism for bosons): $M=1$, and $\ell = 0.1$; $n_{\psi_{\text{metric}}}$ (Bumblebee in \textit{metric} formalism for fermions): $M=1$ and $\ell = 0.1$; $n_{\text{met--aff}}$ (Bumblebee in \textit{metric--affine} formalism for bosons): $M=1$, and $X = 0.3$; $n_{\psi_{\text{met--aff}}}$ (Bumblebee in \textit{metric--affine} formalism for fermions): $M=1$ and $X = 0.3$; }
    \label{nfermmetaff1}
\end{figure}

With the definitions established so far, the particle density for fermions in the bumblebee black hole within the \textit{metric--affine} framework can be expressed as
\ie
n_{\psi_{\text{met--aff}}} = \frac{1}{e^{\frac{8 \pi  M \omega }{\sqrt{\frac{\sqrt{-(X-4)^3}}{\sqrt{4-X} (3 X+4)}}}}+1}.
\fe
Fig. \ref{nfermmetaff} illustrates the behavior of $n_{\psi_{\text{met--aff}}}$ for different values of $X$, along with a comparison to the standard Schwarzschild and Kalb--Ramond cases. Furthermore, Fig. \ref{nfermmetaff1ddd} compares \(n_{\psi}\) for the bumblebee black hole in both the \textit{metric} and \textit{metric--affine} approaches, with $X = \ell = 0.1$. This figure also includes the Schwarzschild and Kalb--Ramond cases for reference. Interestingly, in contrast to the bosonic case, the modifications caused by non--metricity in the fermionic particle density \(n_{\psi}\) for the bumblebee black hole are so small. Lastly, Fig. \ref{nfermmetaff1} provides a detailed comparison of all particle densities.

%%%%%%%%%%%%%%%%%%%%%%%%%%%%%%%%%%%%%%%%%%%%%%%%%%%%%%%%%%%%%%%%%%%%%%%%%%%%%%%%%%%%%%%%%%%%%%%%%%%%%%%%%%%%%%%%%%%%%%%%%%%%%%%%%%%%%%%%%%%%%%%%%%%%%%%%%%%%%%%%%%%%%%%%%%%%%%%%%%%%%%%%%%%%%%%%%%%%%%%%%%%%%%%%%%%%%%%%%%%%%%%%%%%%%%%%%%%%%%%%%%%%%%%%%%%%%%%%%%%%%%%%%%%%%%%%%%%%%%%%%%%%%%%%%%%%%%%%%%%%%%%%%%%%%%%%%%%%%%%%%%%%%%%%%%%%%%%%%%%%%%%%%%%%%%%%%%%%%%%%%%%%%%%%%%%%%%%%%%%%%%%%%%%%%%%%%%%%%%%%%%%%%%%%%%%%%%%%%%%%%%%%%%%%%%%%%%%%%%%%%%%%%%%%%%%%%%%%%%%%%%%%%%%%%%%%%%%%%%%%%%%%%%%%%%%%%%%%%%%%%%%%%%%%%%%%%%%%%%%%%%%%%%%%%%%%%%%%%%%%%%%%%%%%%%%%%%%%%%%%%%%%%%%%%%%%%%%%%%%%%%%%%%%%%%%%%%%%%%%%%%%%%

\subsection{Greybody factors for bosons}

%%%%%%%%%%%%%%%%%%%%%%%%%%%%%%%%%%%%%%%%%%%%%%%%%%%%%%%%%%%%%%%%%%%%%%%%%%%%%%%%%%%%%%%%%%%%%%%%%%%%%%%%%%%%%%%%%%%%%%%%%%%%%%%%%%%%%%%%%%%%%%%%%%%%%%%%%%%%%%%%%%%%%%%%%%%%%%%%%%%%%%%%%%%%%%%%%%%%%%%%%%%%%%%%%%%%%%%%%%%%%%%%%%%%%%%%%%%%%%%%%%%%%%%%%%%%%%%%%%%%%%%%%%%%%%%%%%%%%%%%%%%%%%%%%%%%%%%%%%%%

\subsubsection{Scalar perturbations}

Using the same methodology applied in the \textit{metric} approach, we now derive the effective potential for the \textit{metric--affine} bumblebee black hole
\ie
\begin{split}
\label{scalarpotentialmetricaffine}
&\mathcal{V}^{\,\text{s}}_{{\text{met--aff}}}  = \mathcal{A}(r)\left[\frac{{l(l + 1)}}{{{r^2}}} + \frac{1}{{r\sqrt {{\mathcal{A}(r)}{\mathcal{B}(r)^{ - 1}}} }}\frac{\mathrm{d}}{{\mathrm{d}r}}\sqrt {{\mathcal{A}(r)}\mathcal{B}(r)}\right]\\
& = \frac{4 l^2 (r-2 M)+4 l (r-2 M)-2 M (X-4) \sqrt{\frac{4-X}{3 X+4}} \sqrt{\frac{(r-2 M)^2}{r^2}}}{r^3 \sqrt{-((X-4) (3 X+4))}},
\end{split}
\fe
so that the greybody factors read
\ie
\begin{split}
& T^{\,\text{s}}_{b_{\text{met--aff}}}  \ge {\mathop{\rm sech}\nolimits} ^2 \left[\int_{-\infty}^ {+\infty} \frac{\mathcal{V}^{\,\text{s}}_{{\text{met--aff}}}} {2\omega}\mathrm{d}r^{*}\right] \\
& ={\mathop{\rm sech}\nolimits} ^2 \left[\int_{r_{ h}}^ {+\infty} \frac{\mathcal{V}^{\,\text{s}}_{{\text{met--aff}}}} {2\omega\sqrt{\mathcal{A}(r)\mathcal{B}(r)}}\mathrm{d} r\right] \\
& ={\mathop{\rm sech}\nolimits} ^2 \left[ \frac{1}{2\omega} \left(\frac{1}{4 M} \sqrt{\frac{\sqrt{-(X-4)^3}}{(4-X)^{7/2} (3 X+4)}} \right.\right.\\
& \left.\left. \times \left(8 l (l+1) \sqrt{4-X} \sqrt{\frac{\sqrt{-(X-4)^5} (3 X+4)}{(4-X)^{7/2}}} \right.\right.\right. \\
& \left.\left.\left. +\sqrt{-(X-4)^3}\right) \right)  \right].
\end{split}
\fe

\begin{figure}
    \centering
      \includegraphics[scale=0.55]{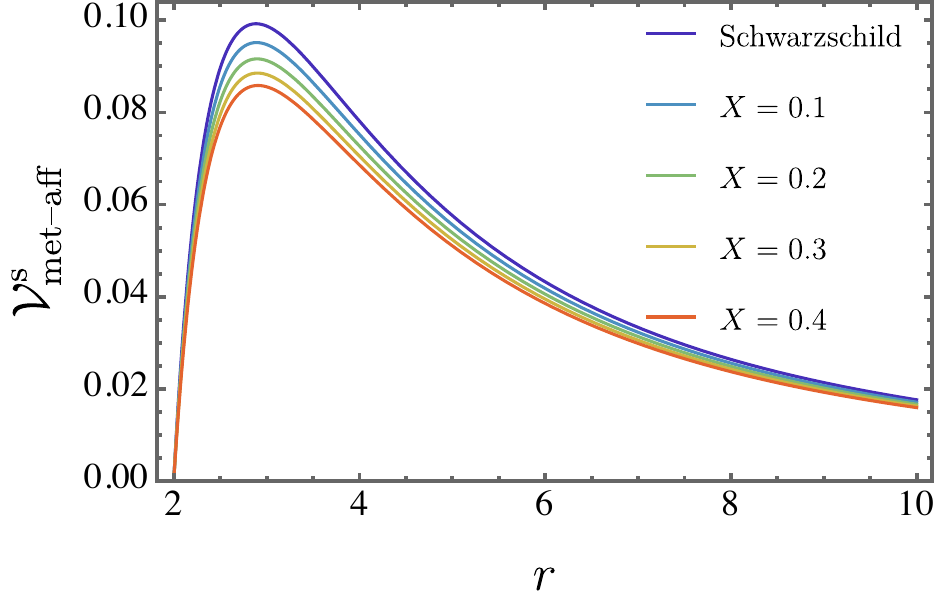}
    \caption{The effective potential $\mathcal{V}^{\,\text{s}}_{{\text{met--aff}}}$ is shown for different values of $X$. Also, the Schwarzschild case is compared in this analysis.}
    \label{scalarpotentialmetaff}
\end{figure}

\begin{figure}
    \centering
      \includegraphics[scale=0.51]{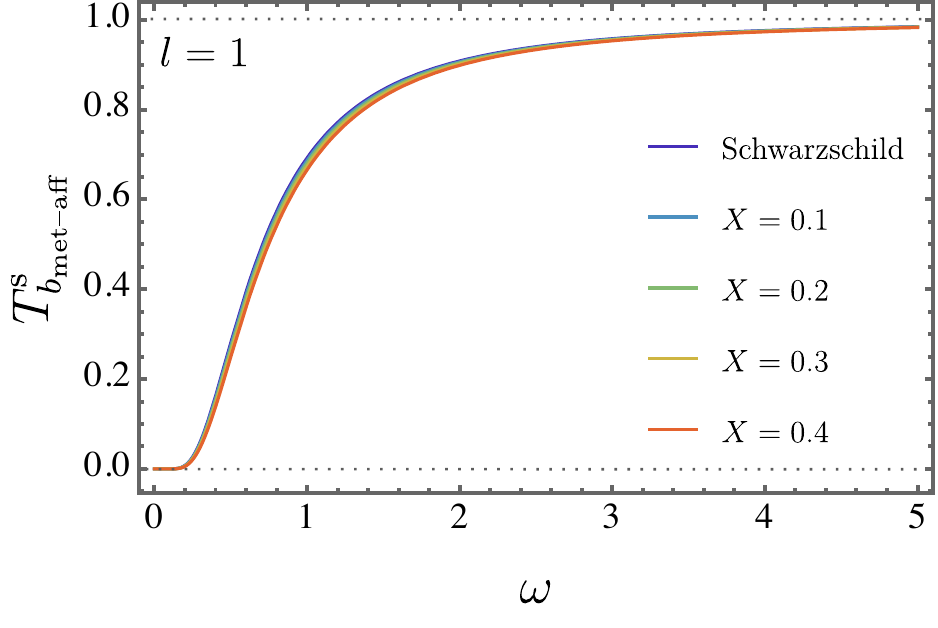}
      \includegraphics[scale=0.51]{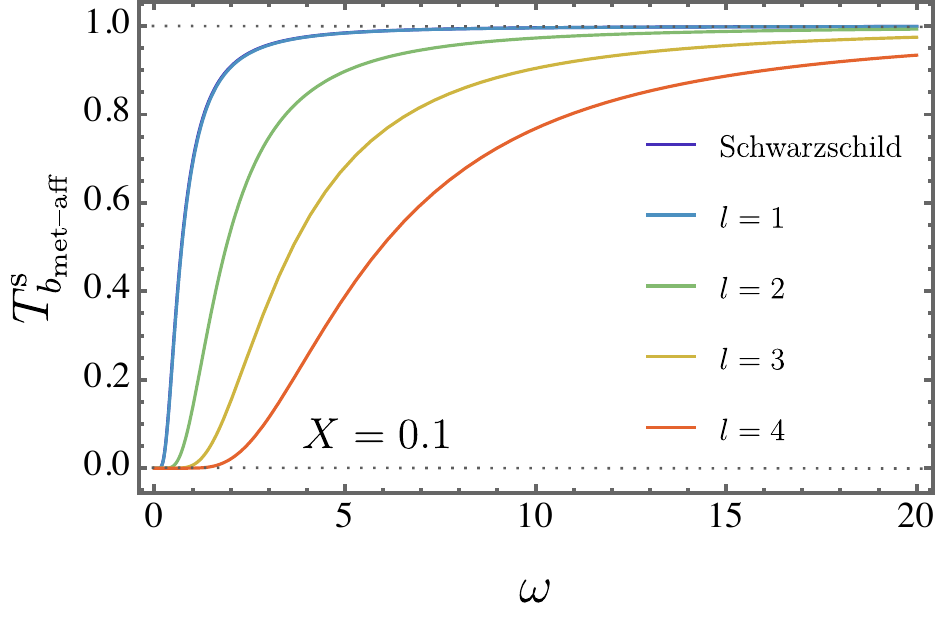}
    \caption{The greybody factors $T^{\,\text{s}}_{b_{\text{met--aff}}}$ is displayed for different values of $X$ when keeping $l=1$ (the the top panel) and for different values of $l$ for a fixed value of $X= 0.1$. For both cases, the Schwarzschild case is compared.}
    \label{greybodymetricaffinebosons}
\end{figure}

Fig. \ref{scalarpotentialmetaff} illustrates the effective potential for scalar perturbations in the \textit{metric--affine} formalism. It is observed that an increase in $X$ results in a reduction of $\mathcal{V}^{\,\text{s}}_{{\text{met--aff}}}$. Meanwhile, the greybody factors are displayed in Fig. \ref{greybodymetricaffinebosons}, considering varying values of $X$ with $l=1$ (top panel) and different values of $l$ for a fixed $X=0.1$. In both scenarios, comparisons are made with the Schwarzschild case.

%%%%%%%%%%%%%%%%%%%%%%%%%%%%%%%%%%%%%%%%%%%%%%%%%%%%%%%%%%%%%%%%%%%%%%%%%%%%%%%%%%%%%%%%%%%%%%%%%%%%%%%%%%%%%%%%%%%%%%%%%%%%%%%%%%%%%%%%%%%%%%%%%%%%%%%%%%%%%%%%%%%%%%%%%%%%%%%%%%%%%%%%%%%%%%%%%%%%%%%%%%%%%%%%%%%%%%%%%%%%%%%%%%%%%%%%%%%%%%%%%%%%%%%%%%%%%%%%%%%%%%%%%%%%%%%%%%%%%%%%%%%%%%%%%%%%%%%%%%%%

\subsubsection{Vector perturbations}

Following the analogy with the scalar perturbation in the \textit{metric--affine} framework, the effective potential is expressed as
\ie
\begin{split}
\label{vectormetricaffinecase}
&\mathcal{V}^{\,\text{v}}_{{\text{met--aff}}}  = -\frac{4 l (l+1) (2 M-r)}{r^3 \sqrt{-((X-4) (3 X+4))}}.
\end{split}
\fe
Therefore, the greybody factors can be presented below
\ie
\begin{split}
& T^{\,\text{v}}_{b_{\text{met--aff}}}  \ge {\mathop{\rm sech}\nolimits} ^2 \left[\int_{-\infty}^ {+\infty} \frac{\mathcal{V}^{\,\text{v}}_{{\text{met--aff}}}} {2\omega}\mathrm{d}r^{*}\right] \\
& ={\mathop{\rm sech}\nolimits} ^2 \left[\int_{r_{ h}}^ {+\infty} \frac{\mathcal{V}^{\,\text{v}}_{{\text{met--aff}}}} {2\omega\sqrt{\mathcal{A}(r)\mathcal{B}(r)}}\mathrm{d} r\right] \\
& ={\mathop{\rm sech}\nolimits} ^2 \left[ \frac{1}{2\omega} \left(\frac{2 l (l+1)}{M \sqrt{4-X} \sqrt{\frac{(4-X)^{5/2}}{\sqrt{-(X-4)^3} (3 X+4)}} \sqrt{3 X+4}} \right)  \right].
\end{split}
\fe

\begin{figure}
    \centering
      \includegraphics[scale=0.55]{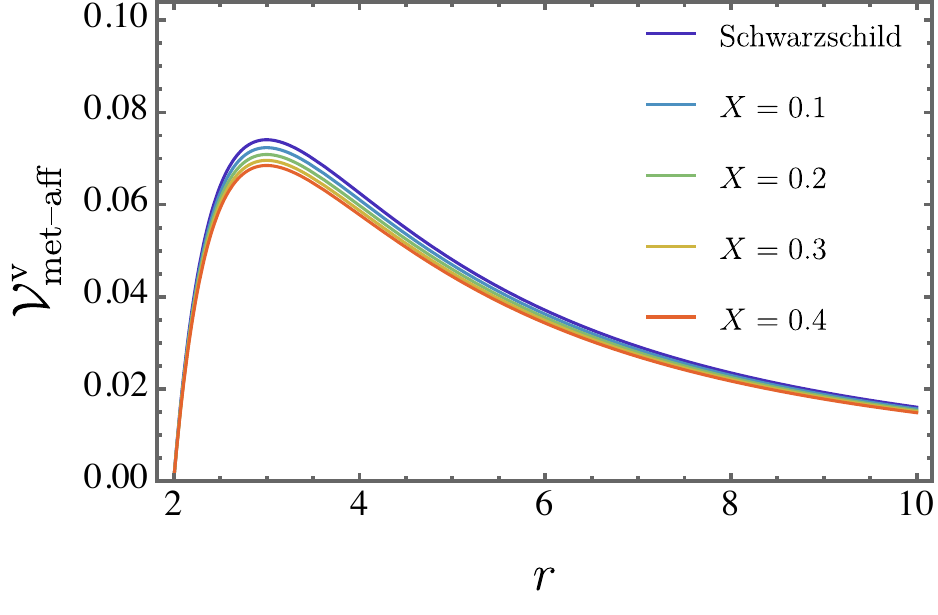}
    \caption{The effective potential $\mathcal{V}^{\,\text{v}}_{{\text{met--aff}}}$ is shown for different values of $X$. Also, the Schwarzschild case is compared in this analysis.}
    \label{scalarpotentialmetaffvectorial}
\end{figure}

\begin{figure}
    \centering
      \includegraphics[scale=0.51]{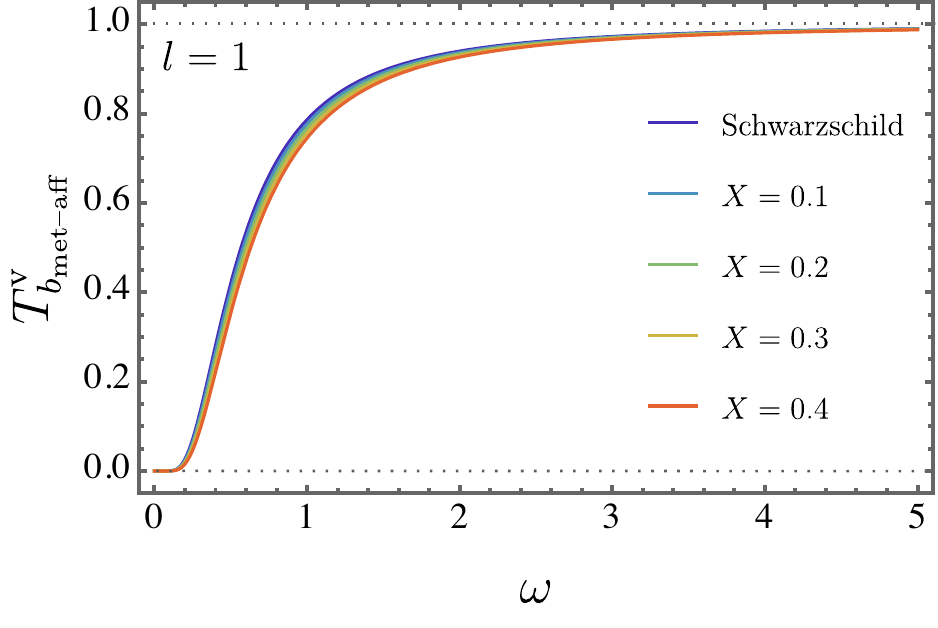}
      \includegraphics[scale=0.51]{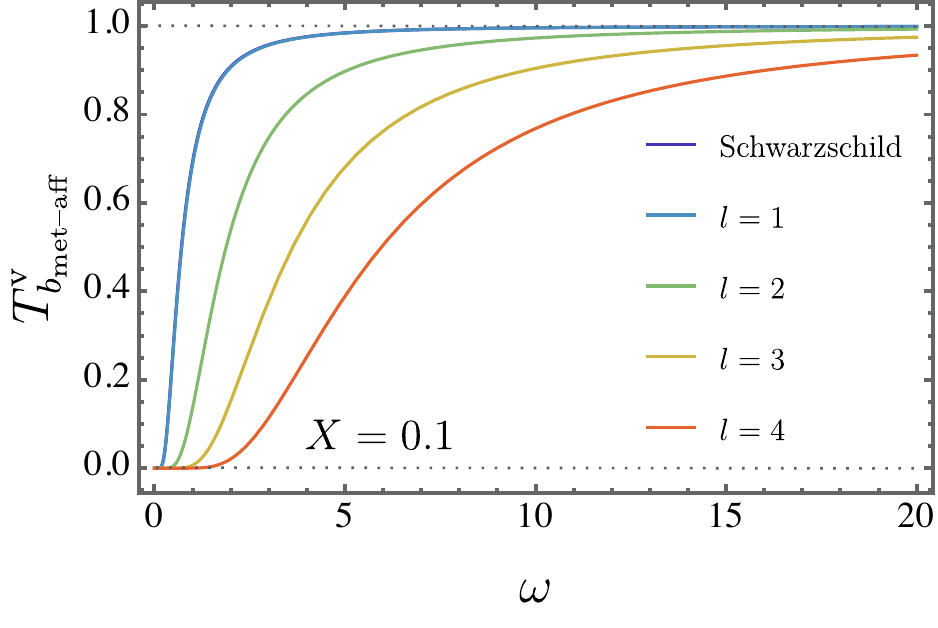}
    \caption{\label{scalarpotentialmetaffvectorial22}The greybody factors $T^{\,\text{v}}_{b_{\text{met--aff}}}$ is displayed for different values of $X$ when keeping $l=1$ (the the top panel) and for different values of $l$ for a fixed value of $X= 0.1$. For both cases, the Schwarzschild case is compared.}
    \label{greybodymetricaffinebosonsvectorial}
\end{figure}

Fig. \ref{scalarpotentialmetaffvectorial} depicts the effective potential for vectorial perturbations in the \textit{metric--affine} formalism. It is observed that increasing $X$ results in a decrease in $\mathcal{V}^{\,\text{v}}_{{\text{met--aff}}}$. Meanwhile, Fig. \ref{greybodymetricaffinebosonsvectorial} presents the greybody factors for varying values of $X$ with $l=1$ (top panel) and different values of $l$ for a fixed $X=0.1$, with both cases compared against the Schwarzschild scenario.

%%%%%%%%%%%%%%%%%%%%%%%%%%%%%%%%%%%%%%%%%%%%%%%%%%%%%%%%%%%%%%%%%%%%%%%%%%%%%%%%%%%%%%%%%%%%%%%%%%%%%%%%%%%%%%%%%%%%%%%%%%%%%%%%%%%%%%%%%%%%%%%%%%%%%%%%%%%%%%%%%%%%%%%%%%%%%%%%%%%%%%%%%%%%%%%%%%%%%%%%%%%%%%%%%%%%%%%%%%%%%%%%%%%%%%%%%%%%%%%%%%%%%%%%%%%%%%%%%%%%%%%%%%%%%%%%%%%%%%%%%%%%%%%%%%%%%%%%%%%%

\subsubsection{Tensor perturbations}

Following the analogy with the scalar and vector perturbations in the \textit{metric--affine} framework, the effective potential is expressed as
\ie
\begin{split}
&\mathcal{V}^{\,\text{t}}_{{\text{met--aff}}}  = \left(1-\frac{2 M}{r}\right) \left(\frac{l (l+1)}{r^2} +\frac{6 M (X-4)^2-2 r (X-4)^2+8 r \sqrt{-((X-4) (3 X+4))}}{r^3 (X-4) (3 X+4)}\right).
\end{split}
\fe
Naturally, if $X \to 0$, we recover the effective potential for the Schwarzschild case odd parity (for the axial perturbations), namely, $\mathcal{V}^{\text{t}}_{\text{Schw}} = \left(1-\frac{2 M}{r}\right) \left(\frac{l (l+1)}{r^2}-\frac{6 M}{r^3}\right)$. Therefore, the greybody factors can be presented below
\ie
\begin{split}
& T^{\,\text{t}}_{b_{\text{met--aff}}}  \ge {\mathop{\rm sech}\nolimits} ^2 \left[\int_{-\infty}^ {+\infty} \frac{\mathcal{V}^{\,\text{t}}_{{\text{met--aff}}}} {2\omega}\mathrm{d}r^{*}\right] \\
& ={\mathop{\rm sech}\nolimits} ^2 \left[ \frac{(1-3 l (l+1)) X^2+8 l (l+1) X+16 l (l+1)-8 \left(X+\sqrt{-((X-4) (3 X+4))}-2\right)}{16 M (3 X+4)}  \right].
\end{split}
\fe

\begin{figure}
    \centering
      \includegraphics[scale=0.55]{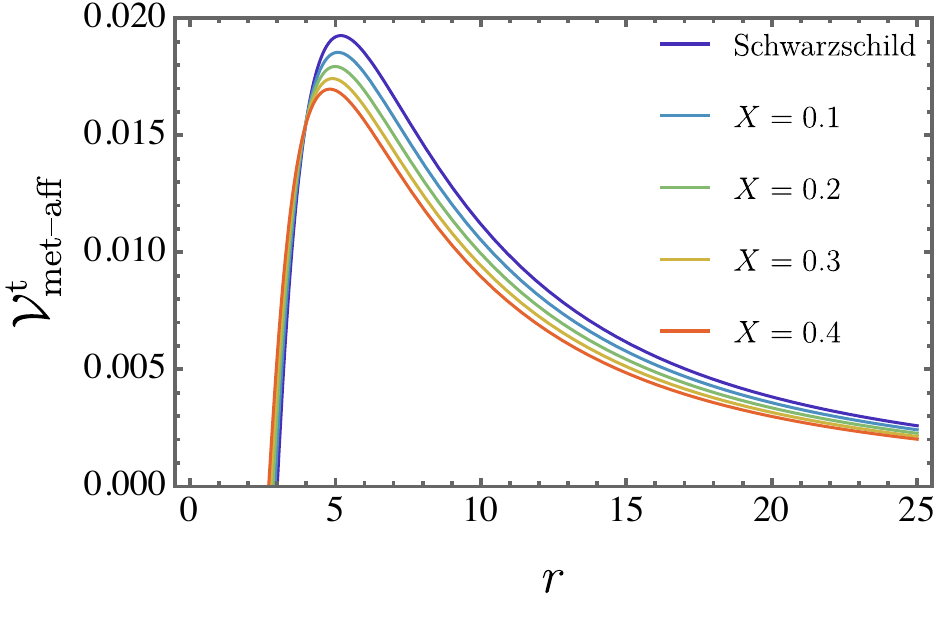}
    \caption{The effective potential $\mathcal{V}^{\,\text{t}}_{{\text{met--aff}}}$ is shown for different values of $X$ for $l=1$. Also, the Schwarzschild case is compared in this analysis.}
    \label{scalarpotentialmetafftensor}
\end{figure}

\begin{figure}
    \centering
      \includegraphics[scale=0.51]{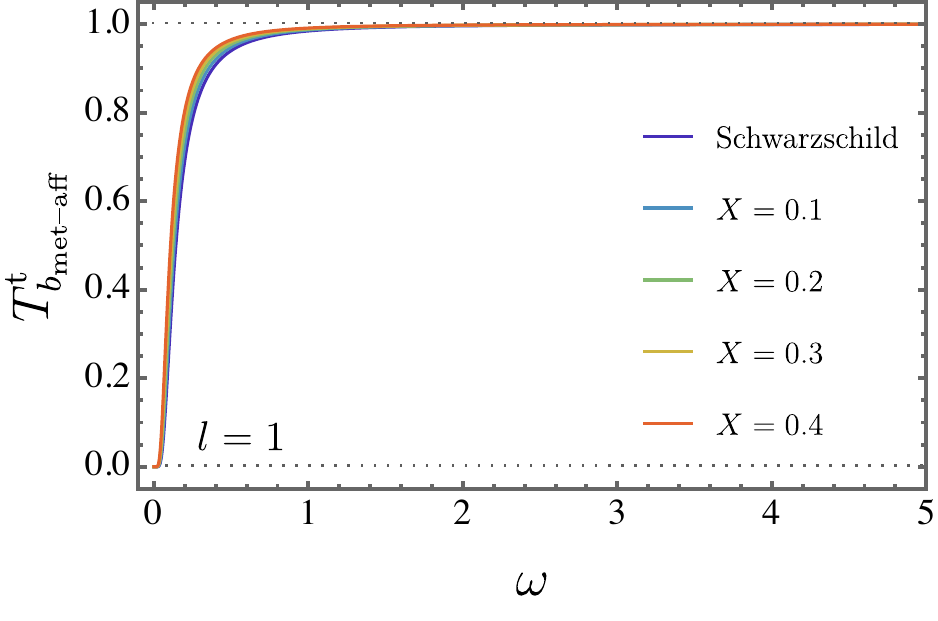}
      \includegraphics[scale=0.51]{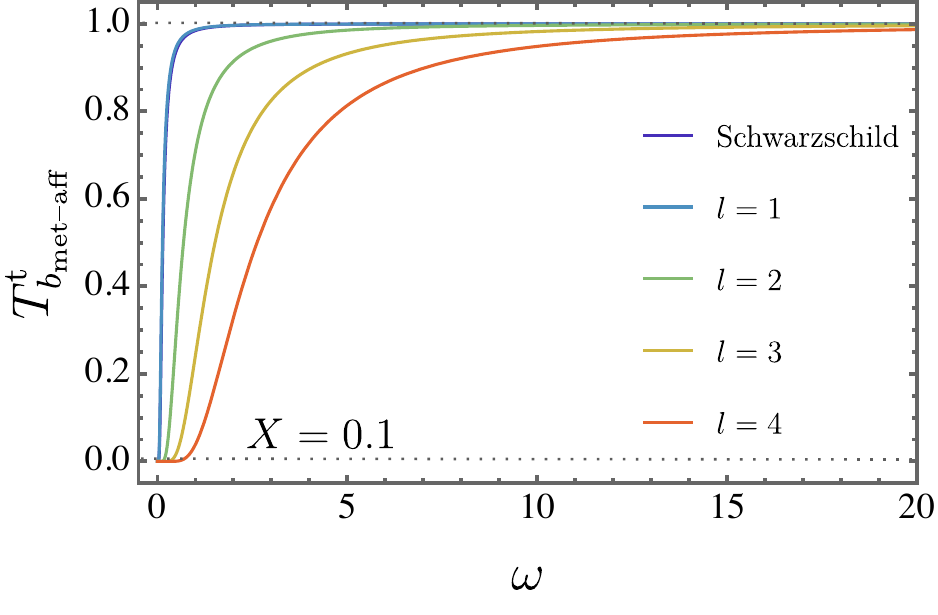}
    \caption{\label{greybodymetricaffinebosonstensor233}The greybody factors $T^{\,\text{t}}_{b_{\text{met--aff}}}$ is displayed for different values of $X$ when keeping $l=1$ (the the top panel) and for different values of $l$ for a fixed value of $X= 0.1$. For both cases, the Schwarzschild case is compared.}
    \label{greybodymetricaffinebosonstensor}
\end{figure}

Fig. \ref{scalarpotentialmetafftensor} depicts the effective potential for tensor perturbations in the \textit{metric--affine} formalism. It is observed that increasing $X$ results in a decrease in $\mathcal{V}^{\,\text{t}}_{{\text{met--aff}}}$. Meanwhile, Fig. \ref{greybodymetricaffinebosonstensor} presents the greybody factors for varying values of $X$ with $l=1$ (top panel) and different values of $l$ for a fixed $X=0.1$, with both cases compared against the Schwarzschild scenario. Differently from the scalar and the vector perturbation, the increase of $X$ (for a fixed $l=1$) leads to an increase of greybody factors for the tensor perturbations.

\begin{figure}
    \centering
      \includegraphics[scale=0.51]{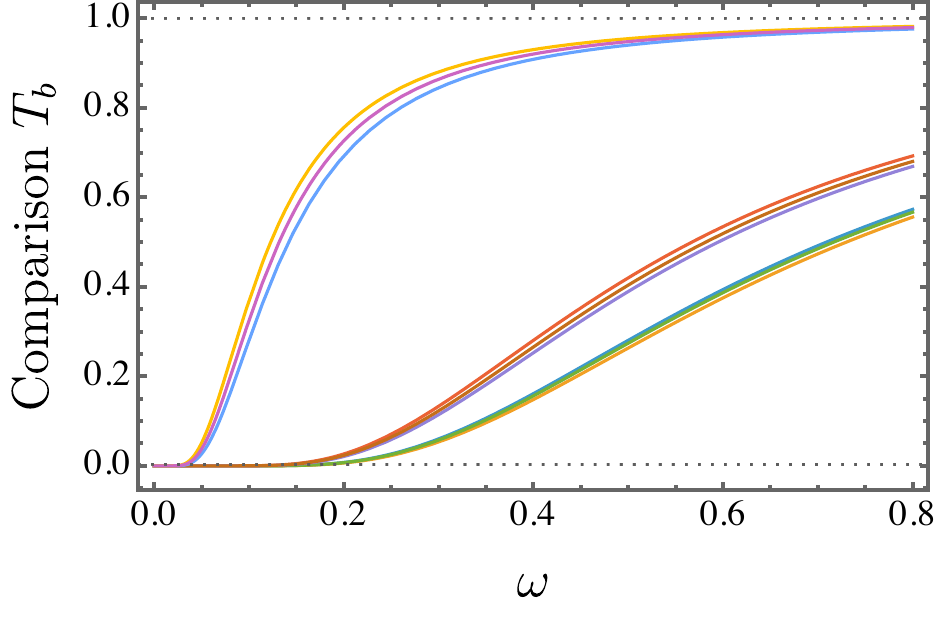}
        \includegraphics[scale=0.51]{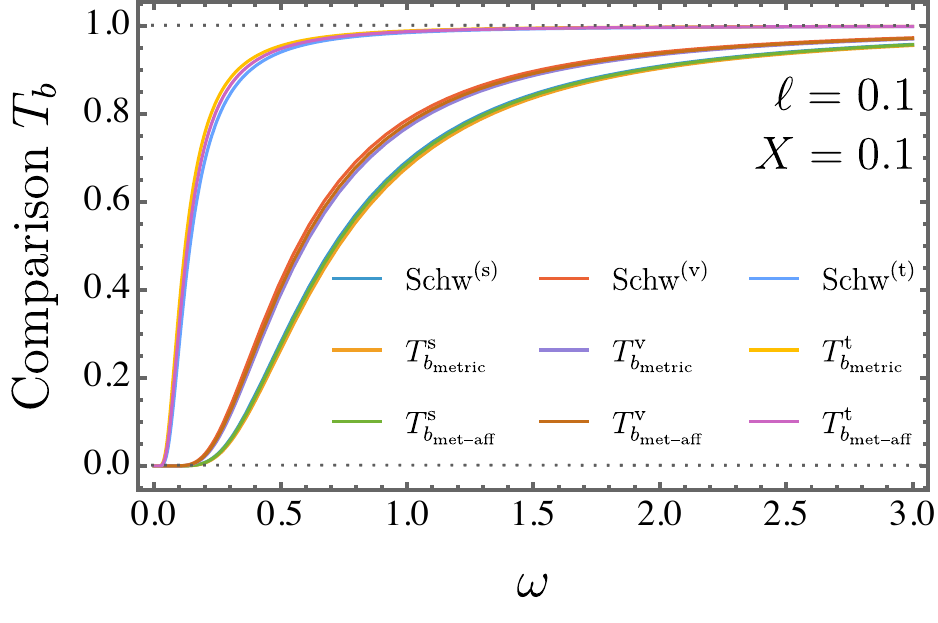}
    \caption{The comparison of the greybody factors for the bosonic case when $X = \ell = 0.1$. The Schwarzschild case (for scalar, vector and tensor perturbations) is also present for the sake of comparison.}
    \label{greybodybosoncomp}
\end{figure}

Finally, in Fig. \ref{greybodybosoncomp}, the greybody factors for all bosonic perturbations are compared, revealing the following hierarchy:  
$T_{b_{\text{metric}}}^{\text{t}} > T_{b_{\text{met--aff}}}^{\text{t}} > \text{Schw}^{(\text{t})} > \text{Schw}^{(\text{v})} > T_{b_{\text{met--aff}}}^{\text{v}} > T_{b_{\text{metric}}}^{\text{v}} > \text{Schw}^{(\text{s})} > T_{b_{\text{met--aff}}}^{\text{s}} > T_{b_{\text{metric}}}^{\text{s}}$.

Additionally, for tensor perturbations, both $T_{b_{\text{metric}}}^{\text{t}}$ and $T_{b_{\text{met--aff}}}^{\text{t}}$ exhibit greater intensity than the corresponding Schwarzschild result—once again contrasting with the behavior observed for the other bosonic modes discussed in this work.

%%%%%%%%%%%%%%%%%%%%%%%%%%%%%%%%%%%%%%%%%%%%%%%%%%%%%%%%%%%%%%%%%%%%%%%%%%%%%%%%%%%%%%%%%%%%%%%%%%%%%%%%%%%%%%%%%%%%%%%%%%%%%%%%%%%%%%%%%%%%%%%%%%%%%%%%%%%%%%%%%%%%%%%%%%%%%%%%%%%%%%%%%%%%%%%%%%%%%%%%%%%%%%%%%%%%%%%%%%%%%%%%%%%%%%%%%%%%%%%%%%%%%%%%%%%%%%%%%%%%%%%%%%%%%%%%%%%%%%%%%%%%%%%%%%%%%%%%%%%%%%%%%%%%%%%%%%%%%%%%%%%%%%%%%%%%%%%%%%%%%%%%%%%%%%%%%%%%%%%%%%%%%%%%%%%%%%%%%%%%%%%%%%%%%%%%%%%%%%%%%%%%%%%%%%%%%%%%%%%%%%%%%%%%%%%%%%%%%%%%%%%%%%%%%%%%%%%%%%%%%%%%%%%%%%%%%%%%%%%%%%%%%%%%%%%%%%%%%%%%%%%%%%%%%%%%%%%%%%%%%%%%%%%%%%%%%%%%%%%%%%%%%%%%%%%%%%%%%%%%%%%%%%%%%%%%%%%%%%%%%%%%%%%%%%%%%%%%%%%%%%%%%%%%%%%%%%%%%%%%%%%%%%%%%%%%%%%%%%%%%%%%%%%%%%%%%%%%%%%%

\subsection{Greybody factors for fermions}

Following the methodology applied in the previous section for the \textit{metric} case, we now compute the greybody factors for fermions by taking into account the black hole solution in bumblebee gravity within the \textit{metric--affine} framework. It is important to note that greybody factors for bosons have been recently explored in the literature \cite{heidari2024scattering}. Using the same approach as in the previous section for determining the greybody factors in the \textit{metric} formalism, we now derive:
\ie
\begin{split}
& V^{+}_{\text{met--aff}}=  \\
= & \frac{2 \left(l+\frac{1}{2}\right)}{r^3 \sqrt{4-X} \sqrt{3 X+4}}  \times \left\{-2 (2 l+1) M \right. \\
& \left. + 2 l r+\frac{(3 M-r) \sqrt{\frac{(4-X)^{5/2} (r-2 M)^2}{r^2 \sqrt{-(X-4)^3} (3 X+4)}}}{\sqrt{\frac{r-2 M}{r \sqrt{4-X} \sqrt{3 X+4}}}}+r\right\}
\end{split}
\fe
in a such way that
\ie
\begin{split}
& T_{b_{\text{met--aff}}}  = \\
&  \text{sech}^2\left(\frac{(2 l+1)^2}{2 \omega  \left(2 M \sqrt{4-X} \sqrt{\frac{(4-X)^{5/2}}{\sqrt{-(X-4)^3} (3 X+4)}} \sqrt{3 X+4}\right)}\right).
\end{split}
\fe

\begin{figure}
    \centering
      \includegraphics[scale=0.55]{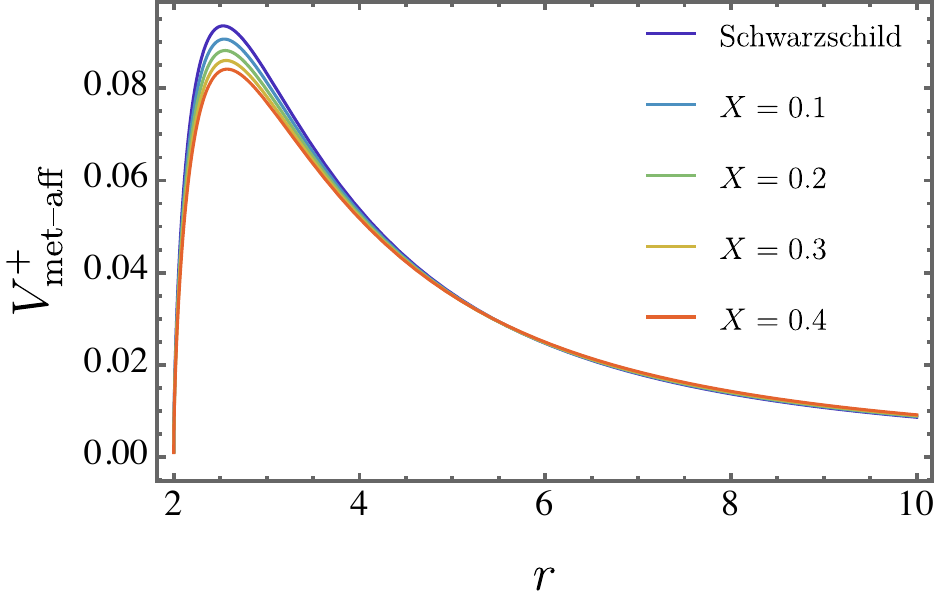}
    \caption{The effective potential $V_{\text{met--aff}}^{+}$ is shown for different values of $X$. Also, the Schwarzschild case is compared in this analysis.}
    \label{vmetaff}
\end{figure}

\begin{figure}
    \centering
      \includegraphics[scale=0.51]{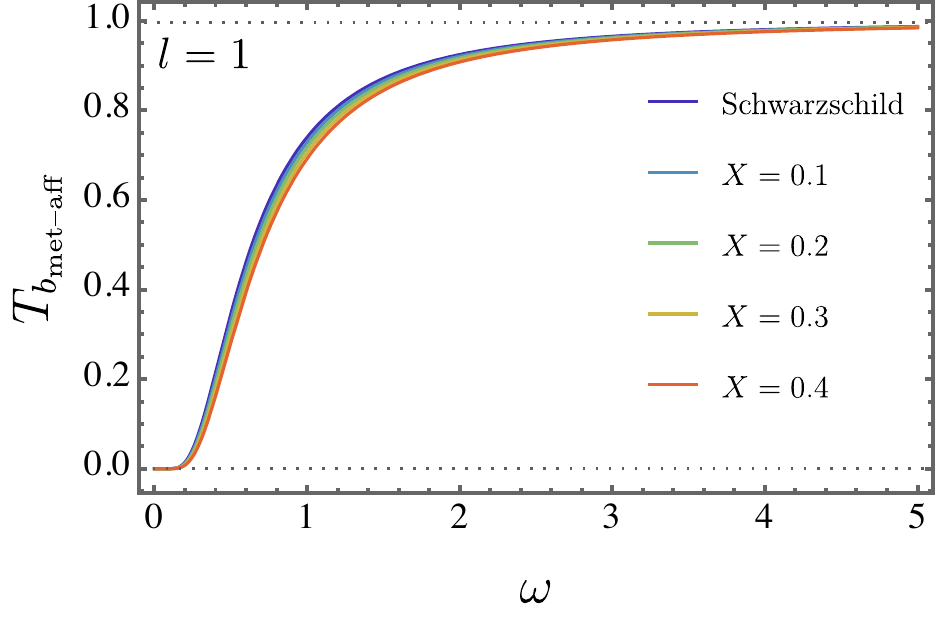}
      \includegraphics[scale=0.51]{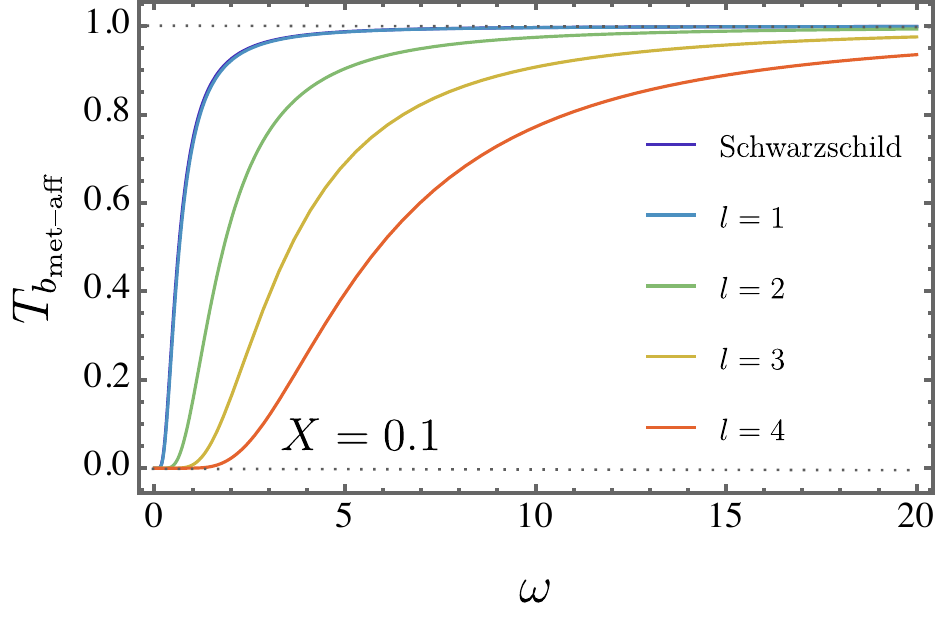}
    \caption{The greybody factors $T_{b_{\text{met--aff}}}$ is displayed for different values of $X$ when keeping $l=1$ (the the top panel) and for different values of $l$ for a fixed value of $X =0.1$. For both cases, the Schwarzschild case is compared.}
    \label{tbmetafffermions1}
\end{figure}

\begin{figure}
    \centering
      \includegraphics[scale=0.55]{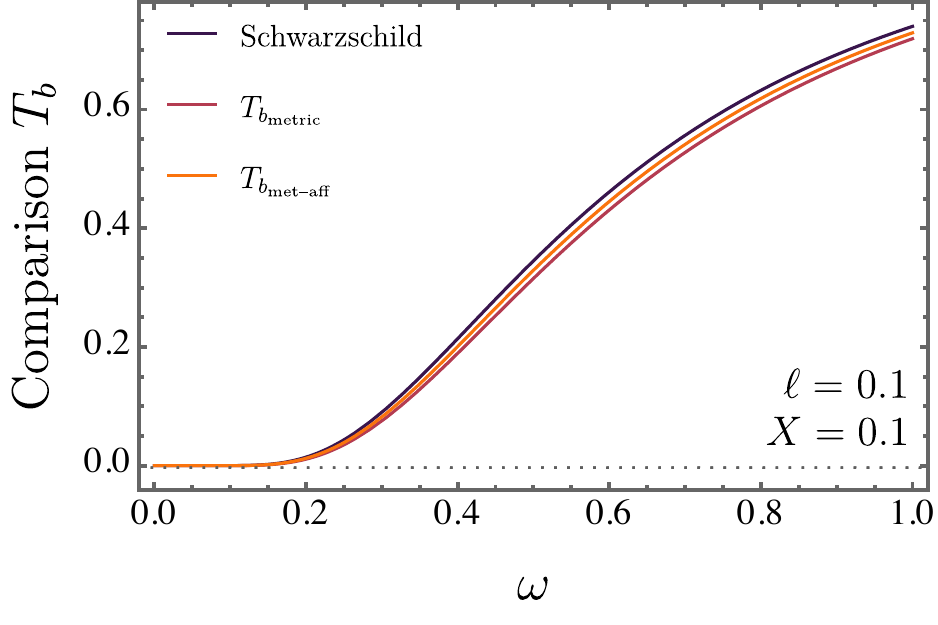}
    \caption{The comparison of $T_{b}$ for the \textit{metric} and \textit{metric--affine} formalisms for fixed values of $\ell$ and $X$, i.e., $X = \ell = 0.1$.}
    \label{tbmetafffermionscomp}
\end{figure}

In Fig. \ref{vmetaff}, we present the behavior of the effective potential $V^{+}_{\text{met--aff}}$ as a function of $r$. As it is expected, it goes to zero when $r \to \infty$. On the other hand, we present Fig. \ref{tbmetafffermions1} to show $T_{b_{\text{met--aff}}}$ as a function of $\omega$ for different values of $X$ for a fixed value of $l=1$ (on the top panel). In contrast, in the same Fig., we exhibit the greybody factors by varying $l$ instead while keeping $X=0.1$ (on the bottom panel). Finally, in order to provide a comparison of the two black holes considered here, we display Fig. \ref{tbmetafffermionscomp}, which highlights the behavior of Schwarzschild case, $T_{b_{\text{metric}}}$ and $T_{b_{\text{met--aff}}}$. In general lines, for our bumblebee black holes, we verify that the non--metricity is responsible for increasing the greybody factors if comparison with the \textit{metric} case. This corroborates the results addressed for the particle density ($n$) in the previous section for the bosonic case.

%%%%%%%%%%%%%%%%%%%%%%%%%%%%%%%%%%%%%%%%%%%%%%%%%%%%%%%%%%%%%%%%%%%%%%%%%%%%%%%%%%%%%%%%%%%%%%%%%%%%%%%%%%%%%%%%%%%%%%%%%%%%%%%%%%%%%%%%%%%%%%%%%%%%%%%%%%%%%%%%%%%%%%%%%%%%%%%%%%%%%%%%%%%%%%%%%%%%%%%%%%%%%%%%%%%%%%%%%%%%%%%%%%%%%%%%%%%%%%%%%%%%%%%%%%%%%%%%%%%%%%%%%%%%%%%%%%%%%%%%%%%%%%%%%%%%%%%%%%%%%%%%%%%%%%%%%%%%%%%%%%%%%%%%%%%%%%%%%%%%%%%%%%%%%%%%%%%%%%%%%%%%%%%%%%%%%%%%%%%%%%%%%%%%%%%%%%%%%%%%%%%%%%%%%%%%%%%%%%%%%%%%%%%%%%%%%%%%%%%%%%%%%%%%%%%%%%%%%%%%%%%%%%%%%%%%%%%%%%%%%%%%%%%%%%%%%%%%%%%%%%%%%%%%%%%%%%%%%%%%%%%%%%%%%%%%%%%%%%%%%%%%%%%%%%%%%%%%%%%%%%%%%%%%%%%%%%%%%%%%%%%%%%%%%%%%%%%%%%%%%%%%%%%%%%%%%%%%%%%%%%%%%%%%%%%%%%%%%%%%%%%%%%%%%%%%%%%%%%%%%%%%%%%%%%%%%%%%%%%%%%%%%%%%%%%%%%%%%%%%%%%%%%%%%%%%%%%

\subsection{The emission rate}

Following the approach taken for the \textit{metric} case in analyzing the emission rate, we now extend the investigation to the \textit{metric--affine} framework. Therefore, we write the emission rate for the \textit{metric--afine} case
\ie
\frac{\mathrm{d}^{2}E}{\mathrm{d}\omega \mathrm{d} t} = \frac{27 \pi ^3 M^2 \sqrt{-((X-4) (3 X+4))} \omega ^3}{2 \left(e^{-\frac{16 \pi  M \omega }{X-2}}-1\right)}.
\fe

\begin{figure}
    \centering
      \includegraphics[scale=0.55]{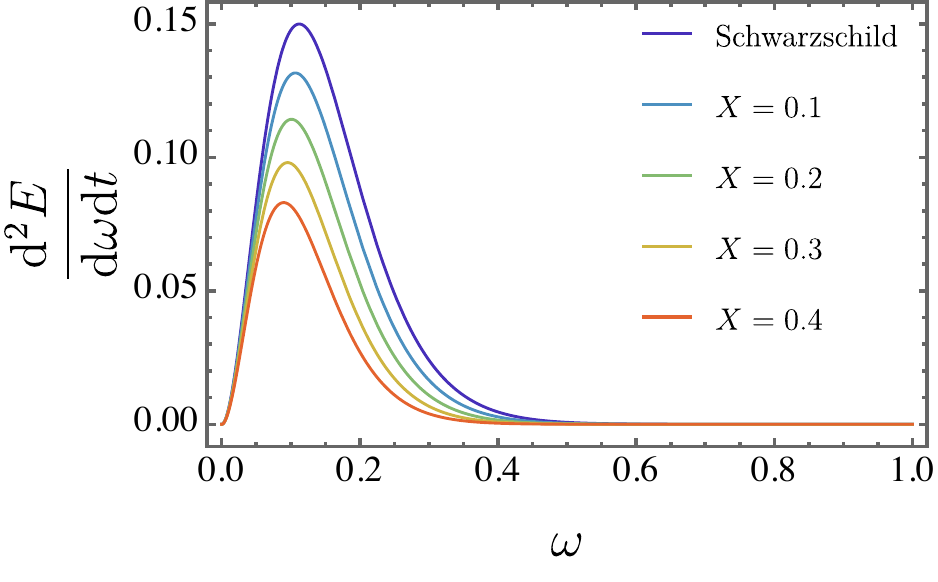}
    \caption{The emission rate for different values of $X$ as a function of $\omega$.}
    \label{emissionratemetricaffine}
\end{figure}

Fig. \ref{emissionratemetricaffine} depicts the emission rate as a function of $\omega$ for different values of $X$. Generally, as $X$ increases, the emission rate diminishes in magnitude. For reference, the Schwarzschild black hole is included in the comparison. These results align with the particle density findings ($n_{\text{met--aff}}$) presented in the earlier subsections.

\begin{figure}
    \centering
      \includegraphics[scale=0.55]{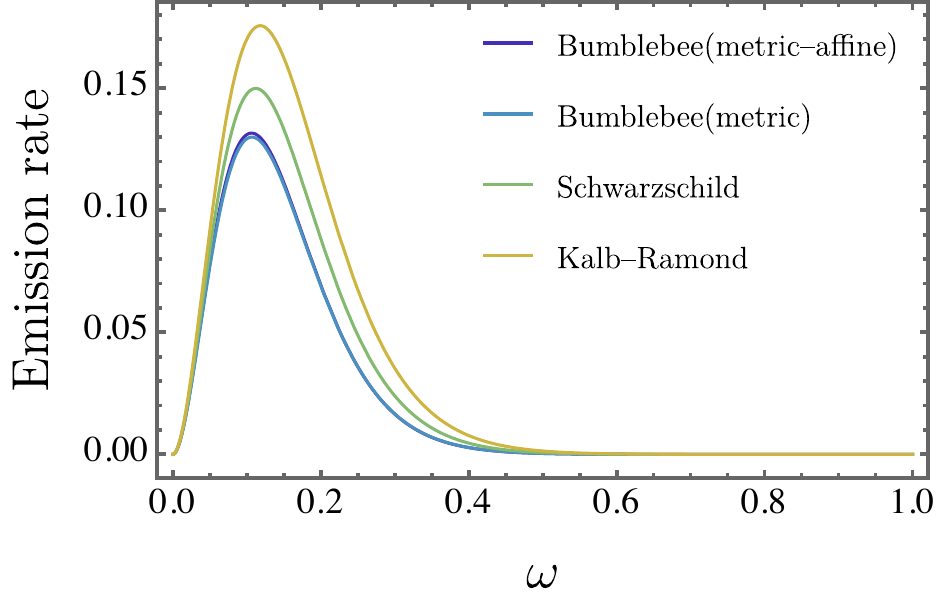}
    \caption{The comparison of the emission rates for the bumblebee in the \textit{metric} and \textit{metric--affine} formalism as well as with the Kalb--Ramond solution.}
    \label{emissionratecomparison}
\end{figure}

Additionally, we compare the emission rates of the two cases examined in this paper: the bumblebee black hole in the \textit{metric} and \textit{metric--affine} frameworks. For further context, the Kalb--Ramond solution is also included to provide a comparison with another recently proposed black hole model that incorporates Lorentz symmetry breaking. These comparisons are illustrated in Fig. \ref{emissionratecomparison}. Notably, non--metricity increases the magnitude of the emission rate relative to the \textit{metric} case.

%%%%%%%%%%%%%%%%%%%%%%%%%%%%%%%%%%%%%%%%%%%%%%%%%%%%%%%%%%%%%%%%%%%%%%%%%%%%%%%%%%%%%%%%%%%%%%%%%%%%%%%%%%%%%%%%%%%%%%%%%%%%%%%%%%%%%%%%%%%%%%%%%%%%%%%%%%%%%%%%%%%%%%%%%%%%%%%%%%%%%%%%%%%%%%%%%%%%%%%%%%%%%%%%%%%%%%%%%%%%%%%%%%%%%%%%%%%%%%%%%%%%%%%%%%%%%%%%%%%%%%%%%%%%%%%%%%%%%%%%%%%%%%%%%%%%%%%%%%%%%%%%%%%%%%%%%%%%%%%%%%%%%%%%%%%%%%%%%%%%%%%%%%%%%%%%%%%%%%%%%%%%%%%%%%%%%%%%%%%%%%%%%%%%%%%%%%%%%%%%%%%%%%%%%%%%%%%%%%%%%%%%%%%%%%%%%%%%%%%%%%%%%%%%%%%%%%%%%%%%%%%%%%%%%%%%%%%%%%%%%%%%%%%%%%%%%%%%%%%%%%%%%%%%%%%%%%%%%%%%%%%%%%%%%%%%%%%%%%%%%%%%%%%%%%%%%%%%%%%%%%%%%%%%%%%%%%%%%%%%%%%%%%%%%%%%%%%%%%%%%%%%%%%%%%%%%%%%%%%%%%%%%%%%%%%%%%%%%%%%%%%%%%%%%%%%%%%%%%%%%%%%%%%%%%%%%%%%%%%%%%%%%%%%%%%%%%%%%%%%%%%%%%%%%%%%%%%

\subsection{The evaporation process}

To extend our earlier findings, this section explores the evaporation process and estimates the lifetime of the bumblebee black hole in the \textit{metric--affine} formalism, with particular attention to the effects of the Lorentz--violating parameter $X$. Additionally, a comparison with the Schwarzschild case is presented to emphasize the differences between these models. As shown in Ref. \cite{araujo2024gravitational}, the Hawking temperature is given by:
\ie
\nonumber
T = \frac{1}{8 \pi M} - \frac{X}{16 (\pi M)}.
\fe
A key factor to consider is the black hole's lifetime. To investigate this, we proceed as in the previous section by employing the \textit{Stefan--Boltzmann} law. The cross--section, \(\sigma\), is expressed as \cite{araujo2024gravitational}:
\ie
\sigma = 27 \pi M^2 \sqrt{\left(\frac{3 X}{4}+1\right) \left(1-\frac{X}{4}\right)},
\fe
and
\ie
\begin{split}
\frac{\mathrm{d}M}{\mathrm{d}\tau} = -\frac{27 \gamma (X-2)^4 \sqrt{-((X-4) (3 X+4))}}{262144 \pi ^3 M^2}
\end{split}
\fe
with $\gamma = a \alpha$. In this manner, we have 
\ie
\begin{split}
&\int_{0}^{t_{\text{evap}}} \Upsilon \mathrm{d}\tau \\
& = - \int_{M_{i}}^{M_{f}} 
\left[   -\frac{27 \gamma (X-2)^4 \sqrt{-((X-4) (3 X+4))}}{262144 \pi^3 M^2} \right]^{-1} \mathrm{d}M.
\end{split}
\fe
Here, $M_{i}$ represents the initial mass, and $M_{f}$ denotes the final mass of the black hole, while $t_{\text{met--aff}}$ refers to the time marking the end of the evaporation process. This integral, as shown below, can be solved analytically \cite{heidari2024scattering}
\ie
\begin{split}
t_{\text{met--aff}}  = \frac{262144 \pi^3 \left( M_{i}^3 -M_{f}^3 \right)}{81 \gamma (X-2)^4 \sqrt{-((X-4) (3 X+4))}}.
\end{split}
\fe
Notice that by taking the limit $\lim\limits_{X \to 0} t_{\text{evap}}$, the evaporation lifetime for the Schwarzschild black hole is recovered
\ie
t_{\text{Schw}} = \frac{4096 \pi^3 \left(Mi^3 -M_{f}^3 \right)}{81 \gamma}.
\fe

\begin{figure}
    \centering
      \includegraphics[scale=0.6]{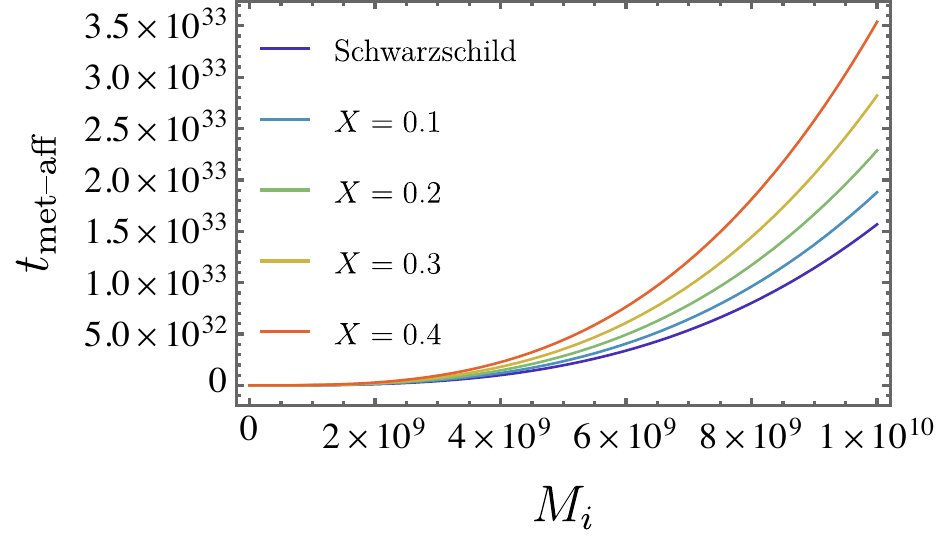}
    \caption{The evaporation time \( t_{\text{met--aff}} \) is shown for different values of \( X \). The Schwarzschild case is also compared.}
    \label{tmetaff}
\end{figure}

Fig. \ref{tmetaff} illustrates how Lorentz violation influences the evaporation process, with natural units applied for simplicity. Overall, as $X$ increases, $t_{\text{met--aff}}$ becomes larger. For reference, the Schwarzschild solution is included in the analysis.

\begin{figure}
    \centering
      \includegraphics[scale=0.55]{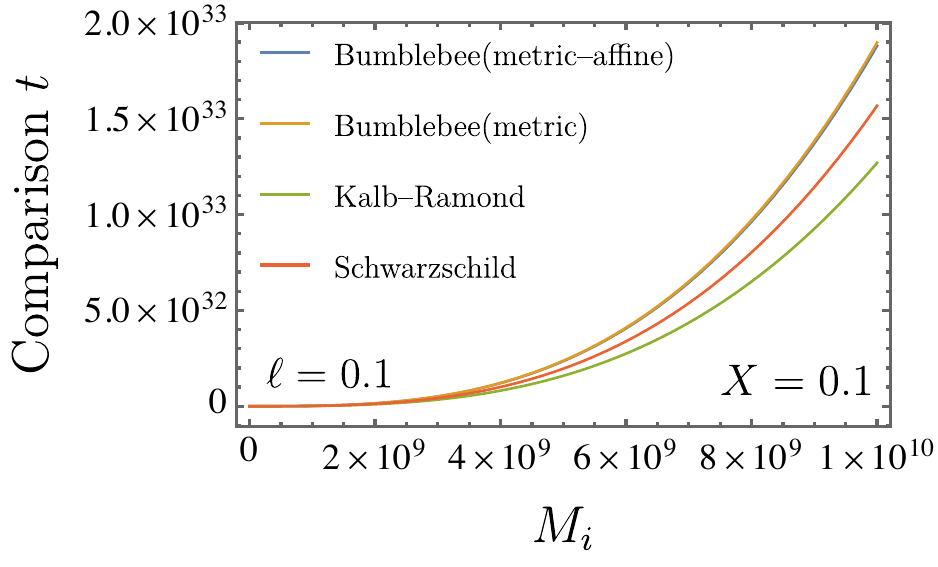}
    \caption{The blackhole lifetime comparison is shown for bumblebee (\textit{metric} and \textit{metric--affine} cases), Kalb--Ramond and Schwarzschild.}
    \label{tcomp}
\end{figure}

The comparison extends to include the \textit{metric} case, the Schwarzschild black hole, and another Lorentz--violating solution proposed in the literature—the Kalb--Ramond black hole. These findings are summarized in Fig. \ref{tcomp}. The results reveal the following order of evaporation times: $t_{\text{KR}} < t_{\text{Schw}} < t_{\text{met--aff}} < t_{\text{metric}}$. This indicates that the Kalb--Ramond black hole evaporates the fastest, while the bumblebee black hole within the \textit{metric} formalism takes the longest to evaporate.

Finally, we compare the lifetimes of the two black hole configurations analyzed in this study, as follows
\ie
\frac{ t_{\text{metric}}}{ t_{\text{met--aff}}} = \frac{(\ell+1)^2 (X-2)^4 \sqrt{-((X-4) (3 X+4))}}{64}.
\fe
To analyze the above expression, we assign specific values of $X = 0.1$ and $\ell = 0.1$. Consequently, we find
\ie
t_{\text{metric}} = 1.00899 \times t_{\text{met--aff}},
\fe
which aligns with Fig. \ref{tcomp}, confirming that $t_{\text{metric}}$ corresponds to a faster evaporation process compared to $t_{\text{met--aff}}$.

%%%%%%%%%%%%%%%%%%%%%%%%%%%%%%%%%%%%%%%%%%%%%%%%%%%%%%%%%%%%%%%%%%%%%%%%%%%%%%%%%%%%%%%%%%%%%%%%%%%%%%%%%%%%%%%%%%%%%%%%%%%%%%%%%%%%%%%%%%%%%%%%%%%%%%%%%%%%%%%%%%%%%%%%%%%%%%%%%%%%%%%%%%%%%%%%%%%%%%%%%%%%%%%%%%%%%%%%%%%%%%%%%%%%%%%%%%%%%%%%%%%%%%%%%%%%%%%%%%%%%%%%%%%%%%%%%%%%%%%%%%%%%%%%%%%%%%%%%%%%%%%%%%%%%%%%%%%%%%%%%%%%%%%%%%%%%%%%%%%%%%%%%%%%%%%%%%%%%%%%%%%%%%%%%%%%%%%%%%%%%%%%%%%%%%%%%%%%%%%%%%%%%%%%%%%%%%%%%%%%%%%%%%%%%%%%%%%%%%%%%%%%%%%%%%%%%%%%%%%%%%%%%%%%%%%%%%%%%%%%%%%%%%%%%%%%%%%%%%%%%%%%%%%%%%%%%%%%%%%%%%%%%%%%%%%%%%%%%%%%%%%%%%%%%%%%%%%%%%%%%%%%%%%%%%%%%%%%%%%%%%%%%%%%%%%%%%%%%%%%%%%%%%%%%%%%%%%%%%%%%%%%%%%%%%%%%%%%%%%%%%%%%%%%%%%%%%%%%%%%%%%%%%%%%%%%%%%%%%%%%%%%%%%%%%%%%%%%%%%%%%%%%%%%%%%%%%%

\section{Correlation between greybody
factors and quasinormal modes}

The WKB approximation offers a semi-analytical framework well--suited for estimating quasinormal mode frequencies. Given the intricate form of the lapse function encountered in the numerical treatment, the analysis employs the third--order WKB technique, which yields the quasinormal modes spectrum through the following expression \cite{iyer1987black,iyer1987black2,konoplya2019higher,konoplya2011quasinormal,cardoso2001quasinormal}
\begin{equation}\label{3wkb}
\begin{split}
    {\omega ^2}=&  \,\,{V_0} + \sqrt { - 2{V_0}^{\prime \prime }} \Lambda (n) - i\left(n + \frac{1}{2}\right)\sqrt { - 2{V_0}^{\prime \prime }} (1 + \Omega (n)),
\end{split}
 \end{equation}
with 
\begin{equation}
\begin{split}
    \Lambda (n) =   \frac{1}{{\sqrt { - 2{V_0}^{\prime \prime }} }}\left[\frac{1}{8}\left(\frac{{V_0^{\left(4\right)}}}{{{V_0}^{\prime \prime }}}\right)\left(\frac{1}{4} + {\alpha ^2}\right) - \frac{1}{{288}}{\left(\frac{{{V_0}^{\prime \prime \prime }}}{{{V_0}^{\prime \prime }}}\right)^2}\left(7 + 60{\alpha ^2}\right)\right],
\end{split}
\end{equation}
and 
\begin{equation}
     \begin{split}
    \Omega (n) = & \left( {\frac{1}{{ - 2{V_0}^{\prime \prime }}}} \right)
  \frac{5}{{6912}}{\left(\frac{{{V_0}^{\prime \prime \prime }}}{{{V_0}^{\prime \prime }}}\right)^4}\left( {77 + 188 \times {\alpha ^2}} \right)  - \frac{1}{{384}} \times \left( {\frac{{{V_0}{{^{\prime \prime \prime }}^2}V_0^{(4)}}}{{{V_0}{{^{\prime \prime }}^3}}}} \right)\left( {51 + 100{\alpha ^2}} \right)  \\
   &+ \frac{1}{{2304}}{\left(\frac{{V_0^{(4)}}}{{{V_0}^{\prime \prime }}}\right)^2}\left( {67 + 68{\alpha ^2}} \right)  + \frac{1}{{288}}\left( {\frac{{{V_0}^{\prime \prime \prime }V_0^{(5)}}}{{{V_0}{{^{\prime \prime }}^2}}}} \right)\left( {19 + 28{\alpha ^2}} \right)  - \frac{1}{{288}}\left(\frac{{V_0^{(6)}}}{{{V_0}^{\prime \prime }}}\right)\left( {5 + 4{\alpha ^2}} \right).
\end{split} 
\end{equation}
Here, $\alpha = n + \frac{1}{2}$, where $n$ denotes the overtone number, subject to the condition $n \leq l$.

An approximate relationship between greybody factors and quasinormal modes was recently proposed in Ref.~\cite{konoplya2024correspondence}, shown to become exact in the high--frequency (eikonal) regime. In this limit, the greybody factors for spherically symmetric black holes simplify considerably and can be characterized using only the fundamental quasinormal modes. For smaller angular momentum values ($l$), however, additional corrections involving overtone contributions become relevant. Meanwhile, within the WKB formalism, the transmission and reflection coefficients are determined through the expression below \cite{iyer1987black}
\begin{equation}
{\left| R \right|^2} = \frac{1}{{1 + {e^{ - 2\pi i{\mathcal{K}}}}}},
\end{equation}
\begin{equation}\label{Trans}
{\left| T \right|^2} = \frac{1}{{1 + {e^{  2\pi i{\mathcal{K}}}}}}.
\end{equation}

As outlined in Ref.~\cite{konoplya2024correspondence}, the quantity $\mathcal{K}$ is defined in terms of the two leading quasinormal frequencies, labeled $\omega_0$ and $\omega_1$, which correspond to the fundamental ($n = 0$) and first overtone ($n = 1$) modes. Each frequency $\omega$ consists of a real part $\omega_R$ and an imaginary part $\omega_I$, representing oscillation and damping, respectively
\begin{equation}\label{Tfactor}
     - i{\mathcal{K}} =  - \frac{{{\omega ^2} - {\omega _{0R}}^2}}{{4{\omega _{0R}}{\omega _{0I}}}} + {\Delta _1} + {\Delta _2} + {\Delta _f},
\end{equation}
with
\ie
{\Delta _1}  = \frac{{{\omega _{0R}} - {\omega _{1R}}}}{{16{\omega _{0I}}}},
\fe
\ie
\begin{split}
    \Delta_2 & =  - \frac{{{\omega ^2} - \omega_{0R}^2}}{{32{\omega _{0R}}{\omega _{0I}}}}\left[\frac{{{{({\omega _{0R}} - {\omega _{R1}})}^2}}}{{4{\omega _{0I}}^2}} - \frac{{3{\omega _{0I}} - {\omega _{1I}}}}{{3{\omega _{0I}}}}\right] + \frac{{{{({\omega ^2} - \omega _{0R}^2)}^2}}}{{16\omega _{0R}^3{\omega _{0I}}}}\left[1 + \frac{{{\omega _{0R}}({\omega _{0R}} - {\omega _{1R}})}}{{4\omega _{0I}^2}}\right], 
\end{split}
\fe
and
\ie
\begin{split}
\Delta_f &=  - \frac{{{{({\omega ^2} - \omega _{0R}^2)}^3}}}{{32\omega _{0R}^5{\omega _{0I}}}}\left\{1 + \frac{{{\omega _{0R}}({\omega _{0R}} - {\omega _{1R}})}}{{4{\omega _{0I}}^2}}  + \omega _{0R}^2\left[\frac{{{{({\omega _{0R}} - {\omega _{1R}})}^2}}}{{16\omega _{0I}^4}} - \frac{{3{\omega _{0I}} - {\omega _{1I}}}}{{12{\omega_{0I}}}}\right]\right\}.
\end{split}
\fe

In the following subsections, we apply the previously developed framework to compute the greybody factors associated with scalar, vector, and tensor perturbations, considering both the \textit{metric} and \textit{metric--affine} formulations. This is achieved by evaluating $\mathcal{K}$ via Eq. (\ref{Trans}). The dominant quasinormal frequencies required for this procedure are determined using the third--order WKB approximation, as described by Eq.~(\ref{3wkb}). For the plots presented in the subsequent analysis, we adopt the notation $\Gamma(\omega) = |T|^2$ to represent the greybody factor.

%%%%%%%%%%%%%%%%%%%%%%%%%%%%%%%%%%%%%%%%%%%%%%%%%%%%%%%%%%%%%%%%%%%%%%%%%%%%%%%%%%%%%%%%%%%%%%%%%%%%%%%%%%%%%%%%%%%%%%%%%%%%%%%%%%%%%%%%%%%%%%%%%%%%%%%%%%%%%%%%%%%%%%%%%%%%%%%%%%%%%%%%%%%%%%%%%%%%%%%%%%%%%%%%%%%%%%%%%%%%%%%%%%%%%%%%%%%%%%%%%%%%%%%%%%%%%%%%%%%%%%%%

\subsection{The \textit{metric} case}

%%%%%%%%%%%%%%%%%%%%%%%%%%%%%%%%%%%%%%%%%%%%%%%%%%%%%%%%%%%%%%%%%%%%%%%%%%%%%%%%%%%%%%%%%%%%%%%%%%%%%%%%%%%%%%%%%%%%%%%%%%%%%%%%%%%%%%%%%%%%%%%%%%%%%%%%%%%%%%%%%%%%%%%%%%%%%%%%%%%%%%%%%%%%%%%%%%%%%%%%%%%%%%%%%%%%%%%%%%%%%%%%%%%%%%%%%%%%%%%%%%%%%%%%%%%%%%%%%%%%%%%%

\subsubsection{Scalar perturbations}

In this subsection, the correspondence between the greybody factors and the quasinormal modes is derived specifically for scalar perturbations in the \textit{metric} case. To this end, we employ Eq.~(\ref{scalarpotentialmetric}) to carry out the calculations.

\begin{figure}
    \centering
      \includegraphics[scale=0.55]{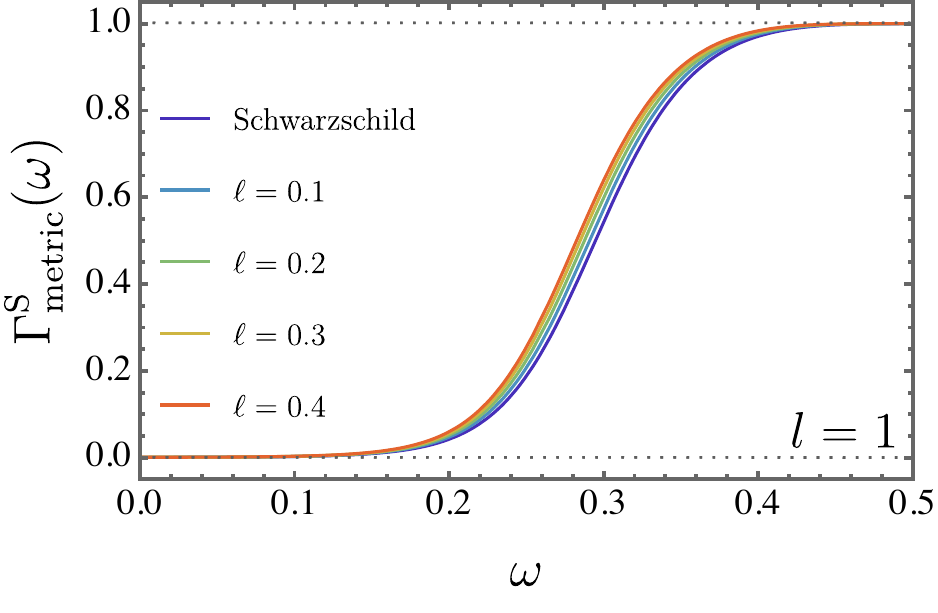}
    \caption{The greybody factors $\Gamma^{\text{S}}_{\text{metric}}(\omega)$ for scalar perturbations, computed from the quasinormal modes, are presented as functions of the frequency $\omega$ for $l = 1$ and $M = 1$.}
    \label{scalarl1metric}
\end{figure}

Fig. \ref{scalarl1metric} presents the greybody factors $\Gamma^{\text{S}}_{\text{metric}}(\omega)$ as functions of the frequency $\omega$ for $l = 1$ and $M = 1$, including a comparison with the Schwarzschild background. When computed via quasinormal modes, notable differences appear. The frequency domain is reduced compared to Fig. \ref{greybodymetricbosons}, and an increase in $\ell$ leads to a corresponding rise in $\Gamma^{\text{S}}_{\text{metric}}(\omega)$, in contrast to the behavior observed earlier. This effect might be associated with the decrease in the real part of the quasinormal frequencies relative to the Schwarzschild case as $\ell$ increases.

%%%%%%%%%%%%%%%%%%%%%%%%%%%%%%%%%%%%%%%%%%%%%%%%%%%%%%%%%%%%%%%%%%%%%%%%%%%%%%%%%%%%%%%%%%%%%%%%%%%%%%%%%%%%%%%%%%%%%%%%%%%%%%%%%%%%%%%%%%%%%%%%%%%%%%%%%%%%%%%%%%%%%%%%%%%%%%%%%%%%%%%%%%%%%%%%%%%%%%%%%%%%%%%%%%%%%%%%%%%%%%%%%%%%%%%%%%%%%%%%%%%%%%%%%%%%%%%%%%%%%%%%

\subsubsection{Vector perturbations}

As discussed in the previous section and consistent with Eq.~(\ref{vectorpotentialmetric}), the Lorentz--violating contribution to the effective potential for vector perturbations vanishes, implying that only the trivial contribution remains.

%%%%%%%%%%%%%%%%%%%%%%%%%%%%%%%%%%%%%%%%%%%%%%%%%%%%%%%%%%%%%%%%%%%%%%%%%%%%%%%%%%%%%%%%%%%%%%%%%%%%%%%%%%%%%%%%%%%%%%%%%%%%%%%%%%%%%%%%%%%%%%%%%%%%%%%%%%%%%%%%%%%%%%%%%%%%%%%%%%%%%%%%%%%%%%%%%%%%%%%%%%%%%%%%%%%%%%%%%%%%%%%%%%%%%%%%%%%%%%%%%%%%%%%%%%%%%%%%%%%%%%%%

\subsubsection{Tensor perturbations}

The analysis now turns to tensor perturbations within the \textit{metric} framework, aiming to establish the connection between greybody factors and quasinormal modes. To carry out this computation, Eq.~(\ref{tensorpotentialmetric}) is employed.

\begin{figure}
    \centering
      \includegraphics[scale=0.55]{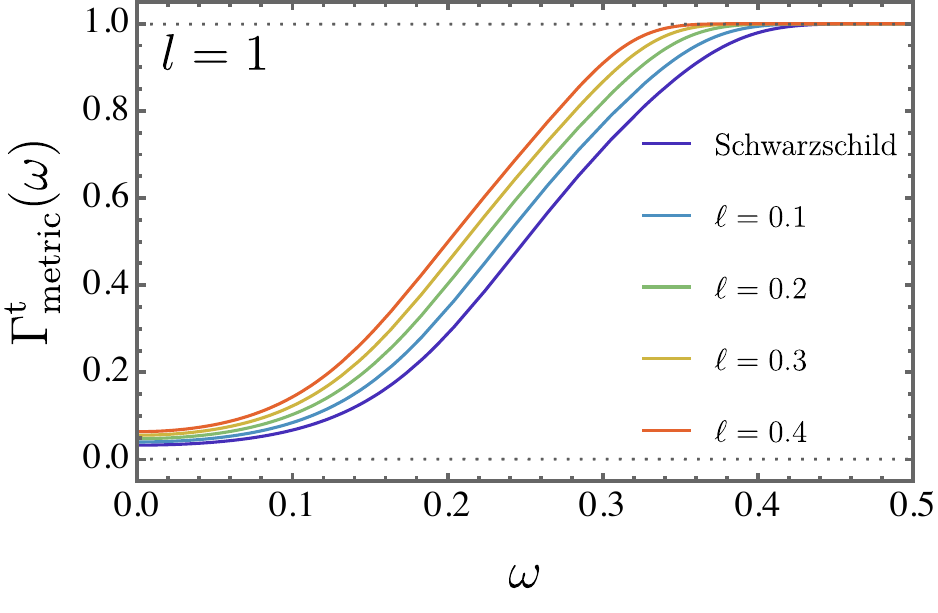}
    \caption{The greybody factors $\Gamma^{\text{t}}_{\text{metric}}(\omega)$ for tensor perturbations, obtained from quasinormal modes, are plotted as functions of the frequency $\omega$ for $l = 1$ and $M = 0.5$.}
    \label{tensor1metric}
\end{figure}

In Fig. \ref{tensor1metric}, the greybody factors $\Gamma^{\text{t}}_{\text{metric}}(\omega)$ are shown as functions of the frequency $\omega$ for $l = 1$ and $M = 0.5$, along with a comparison to the Schwarzschild background. When computed via quasinormal modes, the results reveal significant differences, including a reduced frequency range compared to earlier analyses. A notable feature is the behavior under variations of $\ell$: as $\ell$ increases, the greybody factors also increase—opposite to the trend observed in previous cases. This behavior might be linked to the suppression of the real part of the quasinormal frequencies with increasing $\ell$, relative to the Schwarzschild case.

%%%%%%%%%%%%%%%%%%%%%%%%%%%%%%%%%%%%%%%%%%%%%%%%%%%%%%%%%%%%%%%%%%%%%%%%%%%%%%%%%%%%%%%%%%%%%%%%%%%%%%%%%%%%%%%%%%%%%%%%%%%%%%%%%%%%%%%%%%%%%%%%%%%%%%%%%%%%%%%%%%%%%%%%%%%%%%%%%%%%%%%%%%%%%%%%%%%%%%%%%%%%%%%%%%%%%%%%%%%%%%%%%%%%%%%%%%%%%%%%%%%%%%%%%%%%%%%%%%%%%%%%

\subsection{The \textit{metric--affine} case}

%%%%%%%%%%%%%%%%%%%%%%%%%%%%%%%%%%%%%%%%%%%%%%%%%%%%%%%%%%%%%%%%%%%%%%%%%%%%%%%%%%%%%%%%%%%%%%%%%%%%%%%%%%%%%%%%%%%%%%%%%%%%%%%%%%%%%%%%%%%%%%%%%%%%%%%%%%%%%%%%%%%%%%%%%%%%%%%%%%%%%%%%%%%%%%%%%%%%%%%%%%%%%%%%%%%%%%%%%%%%%%%%%%%%%%%%%%%%%%%%%%%%%%%%%%%%%%%%%%%%%%%%

\subsubsection{Scalar perturbations}

We now explore how greybody factors relate to quasinormal modes in the context of scalar perturbations governed by the \textit{metric--affine} geometry. The starting point for this analysis is the scalar potential given by Eq.~(\ref{scalarpotentialmetricaffine}), which forms the basis for the calculations that follow.

\begin{figure}
    \centering
      \includegraphics[scale=0.55]{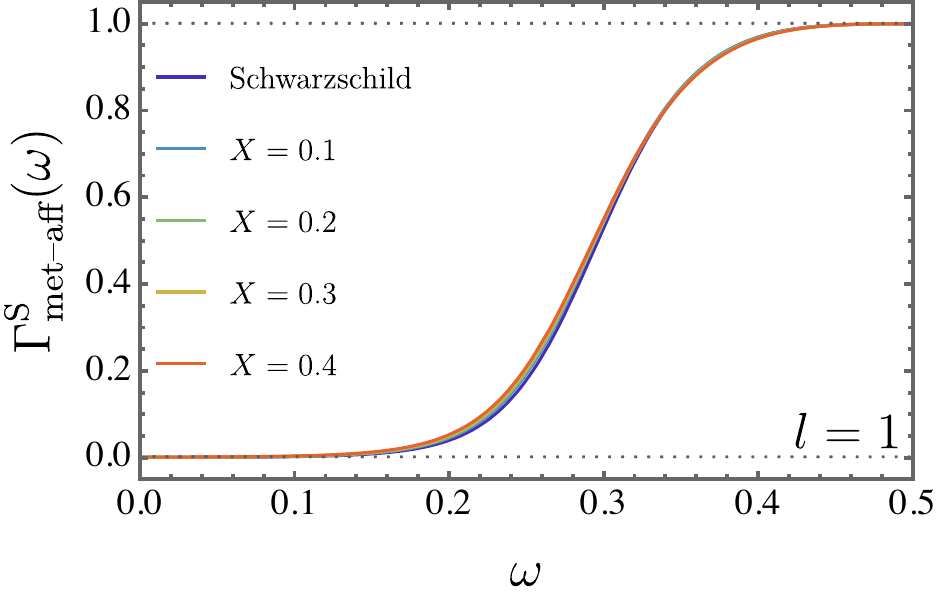}
    \caption{The greybody factors $\Gamma^{\text{S}}_{\text{met--aff}}(\omega)$ for scalar perturbations are obtained from quasinormal modes and plotted as functions of the frequency $\omega$, considering $l = 1$ and $M = 1$.}
    \label{scalarl1metricaffine}
\end{figure}

Fig. \ref{scalarl1metricaffine} displays the greybody factors $\Gamma^{\text{S}}_{\text{met--aff}}(\omega)$ as functions of the frequency $\omega$ for $l = 1$ and $M = 1$, with the Schwarzschild case included for comparison. When calculated through quasinormal modes, noticeable differences arise. The frequency domain, for instance, is reduced compared to the results shown in Fig.~\ref{greybodymetricaffinebosons}. Moreover, the influence of the parameter $X$ on $\Gamma^{\text{S}}_{\text{met--aff}}(\omega)$ depends on the frequency range: for $\omega \lesssim 0.336$, $X$ enhances the magnitude of the greybody factor, while for $\omega \gtrsim 0.336$, it leads to a suppression. This feature marks an inflection point where the effect of $X$ reverses, deviating from the behavior observed earlier for tensor perturbations in the \textit{metric--affine} case and in the Schwarzschild background.

%%%%%%%%%%%%%%%%%%%%%%%%%%%%%%%%%%%%%%%%%%%%%%%%%%%%%%%%%%%%%%%%%%%%%%%%%%%%%%%%%%%%%%%%%%%%%%%%%%%%%%%%%%%%%%%%%%%%%%%%%%%%%%%%%%%%%%%%%%%%%%%%%%%%%%%%%%%%%%%%%%%%%%%%%%%%%%%%%%%%%%%%%%%%%%%%%%%%%%%%%%%%%%%%%%%%%%%%%%%%%%%%%%%%%%%%%%%%%%%%%%%%%%%%%%%%%%%%%%%%%%%%

\subsubsection{Vector perturbations}

%%%%%%%%%%%%%%%%%%%%%%%%%%%%%%%%%%%%%%%%%%%%%%%%%%%%%%%%%%%%%%%%%%%%%%%%%%%%%%%%%%%%%%%%%%%%%%%%%%%%%%%%%%%%%%%%%%%%%%%%%%%%%%%%%%%%%%%%%%%%%%%%%%%%%%%%%%%%%%%%%%%%%%%%%%%%%%%%%%%%%%%%%%%%%%%%%%%%%%%%%%%%%%%%%%%%%%%%%%%%%%%%%%%%%%%%%%%%%%%%%%%%%%%%%%%%%%%%%%%%%%%%

This subsection examines the connection between greybody factors and quasinormal modes in the context of vector perturbations within the \textit{metric--affine} framework. The analysis is based on Eq.~(\ref{vectormetricaffinecase}), which provides the necessary effective potential for carrying out the calculations.

\begin{figure}
    \centering
      \includegraphics[scale=0.55]{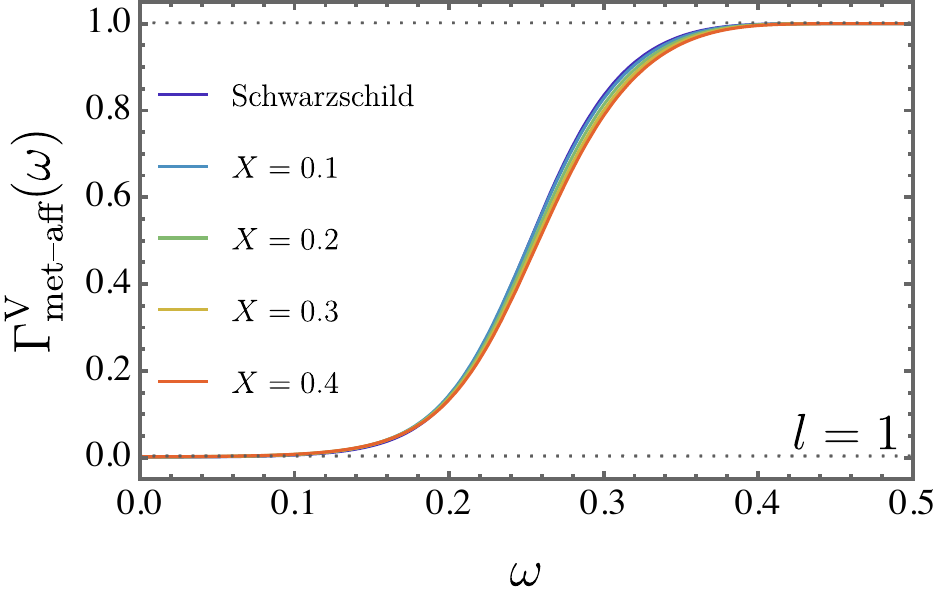}
    \caption{The greybody factors $\Gamma^{\text{V}}_{\text{met--aff}}(\omega)$ for vector perturbations are computed from quasinormal modes and displayed as functions of the frequency $\omega$, considering $l = 1$ and $M = 1$.}
    \label{vectorl1metricaffine}
\end{figure}

Fig. \ref{vectorl1metricaffine} displays the greybody factors $\Gamma^{\text{V}}_{\text{met--aff}}(\omega)$ as functions of the frequency $\omega$ for $l = 1$ and $M = 1$, with the Schwarzschild case shown for comparison. The frequency range is more restricted than in earlier analyses. Nevertheless, the effect of the parameter $X$ remains consistent: increasing $X$ leads to a reduction in the magnitude of the greybody factors. This outcome agrees with the previously obtained results for vector perturbations in the \textit{metric--affine} framework, where larger values of $X$ lower the transmission probability when compared to the Schwarzschild scenario.

\subsubsection{Tensor perturbations}

Finally, the correspondence between greybody factors and quasinormal modes is now investigated in the context of tensor perturbations within the \textit{metric--affine} setting. The calculation relies on the effective potential defined by Eq.~(\ref{scalarpotentialmetafftensor}), which serves as the foundation for the analysis.

\begin{figure}
    \centering
      \includegraphics[scale=0.55]{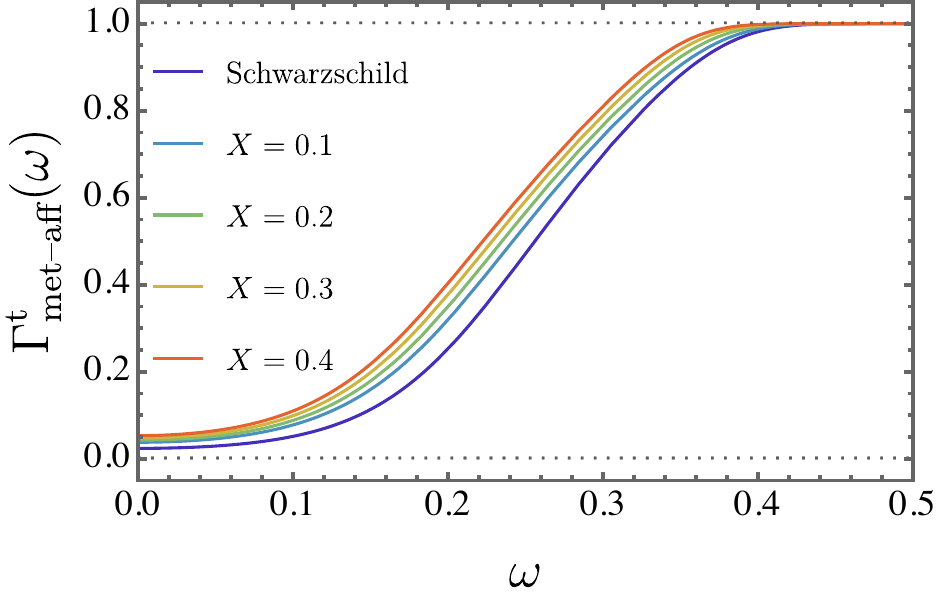}
    \caption{The greybody factors $\Gamma^{\text{t}}_{\text{met--aff}}(\omega)$ for tensor perturbations are computed from quasinormal modes and shown as functions of the frequency $\omega$, with parameters $l = 1$ and $M = 0.5$.}
    \label{tensor1metricaffine}
\end{figure}

Fig. \ref{tensor1metricaffine} presents the behavior of the greybody factors $\Gamma^{\text{t}}_{\text{met--aff}}(\omega)$ as functions of the frequency $\omega$ for $l = 1$ and $M = 0.5$, including a comparison with the Schwarzschild configuration. The spectrum appears more limited in frequency range than the one shown in Fig.~\ref{greybodymetricaffinebosonstensor233}. Another important feature is the influence of the parameter $X$: in contrast to earlier results, increasing $X$ amplifies the greybody factors, indicating a higher transmission probability relative to the Schwarzschild case.

%%%%%%%%%%%%%%%%%%%%%%%%%%%%%%%%%%%%%%%%%%%%%%%%%%%%%%%%%%%%%%%%%%%%%%%%%%%%%%%%%%%%%%%%%%%%%%%%%%%%%%%%%%%%%%%%%%%%%%%%%%%%%%%%%%%%%%%%%%%%%%%%%%%%%%%%%%%%%%%%%%%%%%%%%%%%%%%%%%%%%%%%%%%%%%%%%%%%%%%%%%%%%%%%%%%%%%%%%%%%%%%%%%%%%%%%%%%%%%%%%%%%%%%%%%%%%%%%%%%%%%%%

\section{Constraints based on the evaporation lifetime}

In this section, limits on the Lorentz--violating parameters $\ell$ (associated with the \textit{metric} formulation) and $X$ (corresponding to the \textit{metric--affine} scenario) are inferred by drawing upon recent astrophysical analyses concerning the expected lifetimes of black holes, as explored in Ref.~\cite{Ewasiuk:2024ctc}. To carry out this analysis, we observe that assuming small Lorentz--violating parameters in Eq. (\ref{hawtempmetricase}) (for the \textit{metric} case) and in Eq. (\ref{hawtemp}) (for the \textit{metric--affine} case) leads to similar expressions/structures: $\frac{1}{8\pi M} - \frac{\ell}{16\pi M}$ and $\frac{1}{8\pi M} - \frac{X}{16\pi M}$, respectively. This feature implies that the estimation of bounds based on the evaporation lifetime remains identical (at least for the high--frequency regime) whether one considers either the \textit{metric} or the \textit{metric--affine} formulation within bumblebee gravity.

An important consideration regarding the constraints outlined above is the possibility that black holes might have initially formed with larger masses, gradually losing mass through evaporation to reach the values observed at the time of binary coalescence. Although this requires fine--tuning, the scenario remains plausible and merits examination.

To place bounds on significant evaporation rates using existing observational data, three main strategies can be considered, as discussed in Ref.~\cite{Ewasiuk:2024ctc}. One key approach involves examining the energy balance during binary mergers: if evaporation contributes a power loss $\dot E_{\rm EV}$ comparable to that from gravitational wave emission $\dot E_{\rm GW}$, i.e., $\dot E_{\rm EV} \sim \dot E_{\rm GW}$, the dynamics of the inspiral and merger would be noticeably altered. Such a deviation could manifest as modifications in the gravitational wave signal or might even prevent the merger from occurring. A conservative condition to avoid this issue is to impose $\dot E_{\rm EV} < \dot E_{\rm GW}$.
    
If one assumes that the luminosity of X--ray binaries containing black holes--during their quiescent phase—scales with the Eddington rate, and thus with the black hole mass itself, then any variation in luminosity $\Delta L/L$ observed over a time span $\Delta t$ can be directly linked to a corresponding variation in mass. Under this assumption, the fractional change in mass follows the relation $\Delta m/m \sim \Delta L/L$, leading to the estimate 
\ie
\frac{1}{M} \frac{{\rm d}M}{{\rm d}t}\equiv\frac{\dot M}{M}\sim\frac{\Delta M}{M\Delta t}\lesssim \frac{\Delta L}{L}.
\fe

A more general strategy involves examining black hole mass estimates of the form $M_{\rm BH} = m \pm \Delta m$, derived from diverse observational datasets collected over a period $\Delta t$. By tracking possible shifts in mass across this interval, one can place limits on any underlying mass loss mechanisms or time--dependent deviations from expected behavior
    \begin{equation}
      \frac{\dot M}{M}\lesssim \frac{\Delta m}{m}\frac{1}{\Delta t}.
    \end{equation}

To establish the first constraint, consider a simplified scenario involving a binary system of equal--mass black holes, each with mass $M$, in a circular orbit of radius $R$. In this setup, the energy radiated through gravitational waves is given by the following expression \cite{Ewasiuk:2024ctc}:
\ie
\frac{{\rm d}E_{\rm GW}}{{\rm d}t}=-\frac{2}{5}\frac{G_N^4}{c^5}\frac{M^5}{R^5}.
\fe
In contrast, the energy loss associated with mass evaporation from the black holes is given by:
\ie
\frac{{\rm d}U}{{\rm d}t}=\frac{\rm d}{{\rm d}t}\left(-\frac{G_NM^2}{R}\right)=-\frac{2M^2G_N}{R}\frac{\dot M}{M}.
\fe

By comparing the two expressions for gravitational wave emission and mass evaporation, one obtains the following relation:
\ie
    \frac{\dot M}{M}<\frac{G_N^3M^3}{5R^4}. 
\fe
Utilizing the relation between angular frequency $\Omega$, orbital radius $R$, and total mass $M$ for a circular binary system, namely
\ie
    G_N M=4R^3\Omega^2,
\fe
we therefore get
\ie
    \frac{\dot M}{M}<\frac{\Omega^{8/3}}{c^5}\left(G_N M\right)^{5/3}. 
\fe

By applying the previous relation with $M \sim 10\ M_\odot$ and $\Omega \sim 10^3$ Hz, one obtains:
\ie
\label{GW}
    \left(\frac{\dot M}{M}\right)_{\rm GW}\lesssim 10\ {\rm Hz},
\fe
which yields the following constraint on the parameters $\ell$ and $X$:
\ie
\ell, X \lesssim  10^{-25}.
\fe

The second constraint arises from observational data on BW Cir, as detailed in Table 2 of Ref.~\cite{casares2009refined}. During its quiescent phase, the system’s luminosity remains steady within roughly a 10\% margin over a duration of 14 years, corresponding to a time interval of about $\Delta t \sim 4 \times 10^8$ seconds. This long--term luminosity stability can be used to impose restrictions on
\ie
    \left(\frac{\dot M}{M}\right)_{\rm lum}\lesssim\frac{\Delta L}{L}\frac{1}{\Delta t}\sim 2\times 10^{-10}\ {\rm Hz},
\fe
and, therefore, we have
\ie
\ell, X \lesssim  10^{-36}.
\fe

The final strategy relies on compiling data from the full catalog of confirmed X-ray binaries—both within the Milky Way and beyond—as detailed in Refs.~\cite{filippenko1999black, Plotkin:2021rzl}, and including black holes identified via gravitational wave detections. To identify optimal systems for placing constraints, we focus on those with minimal fractional mass uncertainty, $\Delta m/m$. Among the most precise cases is Sgr A*, with a mass known to better than 0.3\%. Other promising candidates include Gaia BH1 and the GRS 1009--45 (also known as Nova Velorum 1993 or MM Velorum) system, both exhibiting uncertainties below 3\%. While these latter black holes are less massive than Sgr A*, they are still valuable for this analysis despite the shorter observational baselines. Taking, for instance, the Nova Velorum 1993 system—with a reported central mass estimate of $4.3\ M_\odot$ and an observational timescale of approximately one year \cite{Ewasiuk:2024ctc}—we find:
\ie
    \left(\frac{\dot M}{M}\right)_{\rm XRB}\lesssim 10^{-9}\ {\rm Hz},
\fe
which leads, in the \textit{metric} framework, to the following expression:
\ie
\ell, X \lesssim   10^{-35}.
\fe

In contrast, for Sgr A*, assuming an observational baseline of 15 years, we obtain:
\ie
\left(\frac{\dot M}{M}\right)_{\rm SgrA*}\lesssim 6\times 10^{-12}\ {\rm Hz}
\fe
and consequently, in the context of the \textit{metric} formulation, this implies:
\ie
\ell, X \lesssim   10^{-38}.
\fe

A summary of all constraints on Lorentz--violating parameters derived from black hole lifetime considerations in this study is provided in Tab. ~\ref{tab:LVbounds}.

\begin{table}[h!]
\centering
\caption{Bounds on the Lorentz--violating parameters $\ell$ (\textit{metric} case) and $X$ (\textit{metric--affine} case) from astrophysical constraints on black hole evaporation.}
\label{tab:LVbounds}
\begin{tabular}{lc}
\hline\hline
\textbf{Astrophysical Constraints} &   \\
\hline
\textbf{GW} & $\ell, X \lesssim  10^{-25}$, \\[0.5em]
\textbf{Luminosity} & $\ell, X \lesssim 10^{-36}$, \\[0.5em]
\textbf{XRB} & $\ell, X \lesssim 10^{-35}$, \\[0.5em]
\textbf{SgrA$^*$} & $\ell, X \lesssim 10^{-38}$, \\
\hline\hline
\end{tabular}
\end{table}

%%%%%%%%%%%%%%%%%%%%%%%%%%%%%%%%%%%%%%%%%%%%%%%%%%%%%%%%%%%%%%%%%%%%%%%%%%%%%%%%%%%%%%%%%%%%%%%%%%%%%%%%%%%%%%%%%%%%%%%%%%%%%%%%%%%%%%%%%%%%%%%%%%%%%%%%%%%%%%%%%%%%%%%%%%%%%%%%%%%%%%%%%%%%%%%%%%%%%%%%%%%%%%%%%%%%%%%%%%%%%%%%%%%%%%%%%%%%%%%%%%%%%%%%%%%%%%%%%%%%%%%%

\section{Conclusion}

In this work, we examined the influence of non--metricity on particle creation and black hole evaporation within the framework of bumblebee gravity. Specifically, we compared black hole solutions in the \textit{metric} formalism \cite{14} and the \textit{metric--affine} approach \cite{filho2023vacuum} to identify the role of non--metricity in these processes. Furthermore, we contrasted our findings with the Schwarzschild solution and a recently proposed Lorentz--violating solution, the Kalb--Ramond black hole \cite{Liu:2024oas}.

The paper began by analyzing the particle creation properties of the bumblebee black hole in the \textit{metric} formalism introduced in Ref. \cite{14}. The investigation focused on bosonic particle modes, which were studied through the tunneling process. To accomplish this, the metric coordinates were redefined using the Painlevé--Gullstrand form, which removed the divergence at the horizon. As a result, the divergent integrals related to the imaginary parts of the action, \(\mathcal{S}\), specifically \(\text{Im}\,\mathcal{S}_{\text{metric}}\) and \(\text{Im}\,\mathcal{S}_{\text{met--aff}}\), were resolved using the residue method. This approach allowed for the estimation of the bosonic particle densities \(n_{\text{metric}}\) and \(n_{\text{met--aff}}\). In a general panorama, the particle density intensities presented the following relation (for \(X = \ell = 0.1\)): \(n_{\text{KB}} > n_{\text{Schw}} > n_{\text{met--aff}} > n_{\text{metric}}\).

Next, the fermionic particle modes were investigated using the tunneling process as well. In this case, the near--horizon approximation \cite{araujo2023analysis} was employed to perform the calculations, leading to the estimation of the fermionic particle densities \(n_{\psi_{\text{metric}}}\) and \(n_{\psi_{\text{met--aff}}}\). Notably, in contrast to the bosonic case, the effects of non--metricity on them turned out to be so small.

Subsequently, the greybody factors for bosonic and fermionic particles were estimated. In general lines, except for the tensor perturbations, it was found that non--metricity led to an increase in the greybody factors compared to the \textit{metric}, leading to the following relation:  $T_{b_{\text{metric}}}^{\text{t}} > T_{b_{\text{met--aff}}}^{\text{t}} > \text{Schw}^{(\text{t})} > \text{Schw}^{(\text{v})} > T_{b_{\text{met--aff}}}^{\text{v}} > T_{b_{\text{metric}}}^{\text{v}} > \text{Schw}^{(\text{s})} > T_{b_{\text{met--aff}}}^{\text{s}} > T_{b_{\text{metric}}}^{\text{s}}$.

Additionally, the emission rate was calculated, revealing that its magnitude decreased as both $\ell$ and $X$ increased. For reference, the Schwarzschild black hole was included in the analysis. Moreover, when comparing the two black hole models studied here, non--metricity was found to increase the emission rate's magnitude relative to the \textit{metric} case, although the difference between them was small.

Furthermore, the evaporation lifetime of the black holes was analytically derived for both cases. This analysis included a comparison with the Schwarzschild solution and a recently proposed Lorentz--violating black hole, the Kalb--Ramond solution \cite{Liu:2024oas}. The observed hierarchy was as follows: $ t_{\text{KR}} < t_{\text{Schw}} < t_{\text{met--aff}} < t_{\text{metric}} $. In other words, the Kalb--Ramond black hole evaporates the fastest, while the bumblebee black hole in the \textit{metric} approach evaporates the slowest.

Also, the analysis addressed the connection between quasinormal modes and greybody factors. By comparing theoretical predictions with recent astrophysical measurements of black hole lifetimes, the study placed bounds on the Lorentz--violating parameters: $\ell$ for the \textit{metric} scenario and $X$ for the \textit{metric--affine} framework.

As a future perspective, our investigation can be extended to address the entanglement degradation for the bumblebee solution in the \textit{metric--affine} formalism \cite{filho2023vacuum} similar to what was recently accomplished in the literature to the Kalb--Ramond gravity \cite{Liu:2024wpa}.

%%%%%%%%%%%%%%%%%%%%%%%%%%%%%%%%%%%%%%%%%%%%%%%%%%%%%%%%%%%%%%%%%%%%%%%%%%%%%%%%%%%%%%%%%%%%%%%%%%%%%%%%%%%%%%%%%%%%%%%%%%%%%%%%%%%%%%%%%%%%%%%%%%%%%%%%%%%%%%%%%%%%%%%%%%%%%%%%%%%%%%%%%%%%%%%%%%%%%%%%%%%%%%%%%%%%%%%%%%%%%%%%%%%%%%%%%%%%%%%%%%%%%%%%%%%%%%%%%%%%%%%%

\section*{Acknowledgments}
\hspace{0.5cm}

A. A. Araújo Filho acknowledges support from the Conselho Nacional de Desenvolvimento Científico e Tecnológico (CNPq) and the Fundação de Apoio à Pesquisa do Estado da Paraíba (FAPESQ) under grant [150891/2023-7]. Additionally, the author thanks G. J. Olmo, P. J. Porfírio, and I. P. Lobo for their valuable discussions during the revision of this paper. Special thanks are also extended to N. Heidari for the helpful discussions and assistance with the calculations related to the correlation between greybody factors and quasinormal modes.

%%%%%%%%%%%%%%%%%%%%%%%%%%%%%%%%%%%%%%%%%%%%%%%%%%%%%%%%%%%%%%%%%%%%%%%%%%%%%%%%%%%%%%%%%%
\section{Data Availability Statement}

Data Availability Statement: No Data associated in the manuscript

%%%%%%%%%%%%%%%%%%%%%%%%%%%%%%%%%%%%%%%%%%%%%%%%%%%%%%%%%%%%%%%%%%%%%%%%%%%%%%%%%%%%%%%%%%

\bibliographystyle{ieeetr}
\bibliography{main}

\end{document}